# Application of signal processing techniques in the assessment of clinical risks in preterm infants

Ying Zhang

# Abstract

Preterm infants with very low birth weight (VLBW; birth weight ≤1500 g) suffer from a high risk of intra-ventricular hemorrhage (IVH) and other serious diseases. Even for those VLBW infants who survive, they still have a strong possibility of developing long-term neurodevelopmental disabilities.

To improve the clinical risk assessment of preterm infants and develop potential clinical makers for the adverse outcome in preterm infants, the first part of the paper develops the frequency spectral analysis on the non-invasively measured heart rate variability (HRV), blood pressure variability (BPV) and cerebral near-infrared spectroscopy (NIRS) measures. Moderate and high correlations with the clinical risk index for babies (CRIB II) were identified from various spectral measures of arterial baroreflex and cerebral autoregulation functions which provided further justification for these measures as possible markers of clinical risks and predictors of adverse outcome in preterm infants. It was also observed that the cross-spectral transfer function analysis of cerebral NIRS and arterial blood pressure (ABP) was able to provide a number of parameters that were potentially useful for distinguishing preterm infants with IVH from without IVH.

Furthermore, the detrended fluctuation analysis (DFA) that quantifies the fractal correlation properties of physiological signals has been examined, to determine whether it could derive markers for the identification of preterm infants who developed IVH. The results have demonstrated that fractal dynamics embedded in the ABP waveform could provide useful information that facilitates early identification of IVH in preterm infants.

Cardiac output (CO) and total peripheral resistance (TPR) are two important parameters of the cardiovascular system. Measurement of these two parameters can provide valuable information for the assessment and management of patients needing intensive care, including preterm infants in the neonatal intensive care unit (NICU). The results of arterial blood pressure waveform analysis demonstrate that the diastolic decay rate had a significant positive relationship with CO and negative relationship with TPR. To further assess the changes in CO and TPR in the preterm infants, the multivariate regression model based on the useful features from arterial blood pressure waveform analysis was used to improve the accuracy and robustness of the estimation. The combination of signal analysis and multivariate regression model in estimation of CO has produced some outcomes, and in the future, more effort should be involved in this kind of research to improve the prediction of serious diseases in preterm infants.

Overall, this PhD research project has demonstrated that the features extracted from the cardiovascular waveforms by different signal processing methods, including frequency spectrum analysis, DFA and ABP waveform analysis, could provide potentially useful information for assessing clinical risks in the preterm infants. This represents an important step towards continuous monitoring and automatic detection of life threatening episodes in these infants.

# Table of Contents

# Abbreviations

| | |
|---|---|
| ABP | Arterial blood pressure |
| BP | Blood pressure |
| BPV | Blood pressure variability |
| BPV | Blood pressure variability |
| BRS | Baroreceptor reflex sensitivity |
| BW | Birth weight |
| CBF | Cerebral blood flow |
| CBF | Cerebral blood flow |
| CHB | Total haemoglobin change |
| CO | Cardiac output |
| CRIB | Clinical risk index for babies |
| DFA | Detrended fluctuation analysis |
| ECG | Electrocardiogram |
| FiO2 | Fraction of inspired oxygen |
| FTOE | Fractional tissue O2 extraction |
| GA | Gestational age |
| GMSC | Generalized magnitude squared coherence |
| HbD | Hemoglobin difference |
| HbO2 | Oxygenated haemoglobin |
| HF | High frequency |
| HHb | Deoxygenated haemoglobin |
| HR | Heart rate |
| HRV | Heart rate variability |
| IVH | Intra-ventricular hemorrhage |
| LF | Low frequency |
| LVO | Left ventricular outflow |
| NICU | Neonatal intensive care units |
| NICU | Neonatal intensive care unit |
| NIRS | Near-infrared spectroscopy |
| PHH | Posthemorrhagic hydrocephalus |
| PPG | Photoplethysmographic |
| PVL | Periventricular leukomalacia |
| RVO | Right ventricular outflow |
| SBP | Systolic blood pressure |
| SVR | Systemic vascular resistance |
| THI | Normalized tissue haemoglobin index |
| TOI | Tissue oxygenation index |
| TOI | Tissue oxygenation index |
| TPR | Total peripheral resistance |
| ULF | Ultra low frequency |
| VLBW | Very low birth weight |

# List of Tables

# List of Figures

# Chapter 1. Introduction

## 1.1 Research motivation

Preterm infants with very low birth weight (VLBW; birth weight ≤1500 g) suffer from a high risk of serious diseases, such as intra-ventricular hemorrhage (IVH). Whilst the survival of these very low birth weight babies has continued to improve over the years, there is still a substantial risk of developing long term neurodevelopmental disability even with low grade IVH [1]. The timing of IVH indicates most hemorrhages develop after birth with such factors as resuscitation and cardiorespiratory management during the first 24 hours of life being important in the causal pathway [2]. Thus the ability to identify preterm infants with IVH at an early stage is of great importance in terms of minimizing mortality and morbidity in the neonatal intensive care setting.

The need for identifying babies with high risk of critical pathological conditions such as IVH has motivated the development of clinical risk and disease severity scores for use in the neonatal intensive care setting. Amongst these scores, the Clinical Risk Index for Babies (CRIB) has gained wide acceptance internationally as a preferred method of quantifying mortality risk amongst very preterm infants, partly due to its simplicity in computation [3]. Whilst the original CRIB score required fractional inspired oxygen in the first 12 hours of life for its computation, an updated version of CRIB (known as CRIB II) was published in 2003 [4]. CRIB II is scored at the first hour of life, based on a number of clinical factors including birth weight, gestational age, gender, temperature on admission and blood base excess.

The use of cerebral perfusion measurement to identify preterm infants with high risk of mortality or IVH has also been proposed [5-8]. The underlying premise was that impairment of cerebral autoregulation could be a potentially important cause of brain injury such as IVH.A popular technique of assessing cerebral autoregulation is by frequency domain cross-spectral transfer function analysis, with higher coherence (or correlation), reduced phase shift and/or higher gain between arterial BP and Cerebral blood flow (CBF) being regarded as possible indication of impaired autoregulation [9-11]. In this research, I examined how various measures derived from spectral and cross-spectral analysis of arterial blood pressure (BP) and cerebral near-infrared spectroscopy (NIRS) were related to CRIB II, which would help to establish these measures as potential indicators of clinical risk in preterm infants.

Furthermore, the various parameters derived from cross-spectral transfer function analysis of arterial BP and cerebral NIRS have been examined whether they could demonstrate significant difference between preterm infants with and without IVH. In contrast to previous studies which adopted a single NIRS measure: tissue oxygenation index (TOI) or hemoglobin difference (HbD) as a surrogate of CBF, we examined the transfer function relationships of arterial BP to a number of NIRS measures to understand which of the measures might be more useful for the detection of IVH.

With the improvements in neonatal intensive care, the incidence of IVH has fallen in preterm infants with very low birth weight [12]. Rates of particularly severe grades of IVH with extension into cerebral tissue (grade V IVH) in the Papille system are still a major attention on those extremely preterm infants [13]. It was hoped that the new method might provide a better time resolution of the events immediately before and after the severe grades of IVH, allowing clearer views on the precise timing relationship of the multiple clinical parameters during the occurrence of IVH. The early detection of IVH will present an important step in minimizing mortality and morbidity in the neonatal intensive care setting [14], [15].

Detrended fluctuation analysis (DFA) has been regarded as a robust method to quantify the scale-invariant (fractal) correlation property of a time series, especially when the signal class cannot be precisely determined. Specifically, this method characterises the mono-fractal property. Previous studies have adopted this method to evaluate the fractal dynamics of heart rate in various diseases in adult subjects [16, 17]. The DFA method aims at measuring the intrinsic fractal-like correlation properties or self-affinity of the time series, characterized by increasing magnitude of fluctuations with increasing time scale (or decreasing frequency).

Assessment of fractal scaling properties in ABP in adult humans based on DFA method has also been reported [18-20]. It was therefore suggested that fractal dynamics of arterial pressure were largely influenced by the autonomic nervous system activity. However, the fractal scaling of arterial BP in preterm infants has not been investigated previously. In this study, I examined whether there was a difference in fractal scaling exponents derived from DFA method of arterial BP between preterm infants with and without IVH. It was hypothesized that the infants who developed IVH might have higher fractal scaling exponents of arterial BP, similar to that seen in heart rate fluctuations [21]. The findings could help to establish the potential value of fractal dynamics embedded in the arterial pressure waveform for early identification of brain hemorrhage in preterm infants.

CO and TPR are two important parameters of the cardiovascular system. Measurement of these two parameters can provide valuable information for the assessment and management of patients needing intensive care, including preterm infants in the NICU [22, 23]. Furthermore, the measurement of systemic hemodynamics can facilitate the development of better therapeutic strategy for these infants, including the use of volume expansion and inotropes [23]. In the present NICU setting, the assessment of CO and TPR requires the measurement of left ventricular outflow (LVO) or right ventricular outflow (RVO) using ultrasound Doppler echocardiographic waveforms. The clinical method for CO measurement should be accurate, continuous, inexpensive and minimally invasive. However, such conventional monitoring requires specialized skill and equipment, and cannot be performed continuously and invasively. However, none of the CO measurement methods possess all of these characteristics and can be widely used in the clinical diagnosis for preterm infants. For example, the thermodilution method does not operate continuously and requires a pulmonary artery catheterization whose safety and accuracy has been questioned.

In this study, the estimation of LVO and TPR was performed by analyzing the rate of blood pressure decay in the diastolic portion of the arterial pulse wave. This has been associated with TPR in the Windkessel model, and has been utilized in the study of animals and adult patients [24, 25].However, whether this arterial waveform parameter is useful for monitoring preterm infants has not been previously investigated. The results from the current study would help to establish the potential utility of the diastolic decay rate parameter for assessing LVO and TPR in preterm infants. Then, the diastolic decay rate and BP features are added as input features to a multivariate regression model to estimate LVO and TPR. A stepwise feature search algorithm was employed to select statistically significant features. K folded cross validation was used to assess the generalized model performance. The degree of agreement between the estimation method and the real measurement of LVO and TPR was assessed using Bland- Altman analysis.

## 1.2 Objectives

With the motivations highlighted in the previous section, the current work aims to achieve the following three objectives:

- To develop signal processing tools to detect IVH using ECG, NIRS and blood pressure signals in preterm infants.

- To develop signal processing tools to identify parameters that reflect clinical risk (CRIB II score): the relationships between CRIB II and various spectral measures of arterial baroreflex and cerebral autoregulation functions are possible markers of clinical risks and predictors of adverse outcome in preterm infants.

- To identify parameters that can detect change in cardiac output (left ventricular outflow LVO) and TPR over time which cannot be measured continuously in current clinical setting but are useful for identifying clinical risk.
- To develop a mathematical model to estimate CO and TPR using ABP waveform.

## 1.3 Paper contributions

The contributions of this paper are:

- This paper demonstrates that cerebral NIRS measurement was able to provide a number of parameters that were potentially useful for distinguishing between preterm infants with or without brain hemorrhage.
- This study is the first to examine the fractal dynamics of arterial BP in preterm infants using the DFA approach. The rationale was to address whether the fractal scaling of arterial BP could provide potentially useful information for distinguishing between preterm infants with and without brain hemorrhage.
- This paper firstly establishes the link between parameters derived from spectral analysis of systemic and cerebral cardiovascular variabilities and clinical risk of preterm infants, via examining the correlation of these parameters with the CRIB II score.
- This paper provides potentially useful information for continuous monitoring clinical risks in the preterm infants by the utility of arterial pressure waveform analysis for estimating LVO and TPR in preterm infants.

## 1.4 Paper organization

The rest of this paper is organized as follows:

- **Chapter 2** provides the background information associated with the current work and a brief literature review of previously developed

monitoring and signal analysis tools utilized in clinical monitoring of preterm infants.

- **Chapter 3** presents the potential use of spectral analysis of heart rate, blood pressure and cerebral NIRS as a monitoring tool to assess the mortality and mobility in preterm infants. It has two parts: (1) Development of a potentially useful method from transfer function analysis to distinguish between preterm infants with or without IVH; (2) Development of the cross spectrum analysis to find out the potential relationship with CRIB II by linear regression approach.
- **Chapter 4** develops the Detrended fluctuation analysis of blood pressure in preterm infants to compare between preterm infants with and without IVH.
- **Chapter 5** presents a method to continuously and noninvasively estimate CO and TPR in Preterm Infants by Arterial Waveform Analysis and a multivariate regression model.
- **Chapter 6** summarizes the work presented in this paper and presents some future directions which can be pursued to advance the assessment of clinical risks in preterm infants.

## 1.5 Author's publications

The work presented in this paper was published in the following papers, which have been produced by the author during her PhD candidacy:

**Journal publications**

- Y. Zhang, G.S.H. Chan, M.B. Tracy, Q.Y. Lee, M. Hinder, A.V. Savkin, and N.H. Lovell, "Spectral analysis of systemic and cerebral cardiovascular variability in preterm infants: relationship with Clinical Risk Index for Babies (CRIB)" ,Physiological Measurement, 32:1,931 – 1,928, 2011.
- Y. Zhang, G.S.H. Chan, M.B. Tracy, M. Hinder, A.V. Savkin, and N.H. Lovell, "Detrended fluctuation analysis of blood pressure in preterm

infants with intraventricular hemorrhage", Medical & Biological Engineering & Computing, 51:1051-1057, 2013.

**Peer reviewed conference publications**

- Y. Zhang, G.S.H. Chan, M.B. Tracy, Q.Y. Lee, M. Hinder, A.V. Savkin, and N.H. Lovell, "Cerebral near-infrared spectroscopy analysis in preterm infants with intraventricular haemorrhage",33rd Annual International Conference of the IEEE Engineering in Medicine and Biology Society , Boston, USA, pp 1937 – 1940,2011.
- *Y. Zhang, G.S.H. Chan, M.B. Tracy, M. Hinder, A.V. Savkin, and N.H. Lovell,*, "Estimation of Cardiac Output and Systemic Vascular Resistance in Preterm Infants by Arterial Waveform Analysis" 34rd Annual International Conference of the IEEE Engineering in Medicine and Biology Society, Osaka, Japan, pp 2308-2311, 2013.

# Chapter 2. Background and Literature Review

## 2.1 Preterm infants and their clinical management

Very preterm infants are at high risk of death, 53% of the babies born before 24 weeks can survive but the survival rate increases dramatically to 93% after 27 weeks gestation in 2011 [26].The birth weight is another important factor to survival rate, the same tendency is demonstrated in the survival rate rise from less than 60% with birth weight ≤ 500 g to more than 90% at birth weight ≥ 1250 g [26].

A dramatic decline in mortality rates in preterm infants could be attributed to advances in neonatal intensive care and specialized obstetric, including exogenous surfactant administration to babies from the late 1980s, and the growing utilization of antenatal corticosteroids from the early 1990s [26],[27].

Although the survival rate increased because of major advancements in neonatal intensive care, the preterm infants still have high risk of serious permanent brain damage and disturbances in brain maturation, as occurs with intra-ventricular haemorrhage (IVH).The rates of most major neonatal morbidities have remained unchanged in the past decade since the survival rate of smaller and less mature infants initially increased[28].

The brain damage will leads to atypical brain development and long term neurodevelopmental disabilities even with low grade IVH [29].Accordingly, preterm infants have a strong possibility of developing motor disabilities including cerebral palsy persist, as well as less prominent developmental disabilities in cognition, social, language and behaviour [1, 29, 30].With the improvement of intensive care, mortality rate is not the only important adverse consequence of preterm infants, more and more attention has been paid on reducing morbidity including brain injury [31].Improved understanding of the risk factors for acquired brain injury and associated factors that affect brain development in this population is setting the stage for improving the brain health of children born preterm.

## 2.1.1 Intra-ventricular Haemorrhage

IVH is the most commonly diagnosed brain lesion in the preterm infants [28] and potentially relates to significant lifelong sequelaes and serious neonatal complications [5]. There is evidence in the past showing that the periventricular leukomalacia (PVL) and posthemorrhagic hydrocephalus (PHH) are two significant sequelae of IVH [32]. According to the location and severity of the PVL lesions, the clinical presentation of sicked infants may include decreased visual fields, spastic diplegia and cognitive impairment [33, 34]. Similar investigations have also been reported that the white matter injury which follows IVH reflects the leading cause of the neurodevelopmental impairments suffered by preterm infants.

The head ultrasound is generally used to detect signs of IVH during the first week after the babies were born [26]. The commonly used method to define IVH grade was first reported by Papile et al in 1978, in which IVH severity increases with higher grade [13]. The IVH grades was classified into 4 grades, from I to IV, including grade I, haemorrhage confined to the subependymal germinal matrix; grade II, haemorrhage into the lateral ventricles without ventricular dilatation;

grad-e III, IVH with ventricular dilation; and grade IV, IVH with parenehymal haemorrhage into cerebral tissue [28].

With the advancements in neonatal intensive care, the incidence of IVH has fallen from 30% in the 1980's to 6 % [12] in preterm infants with very low birth weight (birth weight ≤1500 g). However recent report has indicated that the rates of IVH have remained at 20% to 70% for extremely preterm infants (<27 weeks gestation) [26]. Rates of IVH and in particularly severe grades of IVH with extension into cerebral tissue (grade V IVH) in the Papille system are still a major concern in those extremely preterm infants (<27 weeks gestation) [13]. Fragile cerebral circulation of preterm infants is affected by fluctuations in the systemic circulation associated with the pathophysiology of preterm birth and the first six hours of life when complex interactions occur between mechanical ventilation and the circulation. Preterm birth <27 weeks is an abnormal event often precipitated by maternal infection, bleeding or hypertension, factors all predisposing to reduced uterine perfusion and thus placental function. Placental dysfunction is a predisposing factor to the fragile blood vessels in the germinal matrix bleeding in a grade I IVH. If this extends into the ventricular system (grade II IVH) and then blocks the needed drainage of ventricular fluid via the third and fourth ventricles (grade III IVH), risks of long-term brain damage rise. Venous occlusion in adjacent areas of the cerebral hemispheres can give rise to infarction and significant injury to major motor tracks leading to cerebral palsy or intellectual impairment [1, 30].

The timing of IVH indicates most haemorrhages develop after birth with such factors as resuscitation and cardiorespiratory management during the first 24 hours of life being important in the NICU [35],[36]. Thus the ability to identify preterm infants with IVH at an early stage is of great importance in terms of minimizing mortality and morbidity in the neonatal intensive care setting.

## 2.1.2 Clinical Risk Index for Babies

The need for identifying babies with high risk of critical pathological conditions such as IVH has motivated the development of clinical risk and disease severity scores for use in the neonatal intensive care setting. Amongst these scores, the CRIB has gained wide acceptance internationally as a preferred method of quantifying mortality risk amongst very preterm infants, partly due to its simplicity in computation [3]. It was implemented with data relating to infants of birth weight ≤1500 g or infants born ≤ 31 weeks, treated in neonatal intensive care between 1988 and 1990 [3].

Survival of VLBW infants depends on birth weight (BW) and gestational age (GA), but also on other physiological conditions and perinatal factors of the individual infants. Disease severity scores were developed to compare the risk of mortality between different NICUs, since the sicker infants may still have higher mortality despite with better care in hospital. CRIB Scores are calculated from gestational age, birth weight, maximum and minimum appropriate fraction of inspired oxygen (FiO2), maximum base excess and presence of congenital malformations at the first 12 hour after birth [37].Therefore, it is more closely related to mortality and impairment after intensive care than the BA.Whilst the original CRIB include the data at in the first 12 hours of life with possible neonatal treatment bias, its appropriateness in prediction of mortality have been questioned. Furthermore, original CRIB score required FiO2 in the first 12 hours of life for its computation, which is not a true clinical measure decided by the neonatal care team [27].

## 2.1.3 Estimating Left Ventricular Outflow

The ability to measure Cardiac Output parameters continuously and noninvasively can provide valuable information for the assessment and monitoring of the potential cardiovascular diseases or other critical illnesses, including preterm infants in the neonatal intensive care unit (NICU) [22, 23].For example, high CO is the symptom of sepsis, which is a common cause of mortality and morbidity in preterm infants, particularly those with VLBW [22]. A substantial drop in CO (>50%) has been associated with a high risk of mortality in these infants with sepsis [22]. Furthermore, the measurement of systemic hemodynamics can facilitate the development of better therapeutic strategy for these infants, including the use of volume expansion and inotropes [23].

In the present NICU setting, the assessment of CO and TPR requires the measurement of left ventricular outflow or right ventricular outflow using invasive or non-invasive methods. In this study , LVO was obtained invasively by measuring the aortic valve annulus diameter in the parasternal long axis view, and then velocity time integrals (VTI) were obtained with a minimum of 5 consecutive aortic pulsed Doppler waveforms from an apical long axis view [38].

## 2.2 Technique for preterm infants monitoring

### 2.2.1 Near-infrared spectroscopy

Near infrared spectroscopy (NIRS) measurements were first utilized in pediatrics for the bedside monitoring of cerebral oxygenation in sick preterm infants by Brazy and Lewis [39]. Then this technology was used for the monitoring of cerebral oxygenation and hemodynamics during congenital heart surgery by Greeley et al. [40] . With the improvements in neonatal intensive care setting, neurologic overcomes have surpassed mortality as the primary problems in preterm infants. Therefore the use of NIRS measurements to monitor cerebral circulation and enhance detection of brain injury has gained widespread acceptance.

Cerebral NIRS data were collected by measuring tissue absorbance of light at different wavelengths from 700-1000 nm [41]. Because near infrared light can penetrates biological tissues and even bone, NIRS measurements can be performed through the intracranial compartment. In this study, cerebral NIRS data were measured by the NIRO-300 system (Hamamatsu Photonics, Hamamatsu City, Japan), with the sensor probe placed over the skin of the temporoparietal region on the head of the infant. The instrument transfers laser diode light by a fiber optic bundle to the head and thereby to the intracranial compartment [42].The sensor probe of the NIRS-300 monitor consists of a fiber optic probe that emits four wavelengths of near-infrared light (775, 825, 850 and 904 nm), and a receiver probe made of separate aligned photodetectors, which permit the measurement of spatially resolved NIRS. The NIRS parameters are computed by the device every 0.167 s (6 Hz).

Cerebral / Brain oxygenation information by NIRS includes the indices as follows:

- Oxygenated haemoglobin change (HbO$_2$)

- Deoxygenated haemoglobin change (HHb)

- Total haemoglobin change (cHb)

- Tissue oxygenation index (TOI): TOI = $\frac{HbO2}{HbO2+HHb}$

- Normalized tissue haemoglobin index (THI)

- Fractional tissue O$_2$ extraction (FTOE): FTOE = $\frac{SaO2-TOI}{SaO2}$

- Hb difference (HbD) : HbD = HbO$_2$ – HHb

Previous clinical study indicated that disturbances in CBF were important in the pathogenesis of IVH. In a healthy brain, the cerebral auto-regulation mechanism normally acts to limit the variation in CBF over a range of perfusion pressure, so as to protect the brain from large fluctuations in CBF [43, 44]. However, impairment of this protective mechanism will increase the risks of cerebrovascular injury and brain haemorrhage [5-8].The preterm infants often suffer from disturbances in CBF for two major reasons: First, because of the high incidence of respiratory disease, they need mechanical ventilation, therefore alterations in mean arterial blood pressure (MAP) are very common in such infants. Second, various studies indicate that cerebral auto-regulation is either defective or absent in some preterm infants [42]. To measure CBF and find out the indication of impaired cerebral auto-regulation for preterm infants, the xenon-133 clearance technique was first proposed by in 1979 by Lou et al. [45]. Subsequent study has shown a clear relationship between impaired autoregulation and the incidence of IVH [46, 47].However, the xenon-133 clearance technique to monitor CBF involves radiation exposure and cannot be used to continuously detect the impaired cerebral autoregulation[42].

For the monitoring of preterm infants, the NIRS has been preferred for deriving surrogate measures of CBF changes, given its ability to provide simple, non-invasive and comfortable continuous measurement. The NIRS technique can quantify changes in cerebral oxygenation and haemoglobin content, such as the oxygenated haemoglobin ($HbO_2$), the deoxygenated haemoglobin (HHb), and the tissue oxygenation index (TOI) )(which is ratio of $HbO_2$ to total hemoglobin $HbO_2$+HHb) that reflects oxygen saturation in brain vasculature (predominantly in the venous compartment, which represents ~75% of total cerebral blood volume [48]). The TOI was found to reflect CBF variations during reduction in perfusion pressure, under the conditions of constant arterial oxygen saturation and cerebral oxygen consumption [49]. Some investigators also derived the Hb difference (HbD [5-7, 50], which was shown to be correlated with CBF during hypotensive episodes [50]. NIRS also provides measures of the fractional tissue oxygen extraction (FTOE), which is the ratio of cerebral oxygen consumption to oxygen delivery.

## 2.2.2 Updated Clinical Risk Index for babies

An updated version of CRIB (known as CRIB II) published in 2003 introduced a simplified and re-adjusted scoring system, which avoids the potential problems because of the use of $FiO2$ and data up to 12 hours after born [27]. The CRIB II score is not only a risk-adjustment instrument widely used in NICUs to measure initial illness severity, but also a predictor of mortality and morbidities until discharge. It was based on the first hour findings, therefore there is less treatment bias compared with CRIB score.

CRIB II score is calculated on a number of clinical factors including birth weight, gestational age, gender, temperature on admission and blood base excess. It has a greater discriminatory ability to predict mortality than do other scoring systems. Extremely preterm infants with higher overall CRIB-II scores are more likely to have major disability and poor outcome.

The initial condition of the neonates within the first hour of life does not remain stable throughout their hospitalization. Three components of the CRIB-II score (gestational age, birth weight and gender) do not change after birth. However, two components (body temperature and blood base excess) are changeable, so a high CRIB-II score can remind caregivers to optimize these factors as early as possible; if rectified in a timely manner, the immediate and longer-term impact may be reduced. Currently the CRIB-II score is mainly used to compare mortality and morbidities within NICUs and does not provide individual prediction accurately enough for diagnosis.

Clinical parameters, such as birth weight, gestational age and cranial ultrasound have been separately used to predict long-term outcomes in preterm infants. However, these individual predictors do not have very good sensitivity and specificity compared with a composite scoring system [37].

The impact of major disability in preterm infants is lifelong. Identifying infants at risk of major disability allows earlier follow-up and intervention that may reduce the impact of the disability. It have been reported that that a CRIB-II score of 13 maximizes sensitivity and specificity for predicting major disability, indicating a high likelihood of identifying infants at risk of major disabilities using this cut-off score. CRIB-II can be a useful tool for identifying preterm infants for follow-up or early-intervention programs.Nevertheless it should be noted that CRIB II score is a static and one-off measurement, it cannot be used to continuously monitor the conditions of infants in real time.

### 2.2.3 Cardiovascular data

In the last decade, significant interest in non-invasive monitoring of the cardiovascular system through the analysis of cardiovascular signals, such as Electrocardiogram (ECG) and ABP, has been generated in biomedical research community worldwide. The variability inherent within the ECG, for instance, has been proposed to implicate the increasing regularity representing pathological changes. Both of ECG and ABP represent cardiovascular control.

Cardiovascular variability, such as heart rate variability and blood pressure variability has emerged as potential non-invasive markers of autonomic nervous modulation (sympathetic and vagal nerve activities) and vasomotor regulation mechanisms [51, 52]. Frequency spectral analysis of beat-to-beat fluctuation of heart rate (HR) has been proven to be a useful non-invasive tool for monitoring the variations of sympathovagal balance controlling HR [52].While similar investigations have also been reported by Andriessen *et al.,* using transfer function analysis of HRV and BPV to quantify baroreflex sensitivity [53]. Compared with the conventional pharmacological method, the transfer function analysis is focused on analysis of spontaneous fluctuations in BP and HR, therefore it can be used to continuously monitor baroreflex function. It has also been suggested that the higher gain or coherence in the cerebral pressure-flow relationship could reflect impairment of CBF control by dynamic cerebral autoregulation, thus potentially linked with brain injury, such as IVH, or other mortality risk factors [5-8]. In this study, particular interest was given to the changes in baroreflex control of HR before and after the occurrence of IVH in preterm infants, which were analysed by power spectrum and transfer function analysis of HRV and BPV.

The sympathetic and vagal nerve fluctuations in HR may be independently affected by some disorders of autonomic nervous system.Disorders of the central and peripheral nervous system have effects on HRV [54]. It has been reported that all normal cyclic fluctuations in HR are decreased in the presence of severe brain damage [55]. Although the accuracy of HRV was questioned, it was easily accessible to provide the neurologic information [56].

## 2.3 Potential application of cardiovascular signals

### 2.3.1 Frequency Spectrum Analysis

Frequency spectral analysis provides the opportunity to decompose spontaneous fluctuations in blood pressure and cardiovascular signals into a power spectrum, and to relate the character of the fluctuations to physiological processes. It has been proposed by Tang et al. to apply frequency spectral method to monitor the hemodynamic changes of cardiovascular system along the progression of sepsis [57]. Frequency spectral method has also been used in beat-to-beat fluctuation of HR, which was proven to be a useful tool for noninvasively monitoring the variations of sympatho-vagal balance [58].

Cross-spectral transfer function analysis can provide information related to the regulation of cerebral blood flow by examining the relationship between arterial blood pressure and CBF fluctuations in the frequency domain [9]. The calculation of coherence was used to identify the potential source of the oscillations seemed in the heart rate variability and photoplethysmographic (PPG) waveforms [59].

Previous studies have suggested that Low-frequency (LF) fluctuations at a frequency of approximately 0.1 Hz related to the Baroreceptor reflex (BR) activity and high-frequency (HF) fluctuations are bound up with vagal modulation and respiratory activity [60] . Cross-spectral analysis between BP and RR interval fluctuations in the LF band (0.04–0.15 Hz) has been shown to be an estimate of the baroreceptor reflex sensitivity (BRS) [61-64]. Previously, the feasibility of using cross-spectral analysis to estimate BRS from spontaneous BP and RR interval fluctuations has been investigated in stable preterm infants [60].

The previous report investigated that the impairment of cerebral auto-regulation could be a potentially important cause of brain injury such as IVH [5-8]. The use of cerebral perfusion measurement to identify preterm infants with high risk of mortality or IVH has been proposed [5-8]. In a healthy brain, the cerebral auto-regulation mechanism normally acts to limit the variation in CBF over a range of perfusion pressure, so as to protect the brain from large fluctuations in CBF [43, 44]. Impairment of this protective mechanism, however, may lead to CBF being more dependent on changes in ABP, or more "pressure passive", such that large changes in BP would be translated to large CBF fluctuations, hence increasing the risks of cerebrovascular injury and brain haemorrhage [5-8].

A popular technique of assessing cerebral auto-regulation is by frequency domain cross-spectral transfer function analysis, with higher coherence (or correlation), reduced phase shift and/or higher gain between arterial BP and CBF being regarded as possible indication of impaired auto-regulation [9-11]. Some investigators also derived the Hb difference (HbD) as $HbO_2$ – HHb [5-7, 50], which was shown to be correlated with CBF during hypotensive episodes [50]. The transfer function analysis of BP and cerebral NIRS measures has demonstrated that alterations in dynamic cerebral auto-regulation properties, such as higher coherence or gain, are linked with IVH or other mortality risk factors [5-8].

The baroreceptor reflex (BR) is another important short-term regulatory mechanism of BP and heart rate (HR), which buffers or opposes sudden changes in systemic BP by regulating HR and peripheral vascular resistance [65]. BR sensitivity (BRS) is a crucial index to describe the BR regulation on BP and is classically defined as reflecting the degree of change in HR for a given unit change in BP [65]. Classical methods to measure BRS implements an external stimulus, such as infusion of vasoactive drugs [66] and mechanical pressure neck chamber [67]. New techniques to evaluate the BRS are based on the frequency and time domain analysis of BPV and HRV.Spectral technique that is commonly used to assess arterial BRS is founded on the cross-spectral transfer function calculation of BPV and HRV. The application of this technique in arterial baroreflex modulation of autonomic tone was widely utilized and has been reported in the past researchers [51, 53, 68]. Analysis of BPV and HRV showed a reduced cardiac BRS in adult patients with acute intra-cerebral haemorrhage, with BRS being able to predict adverse outcomes [69]. Whether BRS was related to clinical risks in preterm infants was not known, although a recent study found positive correlation between LF BRS gain and postmenstrual age (gestational age + postnatal age) [53], suggesting that low BRS in the high risk very preterm infants would be likely.

The Welch method was used, which involved averaging the cross power spectra from multiple sub-segments in a similar fashion as the computation of Power Spectrum density (PSD). With the pre-determined auto spectra of signals x and y in frequency *f* representation (i.e., $S_{xx}(f)$ and $S_{yy}(f)$ respectively), the auto power spectra of signals x and y were each defined as $P_{xx}(f) = |S_{xx}(f)|^2$ and $P_{yy}(f) = |S_{yy}(f)|^2$ (*f* represents frequency). The cross power spectrum of x-y as $P_{xy}(f)$ is calculated as the product of $P_{xx}(f)$ and $P_{yy}(f)$, the transfer function of x-y (from x to y) was computed as follows,

$$H(f) = \frac{P_{xy}(f)}{P_{xx}(f)} \tag{2.1}$$

From the equation 2.1, the magnitude and phase angle of the transfer function could be obtained accordingly and the BRS is the magnitude of the transfer function as follows:

$$BRS = |H(f)| = \sqrt{real[H(f)]^2 + imag[H(f)]^2} \tag{2.2}$$

The magnitude-squared coherence function was defined as follows:

$$\gamma 2(f) = \frac{|Pxy(f)|2}{Pxx(f)Pyy(f)} \tag{2.3}$$

BRS obtained through the transfer function analysis can be readily calculated for the coherence *γ*2 which assesses the linear correlation between spectral components in the two signals, with a value ranging from 0 (lack of linear correlation) to 1 (perfect linear relationship).

### 2.3.2 Nonlinear analysis

There are several factors that influence heart rate, including respiration and mental load. HR regulation is a complex system in humans and the simplified model of HR regulation is shown in Figure 2-1**.**The sinus node, generating the HR is governed by sympathetic and parasympathetic efferents as well as several humoral factors.

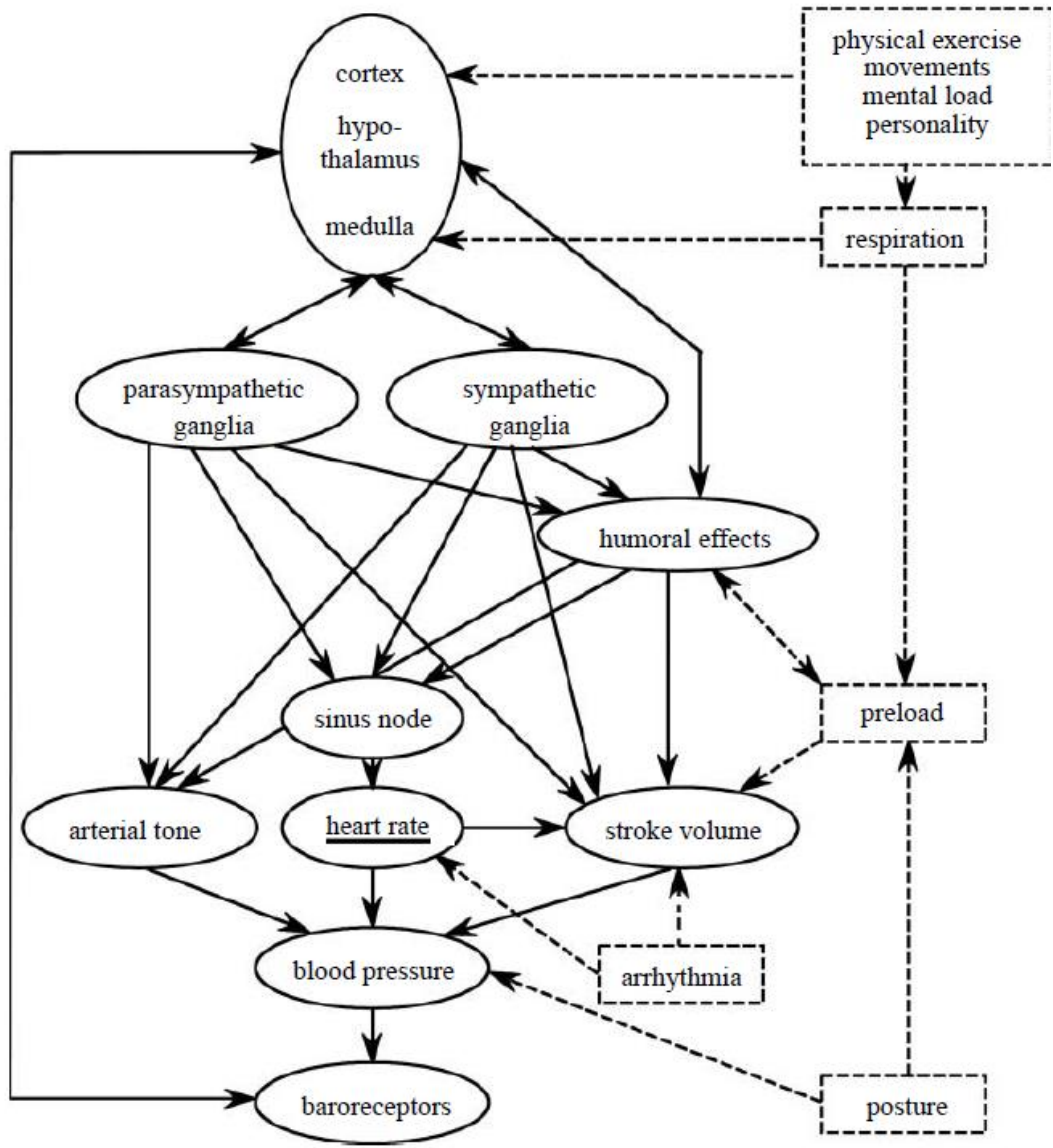

**Figure 2-1:** A simplified model of HR regulation[70]

The sinus node works as the final factor of sympathetically and parasympathetically mediated stimuli and the procedure is reflected in the actual beat-to-beat interval [17]. The regulatory subsystems result in a scale-invariant cardiac control according to different time scales, showing long-range correlations with a typical scaling behavior [17]. Therefore, the classical linear signal analysis methods are not often adequate to present the relevant properties of non-linear systems.

Nonlinear analysis of HRV changes provides the new indices to quantify the complexity of fractal dynamics and stimulates innovative research into cardiovascular dynamics. Methods derived from nonlinear analysis of HRV have been proposed for clinical risk assessment of cardiovascular diseases, like sample entropy, sample asymmetry and detrended fluctuation analysis [17]. Sample entropy analysis of HRV has been used to estimate the complexity of cardiovascular dynamics. It has reported that the entropy falls before clinical signs of neonatal sepsis but the major mechanism is not related to the regularity of the data because of the presence of spikes [71]. The limitation of sample entropy analysis is the requirement of stationarity and the adverse affect of missed beats or artefacts [71].

DFA method has been regarded as a robust method to quantify the scale-invariant (fractal) correlation property of time series, especially when the signal class cannot be precisely determined. Specifically, this method characterises the mono-fractal property. Previous studies have adopted this method to evaluate the fractal dynamics of heart rate in various diseases in adult subjects [16, 17]. The DFA method aims at measuring the intrinsic fractal-like correlation properties or self-affinity of the time series, characterized by increasing magnitude of fluctuations with increasing time scale (or decreasing frequency). The degree of fractal scaling can be depicted by the scaling exponent (α) that quantifies the relationship between fluctuations F($n$) and time scale $n$ based on the mathematical function F($n$) = C$n^α$, with α = 1 corresponding to the classical 1/$f$ fractal signal [16, 17]. In a prior study, preterm infants who later developed

IVH were found to have significantly higher short-term scaling exponent of heart rate compared with those who did not develop IVH [21]. The differing fractal dynamics of heart rate in these two groups of infants have been attributed to the altered autonomic control of the heart associated with the development of IVH.

Assessment of fractal scaling properties of ABP in adult humans, which is based on the DFA method has also been reported [18-20]. By performing autonomic blockade tests, it was shown that vagal blockade by atropine could lead to an increase in the short-term scaling exponent of BP, whereas central blockade of both cardiac and vascular sympathetic outflow by clonidine would result in a decrease of the exponent [20]. It was therefore suggested that fractal dynamics of arterial pressure were largely influenced by the autonomic nervous system activity. However, the fractal scaling of arterial BP in preterm infants has not been investigated previously.

### 2.3.3 Arterial waveform analysis

While the current measurement of CO cannot fulfil all the requirements of patients cohorts, analysis techniques to monitor CO from ABP waveforms may be achieved reliably and continuously with little or no invasiveness via radial artery catheterization and arterial tonometry [72], because ABP is related to CO through the arterial tree [24]. Previous work has also shown that arterial tree is represented by Windkessel model which comprises two parameters: arterial compliance (AC) and TPR of the small arteries as shown in Figure 2-2 (left).ABP decays like a pure exponential during each diastolic interval with a time constant $\tau$, which equals to TPR·AC .Because AC is total compliance of the large arteries and  is nearly constant over a wide pressure range [73-76] , CO could then be measured within a constant scale. Thus, the technique was proposed by Bourgeois et al. to be conducted in fitting an exponent function to the diastolic interval of each ABP pulse in order to measure $\tau$ [24].

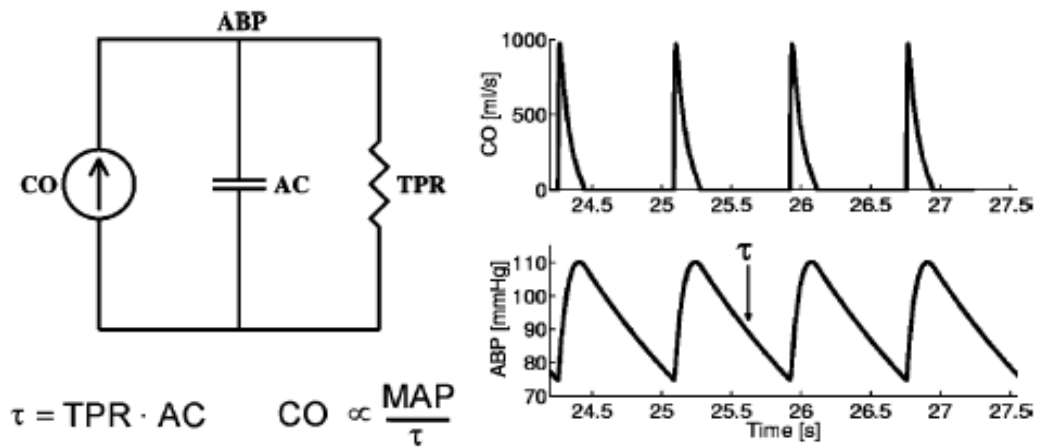

**Figure 2-2:** The technique of Bourgeois *et al.* for monitoring CO from an ABP waveform [24, 77].

## 2.4 Disease Prediction

Increasing survival rate of premature infants has produced a focus change from mortality to morbidity, such as brain injury. The discovery and validation of potential indicators of brain injury and clinical risk is a critical step in the evolution of neonatal care [31].

### 2.4.1 IVH detection

Preterm birth <27 weeks is an abnormal event often precipitated by maternal infection, bleeding or hypertension, factors all predisposing to reduced uterine perfusion and thus placental function. It has been reported that IVH is an important predictor of adverse neurodevelopmental outcome [78]. Both of the gestational age and medical risk factors are equally important predictors of neurologic outcome, especially in Grade 3-4 IVH [79-87]. In the newborn period, 5-10% of preterm infants with Grade 3- 4 IVH suffer seizures and 50% of them have posthemorrhagic hydrocephalus [88]. As a result, mortality is higher in preterm infants with Grade 3-4 IVH have a higher mortality rate than those who have the same gestational without Grade 3-4 IVH [89]. While preterm infants with Grade III-IV IVH have high risk of mental retardation and cerebral palsy [88, 90-96], babies with Grade I - II IVH are also at risk for developmental disability [78]. Furthermore, a recent review found that 45- 86% of preterm children with Grade 3- 4 IVH have suffered from major cognitive handicaps, approximate75% of them need the special education [78, 90, 97]. Placental dysfunction is a predisposing factor to the fragile blood vessels in the germinal matrix bleeding in a grade I IVH. If this extends into the ventricular system (grade II IVH) and then blocks the needed drainage of ventricular fluid via the third and fourth ventricles (grade III IVH), risks of long-term brain damage rise. Venous occlusion in adjacent areas of the cerebral hemispheres can give rise to infarction and significant injury to major motor tracks leading to cerebral palsy or intellectual impairment [1, 30].The principal diagnostic procedure used to identify IVH in preterm infants is cranial ultrasonography. However, the cerebral ultrasound has not been used extensively in clinical settings because of special skills.

IVH is most commonly developed within the first 24 hours after birth with such factors as resuscitation and cardiorespiratory management [35],[36], and haemorrhages can progress over 48 hours or more [78]. By the end of the first week after birth, 90% of the haemorrhages can be detected at their full extent [78]. Thus the ability to identify preterm infants with IVH at an early stage is of great importance in terms of minimizing mortality and morbidity in the neonatal intensive care setting. It is widely accepted that IVH has generally been attributed to variations in CBF to the immature germinal matrix microvasculature [78]. Cross-spectral transfer function analysis can provide information in relation to the regulation of CBF by cerebral auto-regulation mechanisms, by examining the relationship between arterial blood pressure (BP) and CBF fluctuations in the frequency domain [9]. The transfer function analysis of BP and cerebral NIRS measures has demonstrated that alterations in dynamic cerebral auto-regulation properties, such as higher coherence or transfer gain, are linked with IVH or other mortality risk factors [5-8]. Furthermore, fragile cerebral circulation of preterm infants is affected by fluctuations in the systemic circulation associated with the pathophysiology of preterm birth and the first six hours of life when complex interactions occur between mechanical ventilation and the circulation. In the recent study, the application of DFA was proposed to detect early subtle alterations of heart rhythm that could be used to predict impending IVH in preterm infants [21].

The cerebral autoregulation is an important physiologic mechanism to maintain adequate perfusion and oxygenation of the brain and allows concordant oscillation of CBF with blood pressure. The use of cerebral perfusion measurement to identify preterm infants with high risk of mortality or IVH has also been proposed [5-8]. The underlying premise was that impairment of cerebral autoregulation could be a potentially important cause of brain injury such as IVH. Furthermore, the impairment of cerebral blood flow (CBF) controlled by cerebral auto-regulation has been associated with IVH, one popular technique for monitoring cerebral circulation in preterm infants is NIRS measurements.

Thus, dynamic monitoring of CBF can be readily used to detect CBF changes because of rapid and transient MAP variations [98, 99]. Cerebral autoregulation behaves as a high-pass filter [9, 100, 101], which is less effective at higher frequency of blood pressure fluctuation. But the static monitoring of CBF using xenon-133 clearance techniques [102] or NIRS [103, 104] may fail to detect impaired autoregulation because it is a dynamic mechanism. The previous finding suggests that autoregulation plays an important role in mediating the ultralow-frequency range of MAP variations, because the amplitude of spontaneous MAP variations increases when the frequency decreases [105]. Similar investigations have demonstrated that the loss of ultralow frequency (frequency bands of 0.003–0.020Hz) autoregulation is important in the clinical and pathophysiological use [106, 107]. In the previous study, cerebral Doppler ultrasound was used to monitor the fast changes of CBF velocity and NIRS measurements were utilized to detect changes in cerebral oxygenation which reflect fluctuations in CBF. The estimation of the two systems demonstrated that increased coherence in the very low frequency range (frequency bands of 0.020–0.050 Hz) has been an important clinical value [107, 108]. However, cerebral Doppler ultrasound is not so easily applied in clinical settings because of special operation. Thus, NIRS is more feasible to the long term measurement which is needed to estimate autoregulation in the lower frequency range.

The cross-spectral analysis between MAP and HbD has also been used in a previous study to more specifically identify the impaired autoregulation, which was associated with severe cerebrovascular lesions, such as high grade IVH and PVL in preterm infants2. However, that study is questionable as the cranial ultrasonography was not performed continuously from the first day after birth, and thus daily cranial ultrasonography is crucial to identify IVH.

On the other hand, because the alteration of heart rhythm has been related to IVH after its development, thus the techniques used in identifying early subtle alterations of heart rhythm could be predictor of impending IVH in preterm infants [21]. A recent study explored the potential application of detrended fluctuation analysis for identifying early subtle alterations of heart rhythm that could be predictive of impending IVH in very low birth weight babies [21]. The need for identifying babies with high risk of critical pathological conditions such as IVH has also motivated the development of clinical risk and disease severity scores for use in the neonatal intensive care setting.

## 2.4.2 CRIB II correlation

A previous study by Wong *et al.* has suggested that higher CRIB scores in preterm infants with high MAP-TOI coherence (in the ultra-low frequency range), the existence of a direct correlation has not been examined previously. Moreover, previous infant studies only investigated the use of HbD or TOI as a surrogate measure of cerebral perfusion [5-8], but it was not clear whether $HbO_2$ or HHb could also provide useful information for assessing clinical risk. The use of $HbO_2$ to represent cerebral haemodynamics in frequency domain analysis has been reported in adult studies [109, 110]. Recently HHb was found to not only exhibit high coherence with intracranial pressure (ICP) slow waves with good signal-to-noise ratio, but also demonstrate correlated changes with the ICP waves during intraventricular infusion [111]. It has therefore been suggested that HHb may be a useful marker of deoxygenated venous blood volume. The transfer function analysis of BP and cerebral NIRS measures has demonstrated that alterations in dynamic cerebral autoregulation properties, such as higher coherence or gain, are linked with IVH or other mortality risk factors [5-8].From these investigations, the relationships between CRIB II and various spectral measures of arterial baroreflex and cerebral autoregulation functions may potentially reflect impairment of arterial baroreflex and cerebral autoregulation functions, hence providing further justification for these measures as possible markers of clinical risks and predictors of adverse outcome.

## 2.5 Continuous cardiac monitoring

Measurement of CO and TRP can provide valuable information for the assessment and management of patients needing intensive care, including preterm infants in the neonatal intensive care unit (NICU) [22, 23]. However, such measurement requires specialized skill and equipment, and cannot be performed continuously. Many extremely low birth weight infants have impaired cerebral autoregulation (some with a complete absence of autoregulation) and thus a pressure passive cerebral circulation. Intracranial haemorrhage is more likely in this state of pressure passive circulation. Routine methods of estimating cardiac function in these preterm infants are dependent on echocardiographic estimates of volumetric flow or M-Mode estimates of ejection fraction. There are obvious limitations with the need of appropriately trained clinicians to perform these examinations and those measurements are valid for the short time period of the examination. There is thus an urgent need to explore new methods of continuous estimates of cardiac function to support clinical decisions to treat hypotension and support vulnerable cerebral circulation.

## 2.5.1 Estimation of Cardiac Output

Cardiac output, which is defined as the volume of blood ejected by the heart per unit time, is an important parameters evaluating the overall cardiac status both in the seriously ill patients and patients with suspected cardiovascular disease, like acute coronary syndromes, hypovolemia, valvular stenosis, myocarditis cardiomyopathy and arteriosclerosis [112]. CO is determined by many factors, including: cardiac contractility, the nutrient and oxygen requirements of the organs and tissues, blood volume status and systemic vascular resistance (SVR) [112, 113]. Low CO is a symptom of potential adverse cardiovascular conditions such as cardiogenic shock [114].

Thermodilution technique is a most common approach in current clinical settings and is considered as the gold standard for CO measurement [115].However, thermodilution measurement is an invasive procedure requiring the insertion of a pulmonary artery catheter which may involves many risks, such as dysrhythmias, pneumothorax and perforation of the heart chamber, tamponade and arterial valve damage[116]. With the development of new electrical filament wire of the thermodilution catheter, the continuous CO monitoring has been proposed in the 1990s to be as accurate as the single bolus method [117-119], but its accuracy was questioned during periods of hypothermia, such as coming off bypass during cardiac surgery [120, 121]. Furthermore, there is significant time delay of readings when a continuous CO thermodilution catheter is used. In order to reduce noise and improve reliability, the electric filament wire need up to 12 minutes to detect the CO changes. Therefore, the catheters should not be used to monitor rapid changes in CO, except the trend detection [122].

Some alternative methods to estimate CO non-invasively (or with minimal invasiveness) has been proposed recently, like Doppler ultrasound, thoracic bio-impedance and pulse contour analysis [115, 122, 123], however clinical observation still plays a most important role in the clinical decision-making process till now. The utilization of non-invasive measurements requires specialized skill and equipment, and cannot be performed continuously. As an example, although a series of more portable and dedicated Doppler ultrasound probes have been developed to measure blood flow, they still need the skilled personnel to operate [124]. On the other hand, the precision of this measurement was questioned by the uncertainty of the flow profile and diameter of the blood vessel [122], particularly when used in children. Thoracic bio-impedance method, which was used to detect the electrical resistance changes across the thorax during the cardiac cycle, has been controversial for the accuracy and validation in clinical use of critically ill and septic patients [113, 122].CO measurements can also be derived from the pulse contour analysis by continuously monitoring of the arterial pulse pressure waveform from a peripheral arterial pulse wave. The area under the pulse wave is related to CO and the classic Windkessel model of the circulation was used to convert pressure wave to a flow wave [122, 125]. Pulse contour analysis was regarded as a minimally invasive method, because central venous and arterial cannulations are assumed to already have in a critical care setting. However, it is not reliable to monitor the CO trend, since the influence of circulatory changes, such as peripheral resistance [122].

Furthermore, the development of reliable methods for non-invasive estimation of CO plays a crucial role in many problems of control of biomedical systems. For example, non-invasive control of blood flow has been used to provide long term alternative treatment for congestive heart failure patients [126-128]. A switching controller has also been proposed to control the drug delivery for critically ill patients [129, 130].

### 2.5.2 Estimation of Total Peripheral Resistance

Total peripheral resistance (TPR) refers to the resistance to blood flow offered by all of the systemic vasculature except the pulmonary vasculature. TPR is primarily determined by variations in blood vessel diameters, which are controlled by the sympathetic nervous system through changes of vasoconstrictor tone. With the increase of tone, the vessel diameters decrease and the TPR increases. TPR is inversely related to CO and can be calculated by MAP/CO. Although TPR can only be calculated from MAP and CO, it is not determined by either of these parameters. TPR illustrates the potential diagnostic values, as the high TPR may be an indicator of critical illness. For example, an increase in TPR may result in ischemic, hemorrhagic shock and hemorrhagic stroke [131] , while a depression in TPR may be observed in patients with sepsis or anaphylaxis [132].

## 2.6 Past techniques used to detect and prevent disease

Previous studies have suggested that the transfer function analysis of BP and cerebral NIRS measures has demonstrated that alterations in dynamic cerebral autoregulation properties, such as higher coherence or transfer gain, are linked with IVH or other mortality risk factors [5-8]. It has been assumed that the higher gain or coherence in the cerebral pressure-flow could reflect impairment of CBF control by cerebral autoregulation, thus providing the potential link with brain injury such as IVH or other mortality risk factors. The IVH infants had both higher TOI and lower FTOE than the non-IVH. The higher level of cerebral oxygenation in the IVH cases should not be seen as a positive sign, as it could well be a consequence of impaired oxygen extraction capability. In fact, a recent study also showed that cerebral oxygenation on the first day of life was higher in very preterm infants compared with healthy term infants, with the higher TOI again coinciding with lower FTOE in the preterm cases [133]. The reduced oxygen extraction was very likely to have an impact on the relationship between arterial BP and HHb, as the generation of HHb fluctuation was not only dependent on the BP-driven blood flow, but also on the oxygen consumption rate in the cerebral capillaries which governed the conversion of $HbO_2$ to HHb.

Furthermore, cross-spectral analysis was also performed on MAP-RRi to provide information about cardiac baroreflex sensitivity [51, 68], and also on MAP-HbD and MAP-TOI to provide information about cerebral autoregulation [5-8]. In addition, the cross-spectral relationships of MAP-$HbO_2$ and MAP-HHb were analysed to reveal the influence of perfusion pressure on the oxygenated and deoxygenated blood volume in the cerebral circulation. The existence of a correlation in the Low Frequency band (0.04–0.15 Hz) might relate in part to the relatively large contribution of MAP spectral power in this band. The presence of large MAP fluctuations could lead to better signal-to-noise ratio for the study of cerebral pressure-flow relationship, thus also more reliable detection of cerebral autoregulation [134]. A high coherence between arterial BP and NIRS-derived surrogate measures of CBF, such as HbD and TOI, has been interpreted as signs of a pressure-passive cerebral circulation, with blood flow being highly dependent on pressure due to impairment of autoregulation mechanisms [5-8].

Previous studies have utilized the DFA to evaluate the fractal dynamics of heart rate in various diseases in adult subjects [16, 17]. In a prior study, preterm infants who later developed IVH were found to have significantly higher short-term scaling exponent of heart rate compared with those who did not develop IVH [21].Since the altered autonomic control of the heart relates to the development of IVH, it is important to detect the difference in fractal dynamics of heart rate between these infants with and without IVH. ABP was proposed to be applied in the assessment of fractal scaling properties in adult humans based on DFA [18-20].Compared with HRV, there are several advantages from deriving clinical information using the DFA method. Firstly, this method can be performed on an acceptable duration of data (10 min in this case), that is practical in the neonatal intensive care setting. Secondly, the information (namely the scale exponent) is derived from a specific beat range, which can be easily translated to a similar time range or frequency band used in other methods such as power spectrum analysis. This would largely facilitate the physiological interpretation of the derived parameters. Thirdly, it can be applied in conjunction with the commonly used spectral analysis method to provide a more complete picture of the underlying cardiovascular dynamics, as both methods can provide complementary information [20].

There is an urgent need to explore new methods of continuous estimates of cardiac function to support clinical decisions to treat hypotension and support vulnerable cerebral circulation. Bourgeois et al. performed one of the most compelling investigations by using the central ABP waveform analysis technique without the delay of cardiac ejection time. They proposed the arterial tree based on the Windkessel model accounting for the lumped compliance of the large arteries (AC) and the TPR of the small arteries. Furthermore, Windkessel technique proved to be highly accurate, since the diastolic intervals of the central ABP waveform indeed decays like pure exponential. However, central ABP is seldom measured for monitoring in clinical settings. Routine methods of estimating cardiac function in these preterm infants are dependent on echocardiographic estimates of volumetric flow or M-Mode estimates of ejection fraction.

## 2.7 Summary

IVH is diagnosis of a serious brain damage and might lead to long term neurodevelopmental disabilities in preterm infants. The IVH grades was classified into 4 grades, from I to IV, which which represent increasingly severe stages of the IVH with deteriorated clinical outcomes. If the IVH can be detected at the earlier stage, some clinical methods, such as tresuscitation and cardiorespiratory management could be used to pretend the low level IVH extending to serious stage. Current clinical diagnosis of IVH includes frequency spectrum analysis and nonlinear analysis of the corresponding cardiovascular time series and NIRS measurements. To assess underlying risk factors of mortality and morbidity of preterm infants, the correlation between parameters derived from spectral analysis of systemic and cerebral cardiovascular variabilities and the CRIB score were established. On the other hand, LVO and TPR are two important parameters of the cardiovascular system which can help the diagnosis staff to take necessary preventive actions when the patients need intensive care.The estimation of LVO and TPR using arterial blood pressure waveform analysis has also been briefly reviewed in this chapter.

# Chapter 3. Frequency spectrum analysis: Monitoring Tool for IVH detection and adverse outcome in preterm infants

## 3.1 Overview

The mortality and morbidity of the IVH and high possibility of developing long-term neurodevelopmental disabilities [1, 30] have motivated researchers to conduct extensive studies, aiming at better understanding of its pathophysiology and development of new diagnostic and treatment techniques. Many extremely low birth weight infants suffered from impaired cerebral autoregulation (some with a complete absence of autoregulation) and thus a pressure passive cerebral circulation.

This chapter explores the possible ways of using spectral analysis of Near-infrared spectroscopy (NIRS), blood pressure (BP) and heart rate variability (HRV) as a monitoring tool for clinical risk assessment of preterm infants and the detection of IVH. Particular interest was also given to quantifying baroreflex sensitivity of heart rate (HR) by power spectral and transfer function analysis of HRV and BPV [53]. The relationships between CRIB II and various spectral measures of arterial baroreflex and cerebral autoregulation functions have provided these measures as possible markers of clinical risks and predictors of adverse outcome in preterm infants. NIRS for cerebral circulation monitoring has gained popularity in the neonatal intensive care setting, previous study have shown that the analysis of NIRS can provide a wide range of features that may be useful for tracking cerebral circulation in preterm infants, and thus the possibility of identifying preterm infants with IVH.

Section 3.2 starts with the development of the spectral analysis algorithm to assess the baroreflex sensitivity with HRV and BPV signals, and the cerebral autoregulation with NIRS signals. Section 3.3 then explores the potential use of the NIRS, BP and HRV for detecting IVH and examined the relationships between various spectral measures derived from systemic and cerebral cardiovascular variabilities and the CRIB II. The conclusions are drawn in Section 3.4.

## 3.2 Spectral analysis to find out the relationship with CRIB II by Linear regression

The ability to identify preterm infants with high risk of critical pathological conditions at an early stage, such as IVH has motivated the development of clinical risk and disease severity scores for use in the neonatal intensive care setting. Amongst these scores, the CRIB II has gained wide acceptance as an upgrade method of quantifying mortality risk amongst very preterm infants [3].

As mentioned in Section 2.3.1, impairment of cerebral autoregulation mechanism, however, may propose to CBF being more dependent on changes in arterial BP, or more "pressure passive", such that large changes in BP would be translated to large CBF fluctuations, hence increasing the risks of cerebrovascular injury and brain haemorrhage [5-8]. There is evidence showing that cerebral pressure passivity is common in sick preterm infants and may lead to IVH and other potentially serious consequences [5].Thus, spectral analysis of cerebral perfusion measurement have been utilized to identify preterm infants with high risk of mortality or IVH in the previous studies [5-8].

The cross-spectral transfer function analysis is widely used to assess alterations in dynamic cerebral autoregulation properties. For example, higher coherence, reduced phase shift and higher gain between arterial BP and CBF were regarded as possible indication of impaired autoregulation [9-11].In the real clinical setting of preterm infants, the NIRS has been used for deriving surrogate measures of CBF changes, because it can provide simple, non-invasive and comfortable continuous measurement of CBF. As mentioned in Section 2.2.1, the NIRS technique can quantify changes in cerebral oxygenation and haemoglobin content, such as the HbO2, HHb and TOI, which reflects oxygen saturation in brain vasculature [48]. It has been reported that the TOI reflects CBF variations during reduction in perfusion pressure, under the conditions of constant arterial oxygen saturation and cerebral oxygen consumption [49]. A number of investigators have also proposed that the HbD correlated with CBF during hypotensive episodes [5-7, 50]. This log transforms technique allows quantification of cerebral pressure passivity as MAP-HbD gain in premature infants. The transfer function analysis of BP and cerebral NIRS measures has demonstrated that alterations in dynamic cerebral autoregulation properties, such as higher coherence or gain, are linked with IVH or other mortality risk factors [5-8].

In this study, we examined how various measures derived from spectral and cross-spectral analysis of arterial BP and cerebral NIRS were related to CRIB II, an established clinical risk score for preterm infants. Whilst the previous study by Wong et al found higher CRIB scores in preterm infants with high MAP-TOI coherence (in the ultra-low frequency range), the existence of a direct correlation has not been examined previously. Moreover, previous infant studies only investigated the use of HbD or TOI as a surrogate measure of cerebral perfusion [5-8], but it was not clear whether HbO2 or HHb could also provide useful information for assessing clinical risk. The use of HbO2 to represent cerebral haemodynamics in frequency domain analysis has been reported in adult studies [109, 110]. Recently HHb was found to not only exhibit high coherence with intracranial pressure (ICP) slow waves with good signal-to-noise ratio, but also demonstrate correlated changes with the ICP waves during intraventricular infusion [111]. It has therefore been suggested that HHb may be a useful marker of deoxygenated venous blood volume.

In addition, we performed cross-spectral transfer function analysis of blood pressure variability (BPV) and heart rate variability (HRV) to derive measures related to arterial baroreflex modulation of autonomic tone [51, 53, 68]. Analysis of BPV and HRV showed a reduced cardiac baroreflex sensitivity (BRS) in adult patients with acute intra-cerebral haemorrhage, with BRS being able to predict adverse outcomes [69]. Whether BRS was related to clinical risks in preterm infants was not known, although a recent study found positive correlation between LF BRS gain and postmenstrual age (gestational age + postnatal age) [53], suggesting that low BRS in the high risk very preterm infants would be likely. The goal of this study was therefore to determine whether measures derived from spectral analysis of systemic (HRV and BPV) and cerebral NIRS cardiovascular signals were correlated with CRIB, which would help to establish these measures as potential indicators of clinical risk in preterm infants.

### 3.2.1 Methods

#### 3.2.1.1 Subjects

This was a prospective clinical investigation, carried out in the neonatal ICU at the Nepean Hospital in Sydney, Australia. The study was approved by the Sydney West Area Health Service Human Research and Ethics Committee, and informed parental consent was obtained in all cases. The cohort comprised early low birth weight infants with gestational age <30 wks, without significant congenital anomalies who survived at least 2 hours. A single measurement of data with 10 min continuous artefact-free segment from each baby was used for analysis. Out of the 20 infants with full set of measurements obtained, 17 infants had data recordings with sufficiently long artefact-free segment suitable for analysis (Data from 3 infants were rejected due to significant artefacts in the electrocardiogram or arterial blood pressure recordings). The demographic and clinical characteristics of the studied infants are as follows (mean±SD): 10 males 7 females, gestational age 26±2 wks (range 24-29 wks), birth weight 1024±351 g (range 552-1929 g), CRIB II score 10.3±2.5, MAP 34±6 mmHg,

RRi 428±24 ms, SaO$_2$ 95±4%, TOI 66±10%. Out of the 17 infants being studied, 16 infants were mechanically ventilated, 5 infants suffered from IVH and 1 infant died eventually (who did not suffer from IVH).

### 3.2.1.2 Measurements

CRIB II score was computed for each infant based on the previously described algorithm [27], which incorporates birth weight, gestational age, gender, temperature at admission, and base excess. It was scored at the first hour of life.

Physiological data were collected from the infants within 1-3 hours after birth. Electrocardiogram (ECG), arterial BP waveform and arterial blood oxygen saturation (SaO$_2$) were acquired via a bedside patient monitor (Philips Agilent Systems, Philip Healthcare, North Ryde, Australia). Intra-arterial BP was continuously measured via an umbilical or peripheral arterial catheter connected to a transducer calibrated to atmospheric barometric pressure and zeroed to the midaxillary point. Cerebral near-infrared spectroscopy data were measured by the NIRO-300 system (Hamamatsu Photonics, Hamamatsu City, Japan), with the sensor probe placed over the skin of the temporoparietal region on the head of the infant. The sensor probe of the NIRS-300 monitor consists of a fiber optic probe that emits four wavelengths of near-infrared light (775, 825, 850 and 904 nm), and a receiver probe made of separate aligned photodetectors, which permit the measurement of spatially resolved NIRS. The device provides measurement of the changes in relative concentration of HbO$_2$ and HHb, and also the TOI which is given as the ratio of HbO$_2$ to the total haemoglobin (HbO$_2$ + HHb), expressed as a percentage. The NIRS parameters are computed by the device every 0.167 s (6 Hz). HbD was calculated as HbO$_2$ – HHb as described previously [5-7, 50].

The ECG, arterial BP, SaO$_2$ and cerebral NIRS signals were simultaneously recorded by a data acquisition system (ADInstruments, Sydney, Australia) at a sampling rate of 1 kHz. The LabChart software (ADInstruments, Sydney, Australia) provided automatic beat-to-beat detection of R-waves from ECG for the computation of R-R interval (RRi) and heart rate (HR), and also the computation of mean arterial pressure (MAP) from the arterial BP waveform.

### *3.2.1.3 Signal processing*

Analysis was performed on uninterrupted data recordings of 10 minutes (obtained within 1-3 hours after birth from each infant) without noticeable corruption by artefacts. All signal processing was implemented in Matlab (Natick, MA, USA). Spikes in beat-to-beat time series of RRi due to occurrence of abnormal pulse intervals (e.g. prolonged heart period resulting from a missed heart beat), which constituted less than 3% of the total number of beats, were removed and replaced by linear interpolation of the beat values immediately preceding and following the replaced beats.

Movement artifact is a common problem in the biomedical signal processing, because biomedical sensors are sensitive to subject movement and can cause high amplitude short-time perfusions in the signals. These problems will bring the noises especially in the second-order statistics such as spectral analysis and correlation[135]. In heart rate series, for instance, the fluctuations in the normal sinus rhythm of the beat-to-beat interval are much lower than the movement artefact, and thus the HRV will be seriously distorted, especially in the power spectral density.

In this study, an automated method to reject noise spikes (ectopics or artifacts) for the long data set (≥10 mins) of RR interval was needed. In the real collected data, the incidence of artifact is rare and the amplitude of the artifact is very high, it has been suggested that impulsive artifact is more accurately modelled as a white noise process. Although it is highly possible that there is no noise at all, the high amplitude of noise will dramatically increase the mean value of the signal. Thus, the fixed threshold filters will not be suitable in the noise rejection of HRV. An impulse rejection filter for artefact removal in spectral analysis of HRV has been proposed by J. McNames [135] without the arbitrary threshold. It considers deviation from 5 points moving median filter and used an arbitrary threshold factor (1.483) to calculate the following test statistic D (n):

$$D(n) = \frac{|x(n)-x_m|}{1.483 \text{med}\{|x(n)-x_m|\}} \tag{3.1}$$

where $\text{med}\{\cdot\}$ means the median operation over the entire signal and $x_m$ is the median value of the entire signal, thus, $x_m = \text{med}\{x(n)\}$. The denominator is a robust evaluation of the standard deviation of an equivalent Gaussian signal.

For the long records, the median operation is replaced by moving window because of the non-stationary behaviour. The filtered signal was defined as $\hat{f}(n)$ and then was calculated as follows:

$$\hat{f}(n) = \begin{cases} x(n), & D(n) < \tau \\ \hat{f}_l(n), & D(n) \geq \tau \end{cases} \tag{3.2}$$

where $\tau$ is a user-specified threshold and $\hat{f}_l(n)$ is an interpolated value of the signal at the time point n. A median filter was used to calculate the interpolation value as follows:

$$\hat{f}_l(n) = \text{med}\left\{x(n+m): |m| \leq \frac{w_m-1}{2}\right\} \tag{3.3}$$

where $w_m$ is the window length of the segment centered about point n. Based on the comparison of filter results in different value of threshold and window length, the threshold of 4 and a window length of 5 sample points are selected in the final filter. As shown in the middle of Figure 3-1, it appears that most of the noise spikes can be removed with the threshold after the proper window length was selected. The noise spikes can be sufficiently deleted shown in the bottom of Figure 3-1.

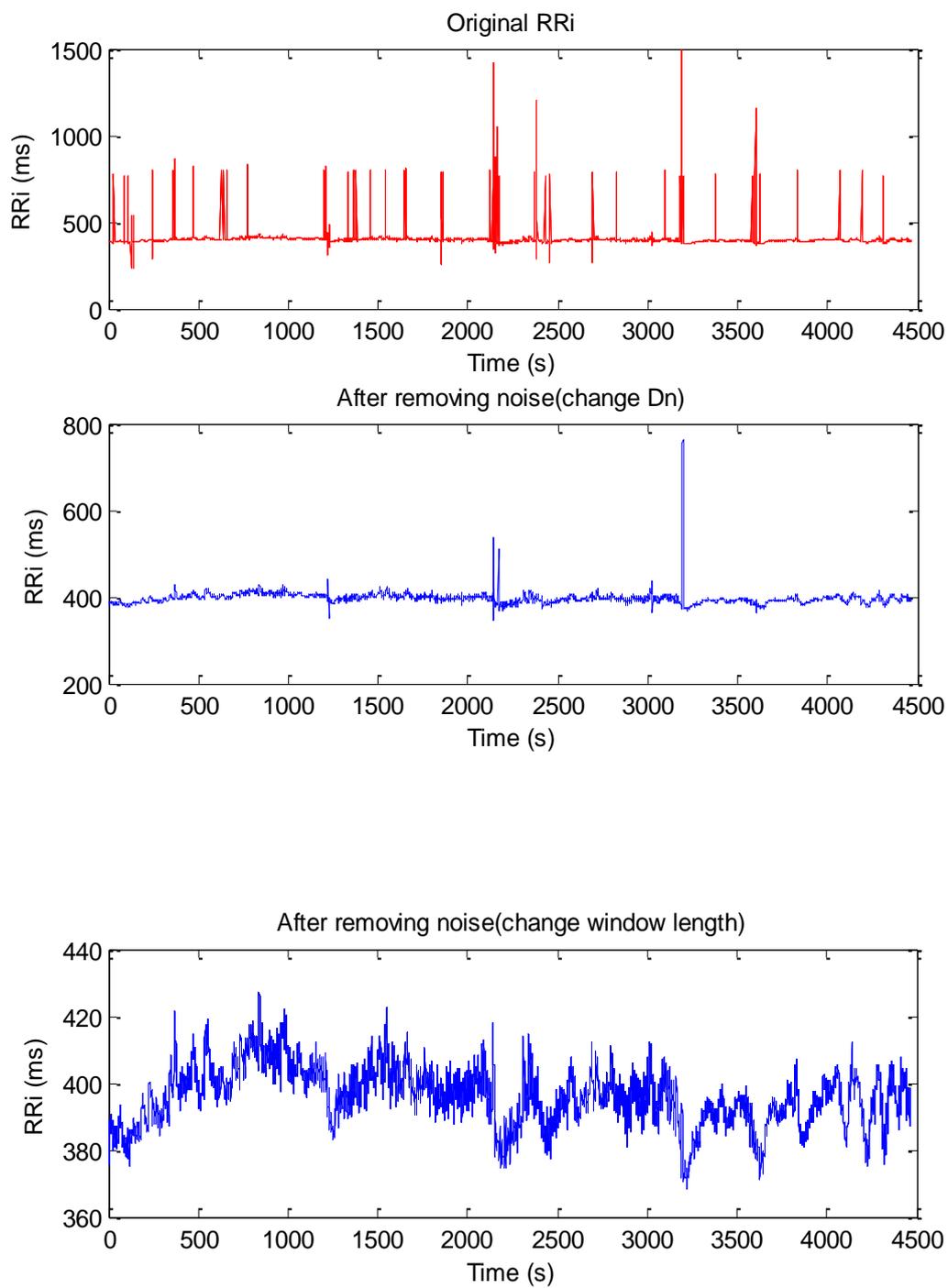

**Figure 3-1**: The results of impulse rejection filter in RRi.

### *3.2.1.4 Frequency spectrum analysis*

The RRi and MAP time series were converted to evenly spaced samples at 8 Hz using a previously proposed method to obtain the HRV and BPV signals [136]. The HbO2, HHb and TOI signals were also resampled to 8 Hz, by performing moving averaging (0.25 s window) with a sampling interval of 0.125 s. Linear detrending was performed on the resampled signals prior to spectral analysis. The power spectral density (PSD) of each signal was computed by the Welch method – this involved dividing each of the 10 min segments into smaller sections of 4 min with 75% overlap, multiplying each section with a Hanning window, performing a 2048-pt fast Fourier transform (FFT) then squaring the magnitude to give the auto power spectrum, and finally averaging the power spectra of all segments to obtain the PSD of the signal.

The Welch method was also used, which involved averaging the cross power spectra from multiple sub-segments in a similar fashion as the computation of PSD described earlier. With the auto power spectra of signals x and y defined as $P_{xx}(f)$ and $P_{yy}(f)$ respectively (*f* represents frequency) and the cross power spectrum of x-y as $P_{xy}(f)$, the transfer function of x-y (from x to y) was computed as follows:

$$H(f) = \frac{Pxy(f)}{Pxx(f)} \tag{3.4}$$

and the magnitude and phase angle of the transfer function could be obtained accordingly. The magnitude-squared coherence function was defined as follows:

$$\gamma 2(f) = \frac{|Pxy(f)|^2}{Pxx(f)Pyy(f)} \tag{3.5}$$

The coherence assesses the linear correlation between spectral components in the two signals, with a value ranging from 0 (lack of linear correlation) to 1 (perfect linear relationship). Cross-spectral analysis was performed on MAP-

$HbO_2$, MAP-HHb, MAP-HbD and MAP-TOI using the Welch method to provide information about cerebral autoregulation [5-8].

In this study, we have performed correlation analysis over a large number of parameters, including information from 3 different frequency bands and a number of NIRS measures. We believe this was necessarily given that each frequency band and each NIRS measure might carry different physiological meanings, and presenting all the information together would provide a more complete picture to the readers. For example, in the study of HRV and BPV, different frequency bands have been attributed to specific physiological mechanisms, and similar argument may apply to the study of cerebral autoregulation (although the precise mechanisms remain to be fully understood). The NIRS measures are also not equivalent, for example $HbO_2$ and HHb represent oxygenated and deoxygenated blood respectively, whose response to MAP perturbation would be different dependent on the tissue metabolic status (a key factor of autoregulation). The value of analysing multiple bands and NIRS measures was highlighted by the fact that TOI VLF percentage power and MAP-HHb LF coherence (two different measures in different bands) both exhibited correlation with CRIB, with TOI VLF percentage power also reflecting similar information as the baroreflex sensitivity measure.

Since we have chosen to perform correlation analysis on all the NIRS measures ($HbO_2$, HHb, HbD and TOI) in all frequency bands a priori, and that each measure and each band is considered different from each other, we did not apply post-hoc correction for multiple tests. However with transfer function analysis of baroreflex, we have removed SBP-RRi and only reported MAP-RRi given the similarity of meaning conveyed by the two measures, and also to be consistent with the cerebral autoregulation analysis that used MAP.

The cardiovascular variability spectra of the infants were divided into three bands, based on those described previously [5, 8, 53]: the very low frequency band (VLF, 0.02–0.04 Hz), the low frequency band (LF, 0.04–0.15 Hz) and the mid frequency band (MF, 0.15–0.25 Hz). The current definition of LF band is the same as that widely used in human adult studies, considered to be under the influences of the autonomic nervous system (sympathetic and vagal nerve activities) and vasomotor regulation mechanisms [51]. A previous infant study has used this band to quantify baroreflex sensitivity via transfer function analysis of BPV and HRV [53]. In another study, a wider frequency range (0.05–0.25 Hz) was used as the LF band in the study of cerebral autoregulation based on transfer function analysis of MAP and HbD [5]. This expanded range includes the "gap" (0.15-0.25 Hz) between the conventional LF band and the breathing frequency of infants (> 0.25 Hz), which has been assigned as the MF band in the current study. The spectral band below 0.04 Hz is typically considered as the VLF band, whose meaning is not totally clear although the influence of local autoregulation mechanism has been implicated [51, 111]. The current definition of VLF band is 0.02-0.04 Hz, similar to that adopted in a previous infant study (0.02-0.05 Hz) for the study of cerebral autoregulation based on transfer function analysis of MAP and TOI [8]. Other studies of cerebral autoregulation in infants also examined the very slow trend at < 0.02 Hz, termed the ultra-low frequency (ULF) band [6-8], although this would not be included in the current analysis given the difficulty in deriving reliable estimate of this band based on a relatively short segment of 10 min.

The relative and percentage powers reflect the relative contribution of different oscillations in the signals, and can help to reduce the large inter-subject variability in the spectral powers that hinders comparison. In the presentation of relative powers, we have made no assumption regarding underlying relationship between oscillations in different bands (e.g. when %LF increases, it merely means that the percentage contribution of LF power in the total power has increased). We have found significant relationships with CRIB for spectral powers expressed in relative terms but not in absolute terms, which reinforced the idea that the relative spectral powers carried important information not apparent from the absolute powers.

Within each band, the spectral power was obtained by integrating the power spectrum over the given frequency range. The percentage power (%VLF, %LF and %MF) was computed by dividing the power in each band by the sum of the three bands, expressed as percentage. The normalised mid frequency power (nMF) was also obtained by dividing the MF power by the sum of LF and MF powers. The transfer function parameters (gain, phase and coherence) in each band were computed by averaging values at frequencies with coherence ≥ 0.5, or from the frequency with the maximum coherence if coherence < 0.5, as this represented the frequency at which the signals were most strongly correlated [68, 137]. The proportion of shared variance in both signals would be at least 50% at coherence ≥ 0.5. In addition, the average coherence of the whole band was computed by averaging all values within each band.

### *3.2.1.5 The coherence spectrum analysis for more than two signals*

The coherence spectrum was used in the section 3.2.1.4 to measure of the linear statistical relationship between two time series, however the coherence spectrum analysis for more than two signals has never been studied. Recently, a method to include more than two processes in the coherence spectrum analysis haw been proposed by David et al. [138] .

In the coherence analysis between two signals, the magnitude squared coherence spectrum (MSC) was often used to provide a frequency-dependent measure of the linear relationship between two stationary random processes. However, when more than two signals are involved in the calculation of MSC, it is not practical to measure all of the pairwise MCS between these signals. In order to include more signals in the coherence analysis, the generalized magnitude squared coherence (GMSC) spectrum is defined as a function of the largest eigenvalue of a matrix including all the complex coherence spectra. The value changes from zero to one, which attains its maximum when the signals are perfectly related with each other and provide a tool for us to quantify the contribution of each signal to the overall correlation.

The M ≥2 signals are $x_1[n]$,…, $x_M[n]$, the complex coherence spectrum between the *i*-th and *j*-th signals are defined as follows:

$$C_{x_ix_j}(w) = \frac{S_{x_ix_j}(w)}{\sqrt{S_{x_ix_i}(w)S_{x_ix_j}(w)}} \quad , \qquad \forall\ i, j = 1, 2, 3 \tag{3.6}$$

Where $S_{x_ix_j}(w)$ is the cross-spectrum and $S_{x_ix_i}$ is the power spectral density of the *i*-th signal. In order to extend the method to general case (M ≥2), a matrix is defined as follows, which contains all pairwise complex coherence spectra:

$$\Sigma_x(w) = \begin{bmatrix} 1 & C_{x_1x_2}(w) & \cdots & C_{x_1x_M}(w) \\ C_{x_2x_1}(w) & 1 & \cdots & C_{x_2x_M}(w) \\ \vdots & & \ddots & \vdots \\ C_{x_Mx_1}(w) & C_{x_Mx_2}(w) & \cdots & 1 \end{bmatrix} \tag{3.7}$$

which can be rewritten as follows:

$$\Sigma_x(w) = D_x^{-1/2}(w) S_x(w) D_x^{-1/2}(w) \tag{3.8}$$

where

$$S_x(w) = \begin{bmatrix} S_{x_1x_1}(w) & S_{x_1x_2}(w) & \cdots & S_{x_1x_M}(w) \\ S_{x_2x_1}(w) & S_{x_2x_2}(w) & \cdots & S_{x_2x_M}(w) \\ \vdots & & \ddots & \vdots \\ S_{x_Mx_1}(w) & S_{x_Mx_2}(w) & \cdots & S_{x_Mx_M}(w) \end{bmatrix} \tag{3.9}$$

and $D_x(w)$ is a diagonal matrix whose entries are $[D_x(w)]_{i,i} = S_{x_ix_i}(w)$.

Then, the GMSC is defined as $\gamma^2$, where

$$\gamma(w) = \frac{1}{M}(\lambda_{MAX}(\Sigma_x(w)) - 1) \tag{3.10}$$

Here *w* is each frequency point and $\lambda_{MAX}(\Sigma_x(w))$ is the largest eigenvalue of the matrix $\Sigma_x(w)$. Then we could get the matrix at every sampling point and calculate the maximal eigenvalue of each matrix to calculate the GMSC. The GMSC spectrum is maximum when the M time series are perfectly pairwise correlated at that frequency point, and minimum when all the signals are uncorrelated.

Based on the linear regression results from section 3.2.1.4 , three to four signals were selected to do the three to four coherence spectrum analysis. For M=3, the matrix is simplified as follows:

$$\Sigma_x(w) = \begin{bmatrix} 1 & c_{x_1 x_2}(w) & c_{x_1 x_3}(w) \\ c_{x_2 x_1}(w) & 1 & c_{x_2 x_3}(w) \\ c_{x_3 x_1}(w) & c_{x_3 x_2}(w) & 1 \end{bmatrix} \tag{3.11}$$

$$\gamma(w) = \frac{1}{2}(\lambda_{MAX}(\Sigma_x(w)) - 1) \tag{3.12}$$

It was hypothesized in this study that the GMSC of three or four signals will increase the value of coherence of the related signals found in the coherence analysis in two signals.

### *3.2.1.6 Data analysis*

The results for the cohort were presented as mean ± SD. Least-squares linear regression analysis was carried out between CRIB II score and each of the physiological/spectral parameters. The Pearson's correlation coefficient (*r*) was computed, and the relationship between two variables was considered significant if $P < 0.05$. In addition, regression analysis was performed between the gestational age and each of the parameters. For a particular type of variables (spectral power, transfer gain, coherence or phase) whose majority of parameters (>50%) did not follow normal distribution (based on the Kolmogorov-Smirnov test), logarithmic transformation would be performed to provide normally distributed parameters, prior to performing regression analysis. We applied transformation on a class of variable rather than on individual variables, to avoid inconsistency in the data being presented and to facilitate comparison of variables within the same class. Based on the normality test, transformation was performed on the absolute spectral power and transfer gain parameters.

## 3.2.2 Results

The values of the spectral powers and transfer function parameters for the cohort are summarised Table 3-1 and Table 3-2 respectively, and the correlations between CRIB II score and various parameters are shown in Table 3-3. Logarithmic transformation was performed on the absolute spectral powers and transfer gains to provide normally distributed parameters, prior to performing regression analysis. A number of significant and moderate correlations were identified; with respect to the percentage/normalised powers, positive correlations with CRIB II were found in MAP nMF ($r = 0.61$, $P = 0.009$), MAP %MF ($r = 0.56$, $P = 0.02$), and TOI %VLF ($r = 0.55$, $P = 0.023$). The transfer function gain (after logarithmic transformation) of MAP-RRi in the LF band, which represents cardiac BRS, was inversely related to CRIB II ($r = -0.61$, $P = 0.01$).

Analysis of percentage and normalised powers from MAP and TOI frequency spectra also revealed interesting correlations with CRIB II. Relative dominance of VLF fluctuations in TOI (as illustrated in Figure 3-5) was associated with higher CRIB II score, and quite a strong negative correlation was seen between TOI %VLF and gestation age ($r = -0.75$, $P = 0.0005$). The suppression of LF and MF powers would be the main reason for this relative dominance of VLF band, given that the absolute powers in these two bands had weak (but non-significant) negative correlation with CRIB (Table 3-3). A possibility existed that the lower BRS (with lower gestational age) might relate to the suppression of LF and MF powers in TOI, given the potential link between baroreflex and cerebral autoregulation mechanisms [139]. As shown in the scatter plots (Figure 3-2), it was clear that the 5 infants with the highest CRIB scores of 12-15 all had low BRS and high TOI %VLF, suggesting possible connection between the two mechanisms. It was noteworthy that significant relationships with CRIB were found when spectral powers were expressed in relative terms, but not in absolute terms, which reinforced the idea that the relative spectral powers carried important information not apparent from the absolute spectral powers.

**Table 3-1:** Spectral powers of systemic and cerebral cardiovascular signals in the cohort of preterm infants.

|  |  | VLF | LF | MF |
|---|---|---|---|---|
| **RRi** | Power (ms$^2$) | 47±87 | 69±120 | 14±22 |
|  | % power | 33±17 | 47±10 | 20±13 |
|  | nMF |  |  | 27±15 |
| **MAP** | Power (mmHg$^2$) | 0.13±0.09 | 0.22±0.18 | 0.06±0.05 |
|  | % power | 35±12 | 53±10 | 13±4 |
|  | nMF |  |  | 19±6 |
| **HbO$_2$** | Power ((µmol/L)$^2$) | 0.047±0.042 | 0.074±0.082 | 0.020±0.050 |
|  | % power | 45±19 | 47±15 | 8±10 |
| **HHb** | Power ((µmol/L)$^2$) | 0.021±0.037 | 0.051±0.120 | 0.028±0.092 |
|  | % power | 41±19 | 47±13 | 12±13 |
| **HbD** | Power ((µmol/L)$^2$) | 0.043±0.026 | 0.047±0.039 | 0.014±0.024 |
|  | % power | 50±22 | 40±15 | 9±12 |
| **TOI** | Power (%$^2$) | 0.11±0.19 | 0.29±0.46 | 0.19±0.35 |
|  | % power | 26±12 | 50±7 | 24±12 |

Values are mean ± SD. VLF, very low frequency (0.02-0.04 Hz); LF, low frequency (0.04-0.15 Hz); MF, mid frequency (0.15-0.25 Hz); RRi, R-R interval; MAP, mean arterial pressure; HbO2, oxygenated haemoglobin; HHb, deoxygenated haemoglobin; HbD, haemoglobin difference; TOI, tissue oxygenation index; % power, the band power divided by sum of powers in all three bands; nMF, normalised mid-frequency power computed from MF/(LF+MF).

**Table 3-2:** Transfer function analysis of systemic and cerebral cardiovascular signals in the cohort of preterm infants.

|  |  | VLF | LF | MF |
|---|---|---|---|---|
| **MAP-RRi** | Gain (ms/mmHg) | 10.9±14.4 | 11.6±11.3 | 13.2±14.9 |
|  | Coherence | 0.56±0.13 | 0.61±0.06 | 0.60±0.06 |
|  | Phase (rad) | 0.16±2.07 | -0.46±1.32 | 0.05±1.45 |
| **MAP-HbO2** | Gain (µmol/L/mmHg) | 0.44±0.17 | 0.41±0.28 | 0.36±0.31 |
|  | Coherence | 0.58±0.15 | 0.67±0.06 | 0.67±0.08 |
|  | Phase (rad) | 0.21±1.69 | -0.65±0.94 | -0.90±0.98 |
| **MAP-HHb** | Gain (µmol/L/mmHg) | 0.22±0.13 | 0.27±0.31 | 0.32±0.41 |
|  | Coherence | 0.57±0.16 | 0.65±0.05 | 0.62±0.06 |
|  | Phase (rad) | -0.35±1.86 | 0.08±0.88 | -0.37±1.17 |
| **MAP-HbD** | Gain (µmol/L/mmHg) | 0.43±0.18 | 0.34±0.21 | 0.36±0.35 |
|  | Coherence | 0.59±0.16 | 0.65±0.08 | 0.64±0.06 |
|  | Phase (rad) | -0.09±1.52 | -0.77±0.81 | -0.98±0.95 |
| **MAP-TOI** | Gain (%/mmHg) | 0.56±0.42 | 0.94±0.89 | 1.39±1.45 |
|  | Coherence | 0.56±0.10 | 0.62±0.04 | 0.60±0.04 |
|  | Phase (rad) | -0.50±1.66 | -0.60±0.79 | -0.12±1.25 |

Values are mean ± SD. VLF, very low frequency (0.02-0.04 Hz); LF, low frequency (0.04-0.15 Hz); MF, mid frequency (0.15-0.25 Hz); RRi, R-R interval; MAP, mean arterial pressure; HbO2, oxygenated haemoglobin; HHb, deoxygenated haemoglobin; HbD, haemoglobin difference; TOI, tissue oxygenation index. Gain, phase and coherence are derived from cross-spectral transfer function analysis of the given pair of signals.

**Table 3-3:** Correlations between CRIB II and spectral powers of systemic and cerebral cardiovascular signals in the cohort of preterm infants.

|  |  | VLF | | LF | | MF | |
|---|---|---|---|---|---|---|---|
|  |  | *r* | *P* | *r* | *P* | *r* | *P* |
| **RRi** | Power | -0.33 | 0.20 | -0.38 | 0.13 | -0.37 | 0.14 |
|  | % power | 0.08 | 0.77 | -0.27 | 0.30 | 0.11 | 0.68 |
|  | nMF |  |  |  |  | 0.11 | 0.68 |
| **MAP** | Power | 0.08 | 0.76 | 0.16 | 0.54 | 0.43 | 0.08 |
|  | % power | -0.17 | 0.52 | -0.05 | 0.86 | 0.56 | 0.020* |
|  | nMF |  |  |  |  | 0.61 | 0.009* |
| **HbO2** | Power | -0.06 | 0.81 | -0.09 | 0.73 | -0.22 | 0.40 |
|  | % power | 0.13 | 0.62 | -0.03 | 0.91 | -0.21 | 0.42 |
| **HHb** | Power | -0.08 | 0.75 | -0.21 | 0.42 | -0.23 | 0.38 |
|  | % power | 0.35 | 0.17 | -0.30 | 0.24 | -0.22 | 0.40 |
| **HbD** | Power | 0.15 | 0.57 | -0.11 | 0.69 | -0.28 | 0.28 |
|  | % power | 0.31 | 0.23 | -0.21 | 0.43 | -0.30 | 0.25 |
| **TOI** | Power | -0.09 | 0.73 | -0.25 | 0.34 | -0.33 | 0.19 |
|  | % power | 0.55 | 0.023* | -0.20 | 0.45 | -0.44 | 0.076 |

Correlation coefficients and p-values (r and P) from linear regression against CRIB II were presented. Logarithmic transformation was performed on the spectral power values prior to regression analysis. VLF, very low frequency (0.02-0.04 Hz); LF, low frequency (0.04-0.15 Hz); MF, mid frequency (0.15-0.25 Hz); RRi, R-R interval; MAP, mean arterial pressure; HbO2, oxygenated haemoglobin; HHb, deoxygenated haemoglobin; HbD, haemoglobin difference; TOI, tissue oxygenation index; % power, the band power divided by sum of powers in all three bands; nMF, normalised mid-frequency power computed from MF/(LF+MF). *P < 0.05.

With respect to the cerebral autoregulation parameters, MAP-HHb LF coherence was positively related to CRIB II with the highest correlation overall ($r = 0.64$, $P = 0.006$). MAP-HbO$_2$ MF phase also had an inverse relationship ($r = -0.58$, $P = 0.014$), although examination of the regression plot showed that this relationship was mainly driven by one subject who had much higher phase and lower CRIB II compared with the rest of the group. The regression plots of the important relationships found are shown in Figure 3-2.

Further analysis showed correlations of gestational age with MAP-RRi LF gain (BRS) ($r = 0.70$, $P = 0.002$ after logarithmic transformation of the gain) and TOI %VLF ($r = -0.75$, $P = 0.0005$), with values higher than those between CRIB II and the parameters. Other parameters, however, had lower correlations with gestational age compared with CRIB II: MAP nMF ($r = -0.42$, $P = 0.09$), MAP-HHb LF coherence ($r = -0.60$, $P = 0.01$) and MAP-HbO2 MF phase ($r = 0.47$, $P = 0.06$).

Figure 3-3 a-b shows the signal traces from two representative infants who had low/high CRIB II scores (5 and 14 respectively) along with low/high MAP-HHb LF coherences. Note the presence of similar low frequency oscillations (~6-10 s) in MAP and HHb in the high CRIB score case, in contrast with the dissimilar fluctuations in the low CRIB score case. The corresponding MAP-HHb coherence plots of these infants are shown in Figure 3-4, which illustrate the high LF coherence (close to 1 at 0.14 Hz) in the high CRIB score case. The TOI variability and its frequency spectrum for these two infants are displayed in Figure 3-5. The presence of oscillations in the LF and MF range was noted in the low CRIB score case, in contrast with the lack of oscillations in the high CRIB score case, with the subject having high %VLF due to the relative dominance of VLF over LF and MF fluctuations. The MAP variability spectrum of the low CRIB score infant is also shown in Figure 3-6a, along with an infant with relatively high CRIB score of 11 and high MAP nMF in Figure 3-6b. Note the occurrence of sharp MAP transients that introduce MF power in the frequency spectrum, contributing to the high nMF.

**Table 3-4:** Coherence analysis of systemic and cerebral cardiovascular signals in the cohort of preterm infants.

|  |  | VLF | LF | MF |
|---|---|---|---|---|
| **MAP-RRi** | Coherence | 0.56 | 0.61 | 0.60 |
| **MAP-HbO$_2$** | Coherence | 0.58 | 0.67 | 0.67 |
| **MAP-HHb** | Coherence | 0.57 | 0.65 | 0.62 |
| **MAP-HbD** | Coherence | 0.59 | 0.65 | 0.64 |
| **MAP-TOI** | Coherence | 0.56 | 0.62 | 0.60 |
| **RRi-HbD-MAP** | Coherence | 0.29 | 0.30 | 0.34 |
| **RRi-HbO$_2$-MAP** | Coherence | 0.41 | 0.48 | 0.09 |
| **RRi-TOI-MAP** | Coherence | 0.33 | 0.27 | 0.15 |
| **RRi-MAP-HbD-HbO2** | Coherence | 0.27 | 0.30 | 0.38 |

Values are mean VLF, very low frequency (0.02-0.04 Hz); LF, low frequency (0.04-0.15 Hz); MF, mid frequency (0.15-0.25 Hz); RRi, R-R interval; MAP, mean arterial pressure; HbO$_2$, oxygenated haemoglobin; HHb, deoxygenated haemoglobin; HbD, haemoglobin difference; TOI, tissue oxygenation index. Gain, phase and coherence are derived from cross-spectral transfer function analysis of the given pair of signals.

For the coherence spectrum analysis multiple signals (3 or 4 signals), the results are summarised in Table 3-4: Coherence analysis of systemic and cerebral cardiovascular signals in the cohort of preterm infants.Table 3-4.The GMSC of 3 to 4 signals did not show higher than the coherence of two signals. While our results were seemingly at odd that the GMSC of RRi-HbO2-MAP in MF band is only 0.09, as the coherence of HbO2-MAP and MAP-RRI both have quite high correlation in MF band. Future studies should examine the longer recordings with multiple measurements to improve the reliability of multiple coherence analysis.

### 3.2.3 Discussion

This study addressed the relationship between CRIB II, an established clinical risk score for preterm infants [27], and various measures derived from spectral analysis of systemic and cerebral cardiovascular variability, which were previously found to be linked with brain haemorrhage (IVH) or other mortality risk factors [5-8]. Significant moderate correlations were found between CRIB II and spectral measures that potentially reflect impairment of arterial baroreflex and cerebral autoregulation functions, hence providing further justification for these measures as possible markers of clinical risks and predictors of adverse outcome.

The MAP-HHb LF coherence, which presumably reflects the level of cerebral autoregulation, was positively related to CRIB II, with the highest overall correlation ($r = 0.64$, $P = 0.006$). The existence of a correlation in the LF band but not the VLF or MF band might relate in part to the relatively large contribution of MAP spectral power in this band (as shown in Figure 3-2). The presence of large MAP fluctuations could lead to better signal-to-noise ratio for the study of cerebral pressure-flow relationship, thus also more reliable detection of cerebral autoregulation [134]. A high coherence between arterial BP and NIRS-derived surrogate measures of CBF, such as HbD and TOI, has been interpreted as signs of a pressure-passive cerebral circulation, with blood flow being highly dependent on pressure due to impairment of autoregulation mechanisms [5-8]. The study of MAP-HHb coherence, however, has not been reported previously. The fact that highest correlation was obtained from MAP-HHb coherence would suggest that this relationship might hold important clinical value, thus should not be neglected. As shown in Figure 3-3 B, it appears that MAP-synchronised LF oscillations are more apparent in HHb compared with $HbO_2$ in the high CRIB score infant. As the oxygen-carrying arterial blood travels through the capillary compartment in the cerebral microvasculature, the oxygen consumption by the brain tissue metabolism would convert oxygenated haemoglobin $HbO_2$ into deoxygenated haemoglobin HHb, that subsequently enters the venous blood. The high MAP-HHb LF coherence might suggest the lack of vascular control by cerebral autoregulation mechanisms operating in the LF range [9], such that arterial pressure change was able to directly drive blood flow through a cerebral microvascular network that consumed oxygen and facilitated the conversion of $HbO_2$ to HHb. The physiological significance of the MAP-HHb relationship and its implication on cerebral autoregulation warrants further investigations in future.

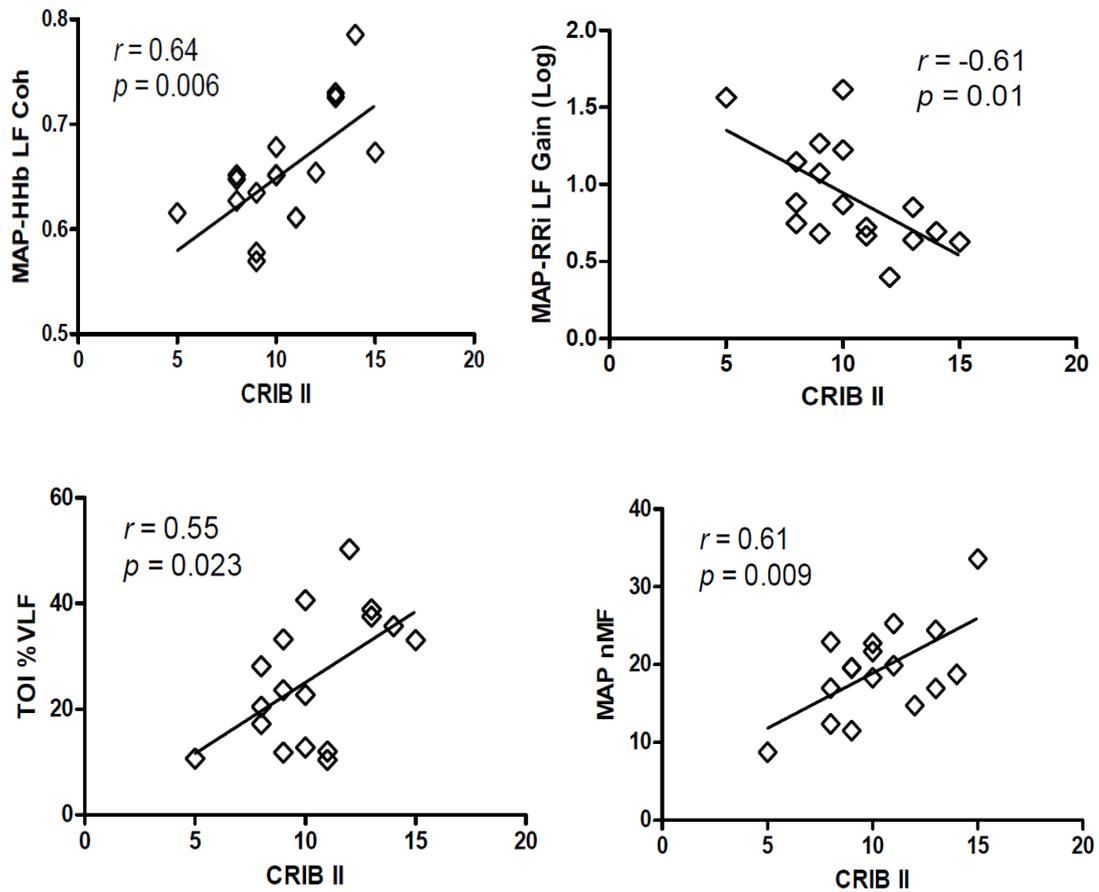

**Figure 3-2:** Linear regression analysis of spectral measures derived from mean arterial pressure (MAP), heart period (RRi) and cerebral near infrared spectroscopy (NIRS) versus Clinical Risk Index for Babies (CRIB II): (a) Top left: low frequency coherence of MAP and deoxygenated haemoglobin (HHb); (b) Top right: low frequency transfer gain of MAP and RRi (after logarithmic transformation); (c) Bottom left: very low frequency percentage power of tissue oxygenation index (TOI); (d) Bottom right; MAP normalised mid-frequency power.

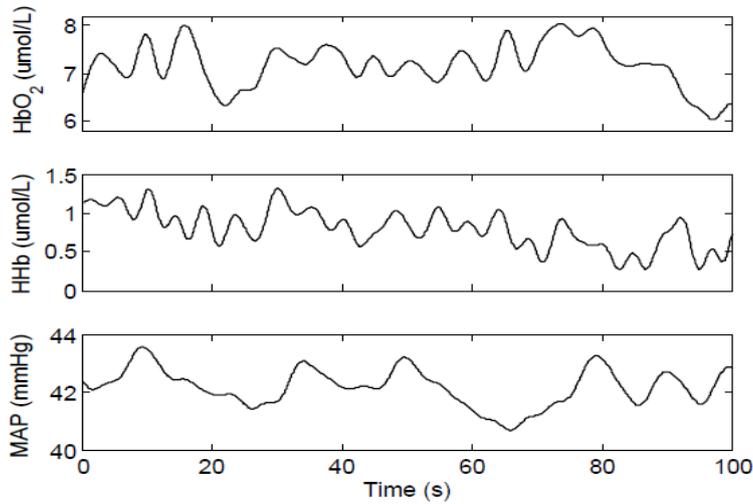

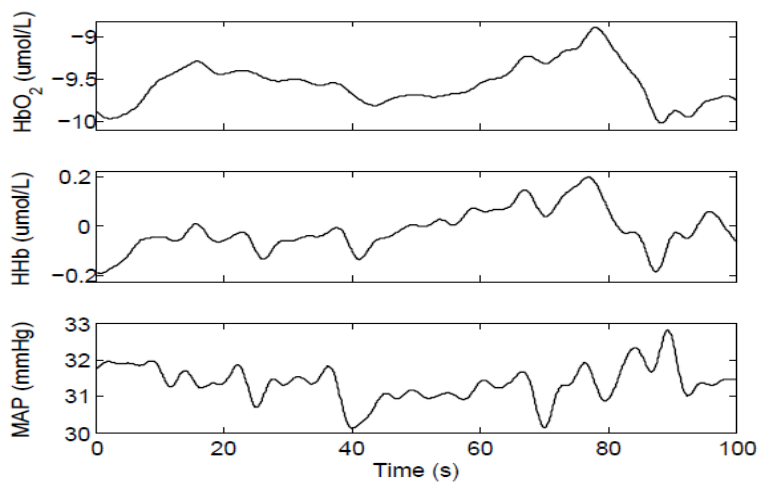

**Figure 3-3:** Snapshots of mean arterial pressure (MAP) and cerebral oxygenated and deoxygenated haemoglobin (HbO2 and HHb) in one infant with a low CRIB II score of 5 (panel A) and another infant with a high CRIB II score of 14 (panel B). The HbO2 and HHb signals shown represent changes in concentration rather than the absolute values. Note the presence of similar low frequency oscillations (~6-10 s) between MAP and HHb in the high CRIB score case (panel B) who had high MAP-HHb coherence (Figure 3-4, bottom), in contrast with the dissimilar fluctuations in the low CRIB score case (upper panel) who had lower coherence (Figure 3-4, top).

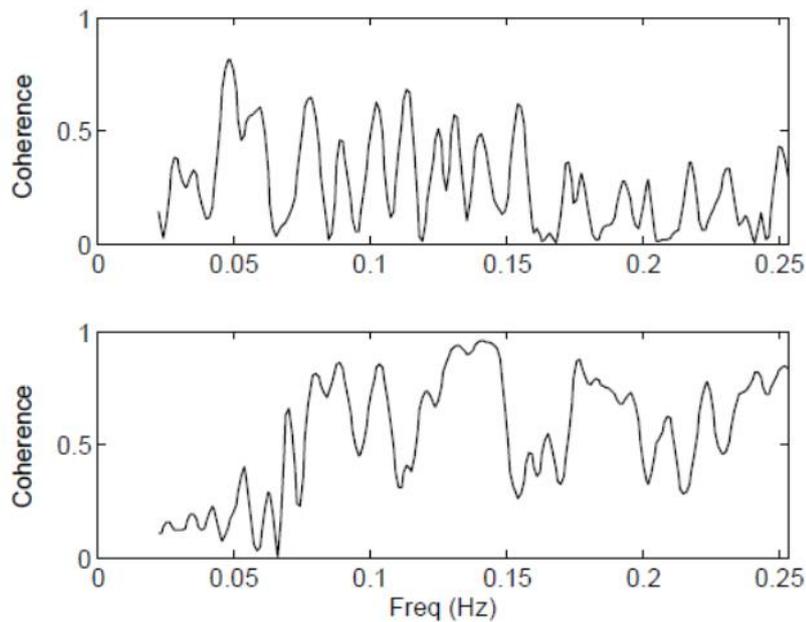

**Figure 3-4**: Cross-spectral coherence of mean arterial pressure (MAP) and cerebral deoxygenated haemoglobins (HHb) of one infant (top) with a low CRIB II score of 5 and relatively low coherence in the low frequency range (LF, 0.04-0.15 Hz), and another infant (bottom) with a high CRIB II score of 14 and high LF coherence (close to 1 at 0.14 Hz). MAP and HHb show similar LF oscillations in the high coherence case and dissimilar fluctuations in the low coherence case, as illustrated in Figure 3-3.

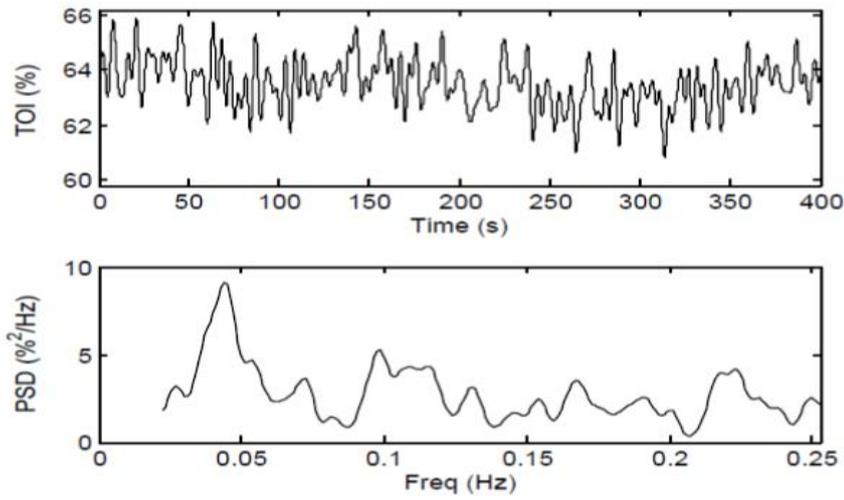

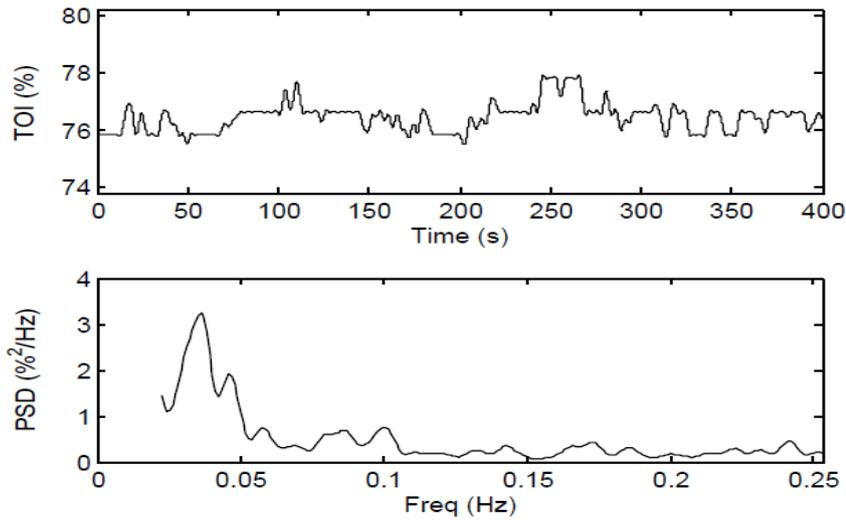

**Figure 3-5:** Snapshots of tissue oxygenation index (TOI) variability and its frequency spectrum in one infant with a low CRIB II score of 5 (panel A) and another infant with a high CRIB II score of 14 (panel B). TOI traces were lowpass filtered to <0.25 Hz to highlight the fluctuations being analysed. Note presence of low frequency (LF) and mid frequency (MF) oscillations in the low CRIB score case, in contrast with the lack of oscillations in the high CRIB score case. This subject also had a comparatively high VLF percentage power (%VLF, in 0.02-0.04 Hz).

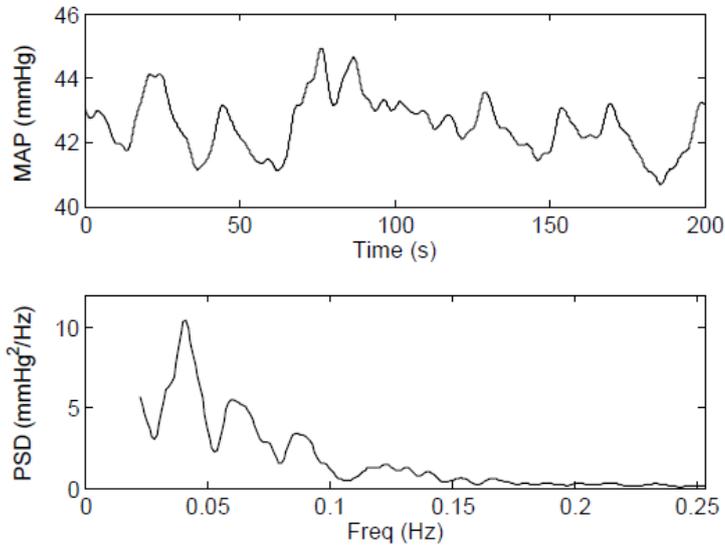

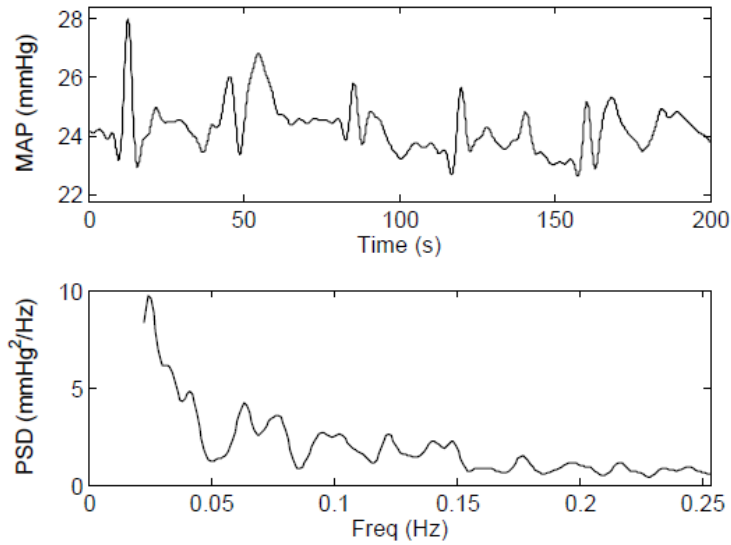

**Figure 3-6:** Snapshot of mean arterial pressure (MAP) variability and its frequency spectrum in one infant with low CRIB II score of 5 (panel A) and another infant with relatively high CRIB II score of 11 (panel B). MAP traces were lowpass filtered to <0.25 Hz to highlight the fluctuations being analysed. Note presence of somewhat periodic sharp MAP transients (panel B) that introduce spectral powers in the mid frequency range (0.15-0.25 Hz), leading to high normalised MF power in the high CRIB score case.

Cardiac baroreflex sensitivity, measured by transfer function gain of BPV and HRV in the LF range, was inversely related to CRIB II. The negative phase (BP leading RRi) and >0.5 coherence in this band justified the possible existence of a baroreflex-driven relationship [68]. The correlation of BRS and CRIB II was believed to be partly due to the relationship between BRS and gestational age ($r = 0.70$, $P = 0.002$ after logarithmic transformation), a major component of the CRIB II score [27]. Previous studies of human infants showed strong positive correlation ($r = 0.80$) between postmenstrual age (gestational age + postnatal age) and BRS (from LF transfer gain), suggesting a progressive maturation of baroreflex vagal function with age [53], which may explain the positive relationship between BRS and gestational age observed in our study. From Figure 3-2b, it seems clear that infants with relatively high CRIB II (> 10) all had low BRS, whilst those with lower CRIB II displayed a wide range of BRS. The depressed BRS may contribute to increased clinical risks, as the study of adult patients showed a reduced cardiac BRS with acute intra-cerebral haemorrhage, with BRS being able to predict adverse outcomes [69].

Dominance of MF over LF fluctuations in MAP was related to a higher CRIB II score. The physiological significance of the MF band in infants has not been addressed previously; in our study, this band (0.15-0.25 Hz) was defined as the "gap" between the respiration frequency of infants (> 0.25 Hz) and the conventional LF band (0.04-0.15 Hz) that was typically adopted in human adult studies. Observations of our infant data showed that the relative increase in MF power seemed to arise from sharp MAP transients, which introduced relatively high frequency spectral components at the MF range (Figure 3-6b). The presence of these BP transients might relate to the lack of compensation by the immature baroreflex mechanism, as in the case of normal adults, the baroreflex should be capable of dampening the BP change by adjustment of heart rate and/or peripheral vascular resistance.

## 3.2.4 Limitations

The rationale of this study was to establish the link between parameters derived from spectral analysis of systemic and cerebral cardiovascular variabilities and clinical risk of preterm infants, via examining the correlation of these parameters with the CRIB score. The existence of a correlation would provide justification that these parameters reflect underlying risk factors in the preterm infants, with potential pathophysiological meaning. The intention, however, was not to derive an estimate of CRIB score using the spectral measures, given that it could be readily computed based on available clinical data at the first hour of life, including birth weight, gestational age, gender, temperature on admission and blood base excess. Nevertheless it should be noted that CRIB score is a static and one-off measure, while information from spectral analysis of cardiovascular signals can be continuously acquired to monitor the conditions of infants in real time. Although this potential benefit could not be demonstrated by the current study, it should be explored in future via better study design.

There were several limitations to the current investigation. Firstly, only a relatively short segment of continuous and artefact-free data was obtainable from the current subject cohort. The 10 min data length permits reliable extraction of spectral parameters down to the VLF range. For the study of the ultra-low frequency range (< 0.02 Hz), which was found to be associated with clinical risks in previous infant studies [6-8], a longer data segment will be more appropriate. Secondly, only a single data segment was used from each subject in the analysis. It has been suggested that long recordings with multiple measurements can improve reliability of coherence analysis for assessing cerebral autoregulation in infants [134]. Future studies should examine the stability of the various spectral measures over time, to determine whether they provide satisfactory precision that permits discrimination between the infants. Thirdly, the sample size was not very large, and a larger sample size will be desirable from the statistical perspectives.

## 3.2.5 Summary

In this study, we examined the relationships between various spectral measures derived from systemic and cerebral cardiovascular variabilities, and CRIB II, an established clinical risk score for preterm infants in the neonatal intensive care setting. Significant and moderate correlations with CRIB II were identified from a number of parameters including MAP nMF, TOI %VLF, MAP-RRi LF gain and MAP-HHb LF coherence. The relationships between CRIB II and various spectral measures of arterial baroreflex and cerebral autoregulation functions have provided further justification for these measures as possible markers of clinical risks and predictors of adverse outcome in preterm infants.

## 3.3 Transfer function analysis to identify infants with and without IVH

In the previous section, we utilized the spectral analysis to examine the relationships between various spectral measures derived from systemic and cerebral cardiovascular variabilities and the CRIB II. Similarly the spectral analysis of NIRS can also be used to identify infants with IVH. Previous studies have reported that the transfer function analysis of BP and cerebral NIRS measures has demonstrated that alterations in dynamic cerebral autoregulation properties, such as higher coherence or transfer gain, are linked with IVH or other mortality risk factors [5-8]. It has been suggested that the higher gain or coherence in the cerebral pressure-flow relationship could reflect impairment of CBF control by cerebral autoregulation, thus potentially linked with brain injury such as IVH.

In this study, we examined whether the various parameters derived from cross-spectral transfer function analysis of arterial BP and cerebral NIRS, including gain, phase and coherence, could demonstrate significant difference between preterm infants with and without IVH. In contrast to previous studies which adopted a single NIRS measure (TOI or HbD) as a surrogate of CBF, we examined the transfer function relationships of arterial BP to a number of NIRS measures including $HbO_2$, HHb, HbD and TOI, to understand which of the measures might be more useful for the detection of IVH. In addition, we also compared the mean values of the NIRS-derived TOI and FTOE between the two groups of infants.

The data and spectral analysis was performed using the same procedure as mentioned in section 3.2.1. Clinical characteristics of included patients are the same as as mentioned in section 3.2.2. Out of the 17 infants, 16 infants were mechanically ventilated, 5 infants suffered from IVH and 1 infant died eventually.

The results for the IVH and non-IVH infants were presented as mean ± SEM. Logarithmic transformation was performed on the transfer gain parameters as they do not follow a normal distribution. Differences between two groups were compared with the Student's t test. A *P* value of <0.05 was considered significant.

### 3.3.1 Results

The mean values of the physiological variables are shown in Table 3-3, Table 3-5. TOI was significantly higher and FTOE was significantly lower in the IVH infants compared with the non-IVH infants (P value = 0.011 and 0.015 respectively).

**Table 3-5:** Physiological variables in the cohort of preterm infants

|  | IVH | Non-IVH | P Value |
|---|---|---|---|
| **RRi (ms)** | 441±11 | 423±6 | 0.160 |
| **SBP (mmHg)** | 44±5 | 43±2 | 0.760 |
| **MAP (mmHg)** | 33±3 | 35±1 | 0.695 |
| **TOI (%)** | 76±5 | 63±2 | 0.011* |
| **SaO2 (%)** | 95±5 | 95±1 | 0.863 |
| **FTOE (%)** | 20±6 | 34±2 | 0.015* |

Values are mean ± SEM. IVH, intraventricular hemorrhage; RRi, R-R interval; SBP, systolic blood pressure; MAP, mean arterial pressure; TOI, tissue oxygenation index; SaO2, arterial oxygen saturation; FTOE, fractional tissue oxygen extraction. * P < 0.05.

No significant difference was found between the IVH and non-IVH infants for the transfer function parameters derived from HbO2, HbD or TOI, but the VLF coherence of MAP-HHb was significantly lower in the IVH compared with the non-IVH infants (P =0.049). Similar results were obtained from the use of average coherence within the VLF band (*P*=0.015) as shown in Table 3-6. The time series of MAP, HbO2 and HHb of an IVH infant with low MAP-HHb coherence in the VLF range and a non-IVH infant with high coherence were illustrated in Figure 3-7 and Figure 3-8.

**Table 3-6:** Transfer function analysis of mean arterial pressure (MAP) and near-infrared spectroscopy (NIRS) parameters in the cohort of preterm infants.

|  |  | IVH | | Non-IVH | | P value | |
|---|---|---|---|---|---|---|---|
|  |  | VLF | LF | VLF | LF | VLF | LF |
| MAP-HbD | Gain | -0.44±0.11 | -0.71±0.12 | -0.40±0.06 | -0.51±0.10 | 0.719 | 0.266 |
|  | Phase | 0.67±0.69 | -1.12±0.09 | -0.41±0.42 | -0.62±0.27 | 0.190 | 0.252 |
|  | Coh | 0.49±0.06 | 0.68±0.02 | 0.63±0.05 | 0.64±0.03 | 0.111 | 0.353 |
|  | Coh$_{AV}$ | 0.32±0.07 | 0.44±0.05 | 0.45±0.06 | 0.37±0.04 | 0.245 | 0.357 |
| MAP-HbO$_2$ | Gain | -0.45±0.15 | -0.61±0.17 | -0.38±0.05 | -0.45±0.09 | 0.559 | 0.355 |
|  | Phase | 0.74±0.69 | -1.17±0.15 | -0.01±0.51 | -0.44±0.30 | 0.428 | 0.149 |
|  | Coh | 0.49±0.08 | 0.67±0.03 | 0.62±0.04 | 0.67±0.02 | 0.128 | 0.892 |
|  | Coh$_{AV}$ | 0.32±0.09 | 0.45±0.06 | 0.45±0.05 | 0.43±0.04 | 0.181 | 0.824 |
| MAP-TOI | Gain | -0.35±0.19 | -0.19±0.15 | -0.40±0.11 | -0.19±0.12 | 0.804 | 0.979 |
|  | Phase | -1.48±0.61 | -0.67±0.47 | -0.10±0.48 | 0.57±0.20 | 0.119 | 0.831 |
|  | Coh | 0.54±0.04 | 0.62±0.02 | 0.56±0.03 | 0.62±0.01 | 0.691 | 0.784 |
|  | Coh$_{AV}$ | 0.27±0.03 | 0.27±0.04 | 0.31±0.04 | 0.30±0.02 | 0.596 | 0.428 |
| MAP-HHb | Gain | -0.90±0.16 | -0.88±0.26 | -0.69±0.08 | -0.73±0.11 | 0.201 | 0.547 |
|  | Phase | -0.05±0.69 | -0.28±0.17 | -0.48±0.59 | -0.23±0.29 | 0.679 | 0.293 |
|  | Coh | 0.46±0.08 | 0.65±0.04 | 0.62±0.04 | 0.65±0.01 | 0.049* | 0.834 |
|  | Coh$_{AV}$ | 0.23±0.04 | 0.38±0.05 | 0.46±0.05 | 0.38±0.03 | 0.015* | 0.983 |

Values are mean ± SEM. HbD, haemoglobin difference; HbO2, oxygenated haemoglobin; TOI, tissue oxygenation index; HHb, deoxygenated haemoglobin; VLF, very low frequency (0.02-0.04 Hz); LF, low frequency (0.04-0.15 Hz); Gain, transfer gain after logarithmic transformation; Phase, transfer phase in rad; Coh, coherence; CohAV, average coherence of whole band. * P < 0.05

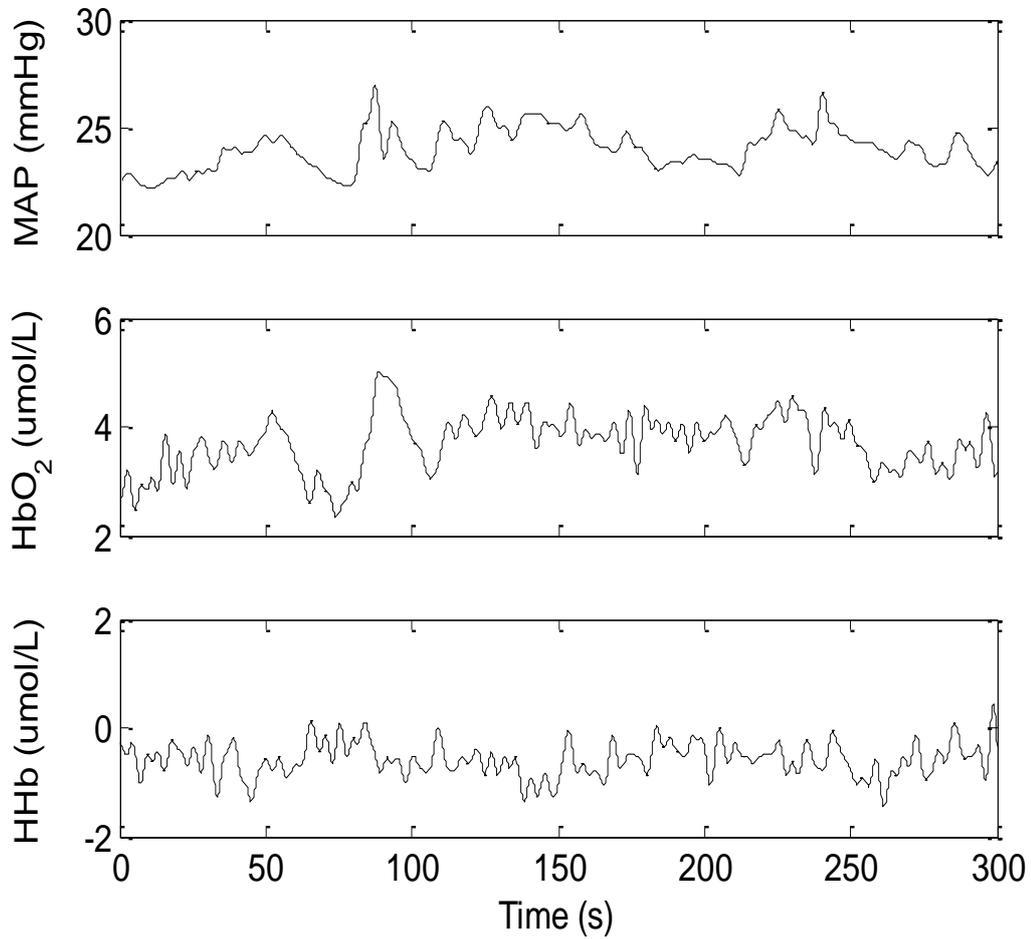

**Figure 3-7 :** Time series of MAP and relative concentrations of HbO2 and HHb in a preterm infant with IVH. Note the highly similar fluctuationsin the very low frequency (VLF) range (25-50 s cycle period) in MAP and HbO2, which were absent in HHb. This IVH infant had a low MAP-HHb coherence of 0.36 in the VLF range.

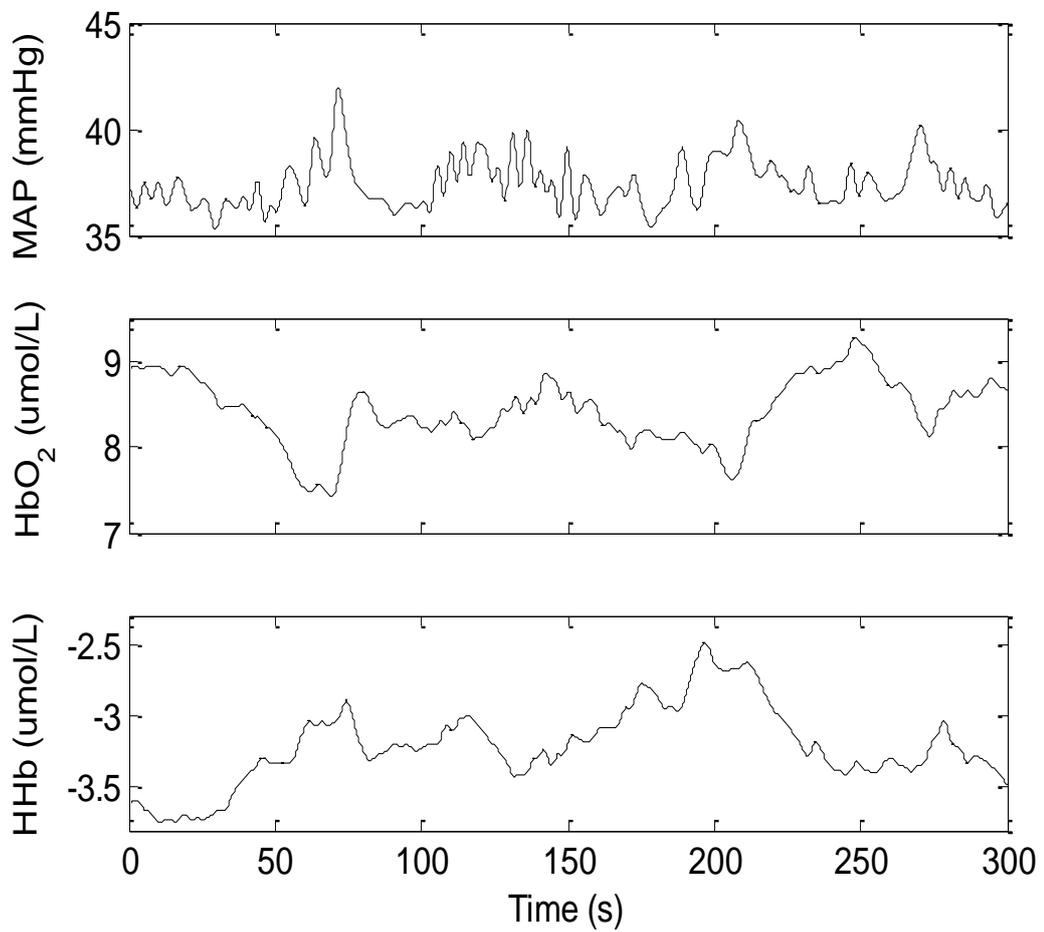

**Figure 3-8:** Time series of MAP and relative concentrations of HbO2 and HHb in a preterm infant without IVH. Note the highly correlated fluctuations in the VLF range (25-50 s cycle period) in MAP, HbO2 and HHb. This non-IVH infant had a high MAP-HHb coherence of 0.86 in the VLF range.

## 3.3.1 Discussion

The main finding of this study was that IVH infants had lower VLF coherence between BP and cerebral HHb derived from NIRS, along with higher TOI and lower FTOE. Parameters derived from transfer function analysis of $HbO_2$, HbD and TOI, however, did not show any significant difference between the two groups. While our results were seemingly at odds with the expectation that IVH infants should have higher coherence [6-8] or gain [5] between BP and NIRS measures due to impairment in cerebral autoregulation as postulated by previous studies, there were two major methodological differences that should not be neglected. First, our findings were derived from the deoxygenated hemoglobin HHb, which had not been examined in the past. Second, studies which found association between high coherence and incidences of IVH [6, 7] were based on the investigation of the very slow trend or the ultra-low frequency (ULF) component (<0.02 Hz) using longer measurements (30 min), whereas our results were obtained from the VLF component using shorter term measurements (10 min). These differences likely accounted for the contrasting results.

Moreover, our findings could be explained based on the physiological conditions of the preterm infants. The IVH infants had both higher TOI and lower FTOE than the non-IVH. The higher level of cerebral oxygenation in the IVH cases should not be seen as a positive sign, as it could well be a consequence of impaired oxygen extraction capability. In fact, a recent study also showed that cerebral oxygenation on the first day of life was higher in very preterm infants compared with healthy term newborns, with the higher TOI again coinciding with lower FTOE in the preterm cases [133]. The reduced oxygen extraction was very likely to have an impact on the relationship between arterial BP and HHb, as the generation of HHb fluctuation was not only dependent on the BP-driven blood flow, but also on the oxygen consumption rate in the cerebral capillaries which governed the conversion of $HbO_2$ to HHb. It was of interest to note that the two IVH infants who had the lowest BP-HHb VLF coherence (0.19 and 0.36) also had rather low FTOE (15% and 8%), suggesting a possible link between the two conditions.

### 3.3.2 Summary

This study demonstrated that cerebral NIRS measurement was able to provide a number of parameters that were potentially useful for distinguishing between preterm infants with or without brain hemorrhage. Specifically, IVH infants were found to have significantly higher cerebral oxygenation level, lower cerebral oxygen extraction rate and lower VLF coherence between arterial BP and deoxygenated hemoglobin. Further studies with larger sample size are warranted for a more complete understanding of the clinical utility of these NIRS measures for early identification of IVH infants. This study suggests that cerebral NIRS measurement may be a suitable monitor of the cerebral autoregulation and can provide a non-invasive tool to detect brain haemorrhage in preterm infants.

# Chapter 4.  Detrended fluctuation analysis of blood pressure in preterm infants with intraventricular haemorrhage

Very preterm infants are at high risk of death and serious permanent brain damage, as occurs with intra-ventricular hemorrhage (IVH). Detrended fluctuation analysis (DFA), that quantifies the fractal correlation properties of physiological signals, has been proposed as a potential method for clinical risk assessment. This study examined whether DFA method of the arterial blood pressure (ABP) signal could derive markers for the identification of preterm infants who developed IVH. This method can also be applied in conjunction with the commonly used spectral analysis method to provide a more complete description of the underlying cardiovascular dynamics, which can help to avoid any unfavourable events in the near future.

In this chapter, we present the DFA approach to evaluate the fractal dynamics embedded in the arterial pressure waveform, which might provide important information in the early detection of IVH in preterm infants. Section 4.1 starts with the basic background about DFA method. Section 4.2 lists the materials and methods used in this chapter. The results from DFA of MAP, SBP and pulse interval are presented in section 4.3. Section 4.4 presents a detailed discussion of results. The limitations are discussed in section 4.5 and conclusions are drawn in section 4.6.

## 4.1 Introduction

With advancements in neonatal intensive care, the incidence of IVH has fallen from 30% in the 1980's to 6 % [12] in preterm infants with very low birth weight (birth weight ≤1500 g). Rates of IVH and in particular severe grades of IVH with extension into cerebral tissue (grade V IVH) in the Papille system are still a major concern in those extremely preterm infants (<27 weeks gestation) [13]. Fragile cerebral circulation of preterm infants is affected by fluctuations in the systemic circulation associated with the pathophysiology of preterm birth and the first six hours of life when complex interactions occur between mechanical ventilation and the circulation. Preterm birth <27 weeks is an abnormal event often precipitated by maternal infection, bleeding or hypertension, factors all predisposing to reduced uterine perfusion and thus placental function. Placental dysfunction is a predisposing factor to the fragile blood vessels in the germinal matrix bleeding in a grade I IVH. If this extends into the ventricular system (grade II IVH) and then blocks the needed drainage of ventricular fluid via the third and fourth ventricles (grade III IVH), risks of long-term brain damage rise. Venous occlusion in adjacent areas of the cerebral hemispheres can give rise to infarction and significant injury to major motor tracks leading to cerebral palsy or intellectual impairment [1, 30]. The timing of IVH indicates most hemorrhages develop after birth with such factors as resuscitation and cardiorespiratory management during the first 24 hours of life being important in the causal pathway [2],[14],[15],[35],[36]. Thus the ability to identify preterm infants with IVH at an early stage is of great importance in terms of minimizing mortality and morbidity in the neonatal intensive care setting.

A recent study explored the potential application of detrended fluctuation analysis (DFA) for identifying early subtle alterations of heart rhythm that could be predictive of impending IVH in very low birth weight babies [21]. DFA has been regarded as a robust method to quantify the scale-invariant (fractal) correlation property of a time series, especially when the signal class cannot be precisely determined. Specifically, this method characterises the mono-fractal property. Previous studies have adopted this method to evaluate the fractal dynamics of heart rate in various diseases in adult subjects [16, 17]. The DFA method aims at measuring the intrinsic fractal-like correlation properties or self-affinity of the time series, characterized by increasing magnitude of fluctuations with increasing time scale (or decreasing frequency). The degree of fractal scaling can be depicted by the scaling exponent (α) that quantifies the relationship between fluctuations F(n) and time scale n based on the mathematical function $F(n) = Cn^{\alpha}$, with α = 1 corresponding to the classical 1/f fractal signal [16, 17]. In a prior study, preterm infants who later developed IVH were found to have significantly higher short-term scaling exponent of heart rate compared with those who did not develop IVH [21]. The differing fractal dynamics of heart rate in these two groups of infants have been attributed to the altered autonomic control of the heart associated with the development of IVH.

Assessment of fractal scaling properties in ABP in adult humans based on DFA method has also been reported [18-20]. By performing autonomic blockade tests, it was shown that vagal blockade by atropine could lead to an increase in the short-term scaling exponent of BP, whereas central blockade of both cardiac and vascular sympathetic outflow by clonidine would result in a decrease of the exponent [20]. It was therefore suggested that fractal dynamics of arterial pressure were largely influenced by the autonomic nervous system activity. However, the fractal scaling of arterial BP in preterm infants has not been investigated previously.

In this study, we examined whether there was a difference in fractal scaling exponents derived from DFA of arterial BP between preterm infants with and without IVH. It was hypothesized that the infants who developed IVH might have higher fractal scaling exponents of arterial BP, similar to that seen in heart rate fluctuations [21]. The findings could help to establish the potential value of fractal dynamics embedded in the arterial pressure waveform for early identification of brain hemorrhage in preterm infants.

## 4.2 Methods

This was a prospective clinical investigation, carried out in a large tertiary neonatal intensive care unit in Sydney, Australia. The study was approved by the Sydney West Area Health Service Human Research and Ethics and conducted according to the World Medical Association Declaration of Helsinki and informed parental consent was obtained in all cases. The cohort comprised early low birth weight infants with gestational age <30 wk, without significant congenital anomalies who survived at least 2 hours. This is a standard caveat for clinical newborn research that infants with significant congenital anomalies are a very heterogeneous group at high risk of death or long-term neuro-disability, both outcomes of interest to the study, hence restriction of entry criteria.The exclusion of infants with significant anomalies is standard in neonatal research to reduce confounding from congenital factors strongly associated with mortality or long-term disability such as trisomy 13. A single measurement of data with 10 min continuous artefact-free segment from each baby was used for analysis. Out of the 46 infants enrolled in the study, 30 infants had arterial BP recordings with sufficiently long artefact-free segments suitable for analysis. The remaining 16 infants either did not have BP recordings (9 infants) or had considerable amount of artefact in the BP data (7 infants).

### 4.2.1 Measurements

Physiological data were collected from the infants within 1-3 hours of birth. Intra-arterial BP was measured via an umbilical or peripheral arterial catheter connected to a transducer calibrated to atmospheric barometric pressure and zeroed to the midaxillary point. The BP signal was continuously acquired via a bedside patient monitor (Philips Agilent Systems, Philip Healthcare, North Ryde, Australia), and recorded by a data acquisition system (ADInstruments, Sydney, Australia) at a sampling rate of 1 kHz. The LabChart software (ADInstruments, Sydney, Australia) provided automatic beat-to-beat computation of systolic blood pressure (SBP), MAP and pulse interval from the arterial BP waveform. Cranial ultrasound measurements for the diagnosis of IVH were taken within the first 2 hours of life and repeated 12 hourly for the first 2 days of life. The scans and assessments were taken in a systematic method with classification scored by a pediatric radiologist blinded to the clinical outcome and study according to the Papille classification [13].

### 4.2.2 Data analysis

DFA was performed on uninterrupted data recordings of 10 minutes (obtained within 1-3 hours after birth from each infant) without noticeable corruption by artefacts. All signal processing was implemented in Matlab (Natick, MA, USA). Spikes in the beat-to-beat time series due to occurrence of abnormal pulse intervals (e.g. prolonged heart period resulting from a missing beat), which constituted less than 3% of the total number of beats, were removed and replaced by linear interpolation of the beat values immediately preceding and following the replaced beats.

The DFA method has been developed to analyze the scale-invariant (fractal) correlation property of a time series [16]. In this case, the time series would correspond to the beat-to-beat values of SBP, MAP and pulse interval. Consider a time series $Y$ with a total length of N, i.e. a total of N beats. For a heart rate of 150 bpm, a 10 minute period would consist of N = 1500 beats. In the application of DFA, the time series was first integrated to improve the fractal property. This is achieved by calculating the cumulative sum of the differences between the $i^{th}$ parameter value $Y(i)$ and the mean parameter value $Y_{mean}$:

$$\mathbf{y(k)} = \sum_{i=1}^{k}[Y(i) - Y_{mean}] \qquad (4.1)$$

Next, the integrated time series $y(k)$ is divided into multiple windows of equal length $n$. For example, for a total of $N$ = 1500 beats and for a window size of $n$ = 15 beats, there would be 100 windows. Within each window, a trend line is fitted to the data, with $y_n(k)$ denoting the $k^{th}$ point on the trend line with a length of $n$. The root-mean-square (RMS) fluctuation $F(n)$ of the detrended integrated time series for a given window size $n$ is calculated by

$$F(n) = \sqrt{\frac{1}{N}\sum_{k=1}^{N}[y(k) - y_n(k)]^2} \qquad (4.2)$$

This procedure is repeated at different window sizes to derive the relationship between $F(n)$ and $n$. Finally, $F(n)$ is plotted against the window length $n$ in a log-log plot, and the slope of the relationship will correspond to the fractal scaling exponent α. For fractal signals, the fluctuations can be expressed as a power-law function of the time scale: $F(n) = Cn^{\alpha}$, in which case $\log_{10}(F(n))$ and $\log_{10}(n)$ are directly proportional with α as the slope of the relationship. The existence of a linear relationship (assessed by the goodness of fit of the regression) between $\log_{10}(F(n))$ and $\log_{10}(n)$ will be considered necessary for the slope α to be a valid measure of fractal scaling. The special case of α = 1 is often termed 1/*f* noise, with the power of fluctuations being linearly related to the time scale. The 1/*f* noise has been frequently observed in nature and in various biological phenomena, although the physiological mechanism behind remains speculative.

Examination of the DFA plot of heart rate has suggested the existence of two linear regions with different fractal slopes separated by a breakpoint $n_{bp}$ [140] – the first region covered the range from 4 to $n_{bp}$ beats, characterized by a short-term scaling exponent $\alpha_1$; the second region covered the range from $n_{bp}$ to 50 beats, characterized by a long-term scaling exponent $\alpha_2$. As the previous infant study has found that the 8-15 beats range of heart period fluctuation appeared to provide most useful information in relation to IVH [21], $n_{bp}$ was chosen to be 15. Therefore, the short-term scaling exponent $\alpha_1$ was calculated from a window size of 4-15 beats, whereas the long-term scaling exponent $\alpha_2$ was calculated from a window size of 15-50 beats.

The results for the cohort were presented as mean ± SD. Differences between the IVH and non-IVH infants were compared with the Student's t test. A *P* value of <0.05 was considered significant.

## 4.3 Results

The demographic and clinical characteristics of the studied infants are presented in Table 4-1: Physiological variables in the cohort of preterm infants. Out of the 30 infants studied, 26 infants were mechanically ventilated, 10 infants suffered from IVH and 2 infants died eventually (one infant did not suffer from IVH). Thus, there were 10 infants with IVH (9 of whom were ventilated) and 20 infants without IVH (17 of whom were ventilated) in our cohort. Note that there was no significant difference in the SBP, MAP, heart rate and CRIB II score (CRIB stands for Clinical Risk Index for Babies [141],[4]) between the two groups of infants. Out of the 16 infants excluded from the analysis, 3 infants suffered from IVH and 2 infants died eventually (one infant did not suffer from IVH).

The results from DFA of MAP, SBP and pulse interval are shown in table 4-2 and in table 4-3 (for the infants with mechanical ventilation). For both the short-term and long-term beat ranges, the regression fit $R^2$ was high (>0.9), which justified the validity of the DFA approach. The group averages of the DFA log-log plot for MAP in the short-term and long-term beat ranges are shown as Figure 1a and 1b respectively. For the IVH infants, the short-term scaling exponents $α_1$ of both MAP and SBP were significantly higher than those of the non-IVH infants ($P = 0.017$ and $P = 0.038$ respectively), and also closer to 1, that signifies $1/f$ fractal scaling. Similar results were seen in the infants with mechanical ventilation only ($P = 0.007$ and $P = 0.020$ for the $α_1$ of MAP and SBP respectively). No significant difference was found between IVH and non-IVH in the long-term scaling exponent $α_2$ of MAP and SBP, and in both cases $α_2$ appeared to be slightly above 1. No difference was found in either the short-term or long term scaling exponents for pulse interval.

## 4.4 Discussion

This study is the first to examine the fractal dynamics of arterial BP in preterm infants using the DFA approach. The rationale was to address whether the fractal scaling of arterial BP could provide potentially useful information for distinguishing between preterm infants with and without brain hemorrhage. The main finding is that the short-term fractal scaling exponent $\alpha_1$ is significantly higher in the IVH infants compared with the non-IVH infants. This finding suggests that fractal dynamics embedded in the arterial pressure waveform might hold important information that could allow early identification of preterm infants at risk of developing significant IVH after birth and led to further studies to examine ways of reducing the risk using this new information. In addition, the high $R^2$ (> 0.9) in the DFA model fit provides validation to the appropriate use of this technique for analyzing fractal scaling in the BP fluctuation of infants.

The study of fractal scaling in arterial BP in adult humans using the DFA approach has revealed important contributions of the autonomic nervous system in the scaling exponents [20]. It was shown that blockade of vagal activity by atropine could lead to an increase in the short-term scaling exponent of BP. The main reason was believed to be the removal of heart rate variation that was being translated to BP fluctuation, specifically those generated by the respiratory modulation of cardiac vagal activity. As the heart beat intervals govern the period of diastolic decay in the arterial pressure pulse (the Windkessel mechanism) and also the diastolic filling of the heart (that determines stroke volume), they would have direct effects on DBP and SBP. Thus, smoother heart rate fluctuations could result in smoother BP fluctuations. Conversely, the administration of clonidine, that causes central blockade of both cardiac and vascular sympathetic outflow and a central increase in cardiac vagal activity, resulted in a decrease of the short-term BP exponent. The higher $\alpha_1$ observed in the IVH infants might therefore be explained by either a reduction of vagal activity or an augmentation of sympathetic outflow compared with the non-IVH infants.

The use of mechanical ventilation was likely to be an important factor for the association between the short-term scaling exponent α1 and IVH, given its influence on the respiratory fluctuation of BP and heart rate. In this cohort of preterm infants, a majority were breathing with mechanical ventilation (26 out of 30), and only a small number were breathing spontaneously (1 IVH and 3 non-IVH). The separate analysis performed on the mechanically ventilated subgroup showed similar results as the whole group (Table 4-3) with the IVH infants having higher α1 than the non-IVH infants. In fact, the P values obtained were even smaller, thus it was possible that the use of mechanical ventilation could be a reason for the observed association between α1 and IVH. Unfortunately, the number of infants without mechanical ventilation was too low in this study for statistical analysis. Whether the ventilatory mode had any direct impact on the results would need to be addressed by further studies with larger sample size. Early studies have also strongly implicated the causal association of lack of patient to ventilator synchrony causing intraventricular haemorrhage [142, 143]. These studies occurred prior to the current methods of patient triggered ventilation and used paralytic agents to prevent adverse patient breathing interaction with the ventilator.

A previous study examined the use of DFA on the fluctuation of heart period (R-R) interval measured by electrocardiogram (ECG) in preterm infants suffering IVH [21]. The short-term scaling exponent ($α_1$) computed from the 8-15 beat range was found to be significantly higher in the infants who later developed IVH compared with those who did not. In the current study, the pulse interval extracted from the arterial BP waveform was used as a measure of heart period fluctuation, but no significant difference was found between the IVH and non-IVH infants. A possible reason could be the difference between R-R interval variability and pulse interval variability, particularly in the high frequency range (respiratory frequency) [144], that corresponded to the time scale of the short-term exponent $α_1$.

Based on a previous study using pharmacological intervention, our observation of an increase in the scaling exponent of BP can be best explained by a suppression of vagal activity. The reduction in heart rate fluctuation due to lack of cardiac vagal activity may lead to the reduction in BP fluctuation, as heart period can have direct effect on BP via the Windkessel mechanism. Vagal activity is known to be involved in the regulation of cerebral blood flow, but the exact mechanism through which IVH is linked with vagal activity remains unclear thus far.

Nevertheless, the higher $α_1$ found in the BP fluctuation of IVH infants was in agreement with the previous finding, both of which suggested a lack of vagal activity or an augmentation of sympathetic outflow compared with the non-IVH infants. Vagal nerve activity has been shown to play an important role in the regulation of cerebral blood flow during hemorrhage and cerebral ischemia [145]. Results from the current work have provided further evidence that development of IVH could involve impairment in parasympathetic outflow. However, the specific mechanism through which IVH is linked with vagal activity remains unclear thus far.

While it has been previously demonstrated that DFA of heart period (derived from ECG) could provide useful information for the identification of IVH [21], there are still potential benefits from the use of arterial BP. Firstly, it is not completely clear whether the fractal dynamics of arterial BP simply follow the heart period, or represent additional mechanism that is not assessable from the DFA of heart period alone. Secondly, from the clinical viewpoint, it is not yet known which method may provide more relevant information for the detection of IVH, or whether the two methods can be applied together to improve the accuracy of detection. This would require further studies with simultaneous ECG and BP recordings in a larger cohort. Thirdly, it may be preferable to use arterial BP instead of ECG in situations where movement artefacts from the infants have led to degradation of ECG quality, such that sufficiently long artefact-free segments are not obtainable.

There are several advantages from deriving clinical information using the DFA method. Firstly, this method can be performed on an acceptable duration of data (10 min in this case), that is practical in the neonatal intensive care setting. Secondly, the information (namely the scale exponent) is derived from a specific beat range,that can be easily translated to a similar time range or frequency band used in other methods such as power spectrum analysis. This would largely facilitate the physiological interpretation of the derived parameters. Thirdly, it can be applied in conjunction with the commonly used spectral analysis method to provide a more complete picture of the underlying cardiovascular dynamics, as both methods can provide complementary information [20].

**Table 4-1:** Physiological variables in the cohort of preterm infants

|  | IVH | Non-IVH | P Value |
| --- | --- | --- | --- |
| Gestational age (wk) | 26.7±1.7 | 26.9±1.9 | 0.79 |
| Birth weight (g) | 1023±357 | 1025±340 | 0.99 |
| CRIB II score | 10±3 | 10±3 | 1 |
| MAP (mmHg) | 35.1±6.0 | 35.2±5.3 | 0.97 |
| SBP (mmHg) | 45.1±9.6 | 43.6±6.5 | 0.62 |
| Heart rate (bpm) | 139±8 | 144±10 | 0.11 |
| Pulse interval (ms) | 353±146 | 402±65 | 0.21 |

Values are mean ± SD. CRIB, Clinical Risk Index for Babies; SBP, systolic blood pressure; MAP, mean arterial pressure.

**Table 4-2:** Detrended fluctuation analysis (DFA) of mean arterial pressure (MAP), systolic blood pressure (SBP) and pulse interval (PI) in the cohort of preterm infants (N = 30).

|     |       | IVH (N = 10) | | Non-IVH (N = 20) | | P Value | |
| --- | ----- | ------------ | ------------ | ------------ | ------------ | -------- | -------- |
|     |       | $\alpha_1$   | $\alpha_2$   | $\alpha_1$   | $\alpha_2$   | $\alpha_1$ | $\alpha_2$ |
| **MAP** | α     | 1.06±0.18    | 1.21±0.16    | 0.84±0.25    | 1.11±0.19    | 0.017*   | 0.152    |
|     | $R^2$ | 0.98±0.02    | 0.97±0.02    | 0.99±0.01    | 0.98±0.02    | 0.362    | 0.386    |
| **SBP** | α     | 0.98±0.20    | 1.17±0.15    | 0.78±0.25    | 1.06±0.21    | 0.038*   | 0.145    |
|     | $R^2$ | 0.98±0.01    | 0.97±0.03    | 0.99±0.01    | 0.97±0.03    | 0.538    | 0.461    |
| **PI**  | α     | 0.73±0.35    | 1.06±0.25    | 0.64±0.37    | 0.96±0.38    | 0.546    | 0.459    |
|     | $R^2$ | 0.99±0.01    | 0.97±0.02    | 0.97±0.03    | 0.95±0.06    | 0.190    | 0.166    |

Values are mean ± SD. $\alpha_1$, short range scaling exponent (4-15 beats); $\alpha_2$, long range scaling exponent (15-50 beats); α, DFA scaling exponent; $R^2$, goodness of fit of the linear regression in DFA. * P < 0.05 between IVH and non-IVH.

**Table 4-3:** Detrended fluctuation analysis (DFA) of mean arterial pressure (MAP), systolic blood pressure (SBP) and pulse interval (PI) in the cohort of preterm infants with mechanical ventilation (N = 26).

|     |       | IVH ventilated (N = 9) | | Non-IVH ventilated (N = 17) | | P Value | |
| --- | ----- | --------- | --------- | --------- | --------- | ------ | ------ |
|     |       | $\alpha_1$ | $\alpha_2$ | $\alpha_1$ | $\alpha_2$ | $\alpha_1$ | $\alpha_2$ |
| MAP | α     | 1.05±0.18 | 1.22±0.17 | 0.79±0.23 | 1.09±0.18 | 0.007* | 0.093 |
|     | $R^2$ | 0.99±0.01 | 0.99±0.01 | 0.99±0.01 | 0.99±0.01 | 0.631  | 0.756 |
| SBP | α     | 0.94±0.17 | 1.16±0.16 | 0.73±0.22 | 1.03±0.20 | 0.020* | 0.092 |
|     | $R^2$ | 0.99±0.01 | 0.99±0.01 | 0.99±0.01 | 0.99±0.02 | 0.928  | 0.871 |
| PI  | α     | 0.66±0.30 | 1.04±0.26 | 0.66±0.40 | 0.93±0.41 | 0.970  | 0.482 |
|     | $R^2$ | 0.99±0.01 | 0.99±0.01 | 0.99±0.01 | 0.97±0.03 | 0.201  | 0.146 |

Values are mean ± SD. $\alpha_1$, short range scaling exponent (4-15 beats); $\alpha_2$, long range scaling exponent (15-50 beats); α, DFA scaling exponent; $R^2$, goodness of fit of the linear regression in DFA. * P < 0.05 between IVH and non-IVH.

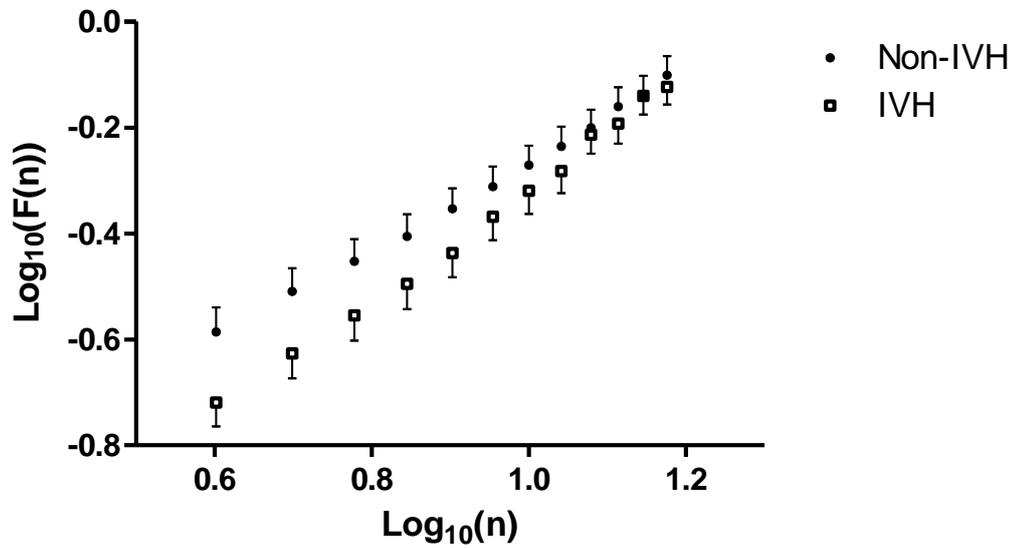

**Figure 4-1:** The plot of $Log_{10}(F(n))$ vs $Log_{10}(n)$ from detrended fluctuation analysis (DFA) of mean arterial pressure (MAP) for infants with intraventricular hemorrhage (IVH) and those without (non-IVH), in the short-term beat range (n = 4-15 beats). The mean and SEM of the groups are shown. The short-term scaling exponents ($α_1$) are computed as the linear slopes of the individual plots.

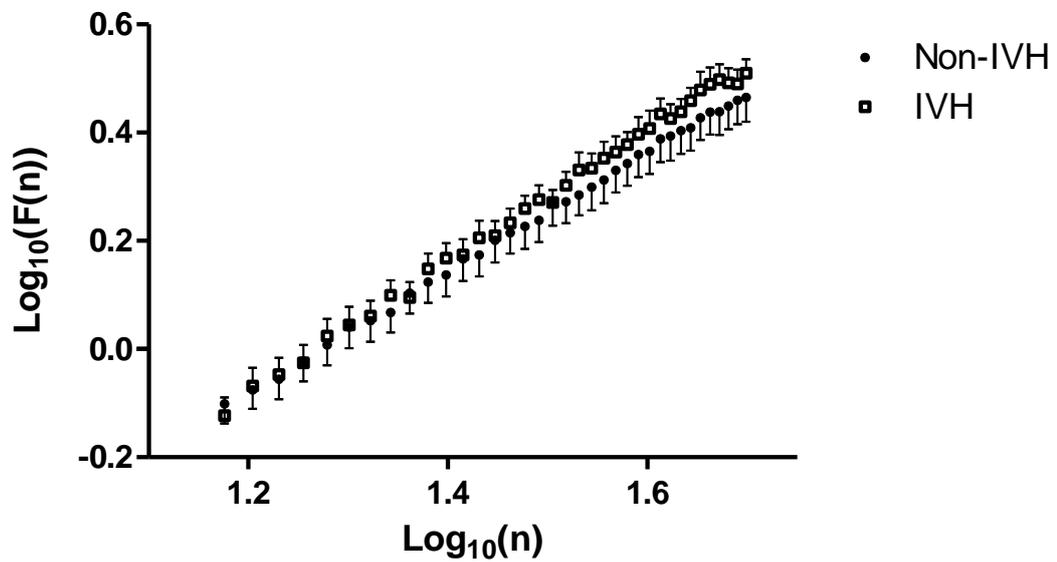

**Figure 4-2:** The plot of $Log_{10}(F(n))$ vs $Log_{10}(n)$ from detrended fluctuation analysis (DFA) of mean arterial pressure (MAP) for infants with intraventricular hemorrhage (IVH) and those without (non-IVH), in the long-term beat range (n = 15-50 beats). The mean and SEM of the groups are shown. The long-term scaling exponents ($α_2$) are computed as the linear slopes of the individual plots.

## 4.5 Limitations

There are some limitations to this study. Firstly, due the relatively low incidence of IVH in our cohort it was not possible to assess how the DFA exponents relate to the severity of hemorrhage. Secondly, the data duration of 10 min placed a limitation on the longest beat range that can be assessed by DFA method. It would be desirable that these limitations could be overcome in future by improvement in the experimental design.

## 4.6 Summary

In conclusion, this study examined whether the application of DFA method on the arterial BP signal could derive markers for the identification of preterm infants who developed IVH. The results showed that the IVH infants had higher short-term scaling exponents of both MAP and SBP compared to the non-IVH infants. Thus, fractal dynamics in arterial BP could provide potentially useful information that facilitates early identification of IVH in preterm infants. Further studies with larger sample size are warranted for a more complete understanding of the clinical utility of these ABP measures for early identification of IVH infants.

# Chapter 5. Estimation of Cardiac Output and Total Peripheral Resistance by Arterial Waveform Analysis

The purpose of monitoring tools of cardiac output (CO) and total peripheral resistance (TPR) in this chapter is to continuously provide valuable information about the overall haemodynamic state of the preterm infants. Based on this information, the doctor can decide whether the patients need intensive care and monitoring. However, in the present NICU setting, the assessment of CO and TPR requires specialized skill and equipment, and cannot be performed continuously.

In section 5.2, we firstly explored the possible ways of using the diastolic decay rate parameter for assessing LVO and TPR in preterm infants. In section 5.3 to further improve the accuracy and robustness of the estimation, a multivariate regression model with the inputs of features from the arterial BP waveform analysis were used to estimate the LVO and TPR.

## 5.1 Introduction

Cardiac output and total peripheral resistance are two important parameters of the cardiovascular system. Measurement of these two parameters can provide valuable information for the assessment and management of patients needing intensive care, including preterm infants in the NICU [22, 23]. For example, high CO and low TPR are symptoms of sepsis, which is a common cause of mortality and morbidity in preterm infants, particular those with very low birth weight (VLBW) [22]. A substantial drop in CO (>50%) has been associated with a high risk of mortality in these infants with sepsis [22]. Furthermore, the measurement of systemic hemodynamics can facilitate the development of better therapeutic strategy for these infants, including the use of volume expansion and inotropes [23]. In the present NICU setting, the assessment of CO and TPR requires the measurement of left ventricular outflow (LVO) or right ventricular outflow (RVO) using ultrasound Doppler echocardiographic waveforms. However, such measurement requires specialized skill and equipment, and cannot be performed continuously.

In this study, the estimation of LVO and TPR was performed by analyzing the rate of blood pressure decay in the diastolic portion of the arterial pulse wave. This has been associated with TPR in the Windkessel model, and has been utilized in the study of animals and adult patients [4, 5]. However, whether this arterial waveform parameter is useful for monitoring preterm infants has not been previously investigated. The results from the current study would help to establish the potential utility of the diastolic decay rate parameter for assessing LVO and TPR in preterm infants.

The arterial tree is a complex system not only because of its geometry but also because of its nonlinear mechanical properties. Thus biophysicists and physiologists have been trying to develop simplified representations of this system to describe some of its functional aspects. Some characteristics of the mechanical properties of the whole arterial tree can be derived from the continuous measurements of aortic arterial pressure (AP) and blood flow. Among the parameters that can be estimated in the most straightforward way for each cardiac cycle, the vascular resistance (R), computed as the ratio of MAP to mean aortic flow, and the time constant of the diastolic AP decrease according to a monoexponential model (t) are the simplest and the most interesting ones. According to Poiseuille's law with the assumptions of a Newtonian fluid and a laminar flow in a cylindrical tube, R contributes to the dissipation of the circulatory energy in the arterial tree during each cardiac cycle. The time constant signals the restoration of energy stored by the arterial wall during the elastic distension due to ventricular ejection. The determination of this time constant depends on the accuracy of the exponential fit of the diastolic part of the AP curve. Such a model based on the windkessel effect has been widely used, and numerous studies have been performed to improve it and to better estimate and interpret the model parameters.

## 5.2 Arterial Waveform Analysis of Arterial BP waveform

### 5.2.1 Methods

#### *5.2.1.1 Subjects*

This was a prospective clinical investigation, carried out in the neonatal ICU at the Nepean Hospital in Sydney, Australia. The study was approved by the Sydney West Area Health Service Human Research and Ethics Committee, and informed parental consent was obtained in all cases.

The cohort comprised early low birth weight infants with gestational age <30 wks, without significant congenital anomalies who survived at least 2 hours. A single measurement of data with 10 min continuous artefact-free segment from each baby was used for analysis. The study cohort comprised a convenience sample of 27 early low birth weight infants with gestational age <30 wks. The demographic characteristics of the studied infants are as follows (mean±SD): 17 males 12 females, gestational age 26.7±1.8 wks (range 24-30 wks), birth weight 1010.7±346.3 g (range 520-1929 g).

Physiological data were collected from the infants at four different time periods: 1-3 hours after birth (27 babies), 12 hours after birth (24 babies), 24 hours after birth (20 babies) and 36 hours after birth (18 babies). Thus a total of 89 measurements were used in the modeling.

### 5.2.1.2 Measurements

Arterial BP waveform was acquired via a bedside patient monitor (Philips Agilent Systems, Philip Healthcare, North Ryde, Australia). Intra-arterial BP was continuously measured via an umbilical or peripheral arterial catheter connected to a transducer calibrated to atmospheric barometric pressure and zeroed to the midaxillary point. A Vivid 7 Dimension ultrasound scanner (GE Health Care, Rydalmere, Australia) was used with a 5–7 MHz sector cardiac probe incorporating pulsed wave and continuous Doppler echocardiographic waveforms. All ultrasounds were performed by the same investigator (M.T.).

LVO was obtained by measuring the aortic valve annulus diameter in the parasternal long axis view, and then velocity time integrals (VTI) were obtained with a minimum of 5 consecutive aortic pulsed Doppler waveforms from an apical long axis view [38].

### 5.2.1.3 Signal processing and feature extraction

ABP signals were low-pass filtered at cut-off frequencies of 50 Hz respectively to remove undesirable high-frequency noise.

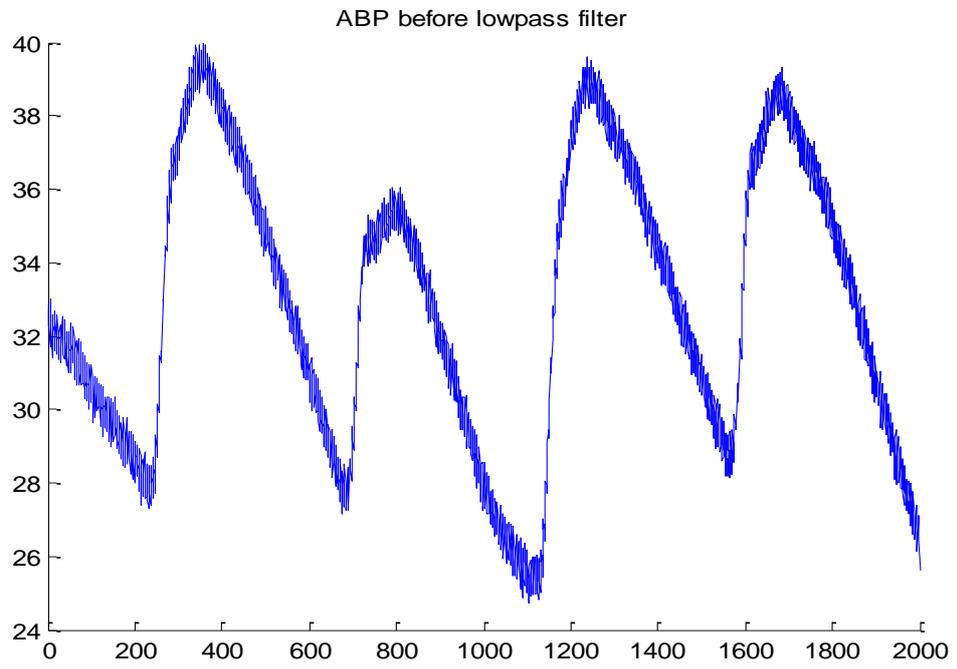

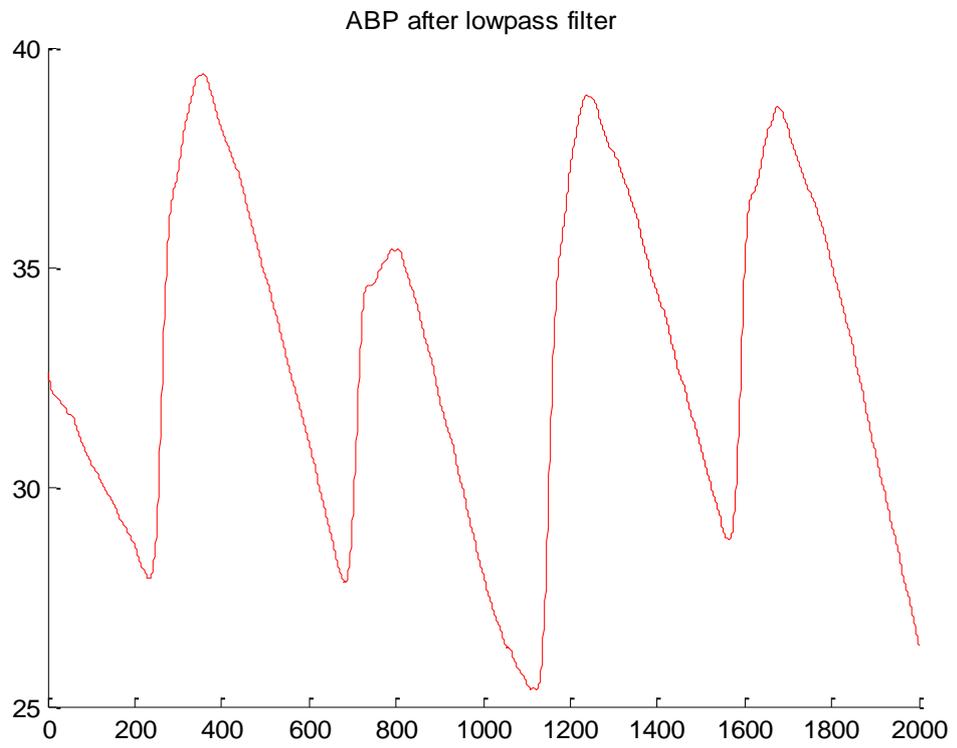

**Figure 5-1:** ABP signals before and after low-pass filtering.

The ABP pulses of preterm infants typically follow an exponential decay during each diastolic interval. As mentioned in section 2.5.1, the Windkessel model of the systemic circulation comprises elements of arterial resistance (R) and arterial compliance (C), the time constant of this exponential decay (τ) corresponds to RC [146, 147]. Assuming that C was similar amongst the infants (as C changed mainly with age and the infants should have very compliant arteries at birth), while R could vary greatly depending on the infants' condition, and thus the main determinant of τ would be R. For each cardiac cycle, CO (measured as LVO here) was related to R via the relationship $R = \frac{MAP}{CO}$. Therefore, it was hypothesized in this study that there is a relationship between CO and 1/τ. In addition, there may be a stronger relationship between CO and MAP/τ, assuming that τ represents R.

All signal processing and feature extraction was implemented in Matlab (Natick, MA, USA). The time constant $\tau$ was estimated beat to beat using the exponential decay time method [148-150] by the model shown as follows:

$$\boldsymbol{AP(t) = AP_0 e^{-\frac{t}{\tau}}} \tag{5.1}$$

where $t$ is time and $AP_0$ is the AP value at time $0$ which is the initial time for application of the model. The last one-third of the AP curve was utilized in this model and a linear regression was developed after logarithmic transformation of AP values, which yielded a correlation coefficient ($r_\tau$) that indicated the quality of the exponential regression. The slope of the regression line was $-1/\tau$ and systemic arterial compliance was calculated using the windkessel relationship $c = \tau/R$.

From each cardiac pulse of the pressure waveform, the diastolic decay time constant τ of the Windkessel model was determined by fitting an exponential function to the diastolic period [146]. The fitted curves are illustrated in Figure 5-2 and Figure 5-3. A direct way of extracting the exponential time constant was to take the logarithm of both pressure and time and carry out linear regression, then the slope of the regression would be $1/\tau$, termed the diastolic decay rate. Figure 5-2 shows an example of a relatively slow decay rate (long time constant) whereas Figure 5-3 shows an example of a fast decay rate (short time constant). This slope parameter was first determined in each set of measured data by taking the mean value of 50 pulses. Finally, the relationship of LVO (and TPR) with $1/\tau$ was examined by linear regression using the 89 sets of measurements, and the corresponding correlation coefficient r was computed.

As a comparison, another parameter given by the product of pulse pressure and heart rate (PP × HR) was derived, and its relationship with LVO was examined. This parameter has previously been used to track LVO changes in adult subjects [151].

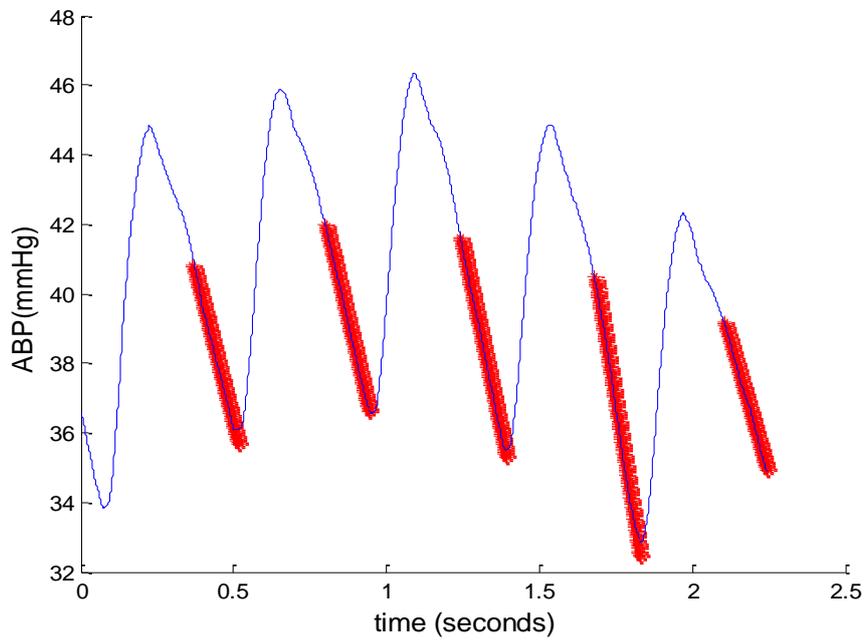

**Figure 5-2:** Arterial blood pressure (ABP) pulses with the fitted exponential functions, for one subject with a diastolic decay rate of 0.44 Hz (time constant of 2.3 s).

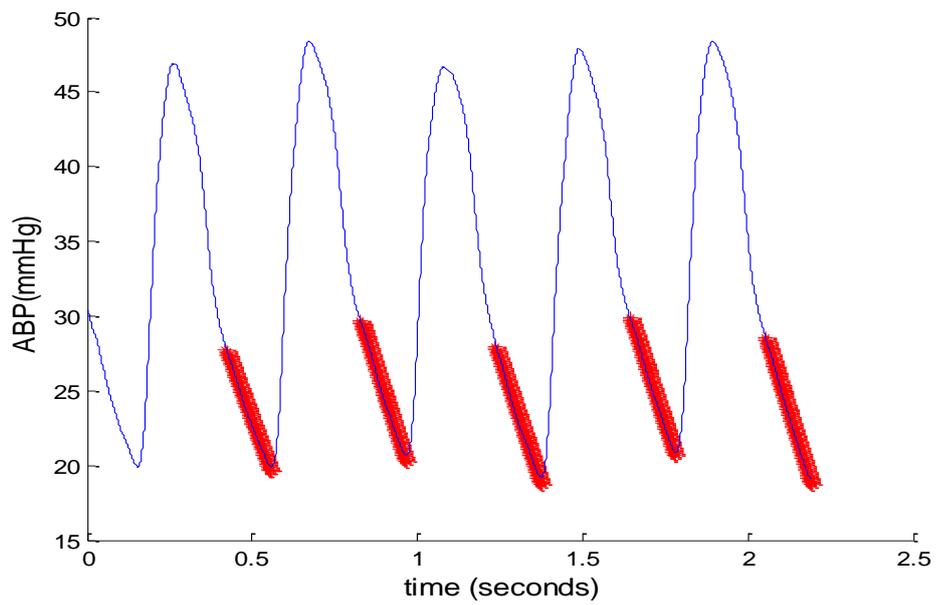

Figure 5-3: Arterial blood pressure (ABP) pulses with the fitted exponential functions, for one subject with a diastolic decay rate of 1.24 Hz (time constant of 0.8 s).

## 5.2.2 Results

The results of regression analysis demonstrate that the diastolic decay rate 1/τ had a significant positive relationship with LVO (r = 0.383, P = 0.0002), and negative relationship with TPR (r = -0.379, P = 0.0002). The corresponding scatter plots are shown in Figure 5-4 and Figure 5-5. It was noted that one subject had a much higher decay rate (1/τ = 1.8 Hz) compared to others, and if this subject was neglected the correlations would become stronger (for LVO, r = 0.435, P = 0.00002; for TPR, r = -0.437, P = 0.00002). Overall, the results have shown that the diastolic decay rate parameter was potentially useful for providing information about LVO and TPR in the preterm infants. Although the correlations were not high, the scatter plots (Figure 5-4and Figure 5-5) seem to suggest that the relationships between the variables may be better described by a more complex nonlinear function rather than a simple linear one. This could be further explored in future to improve the accuracy of estimation.

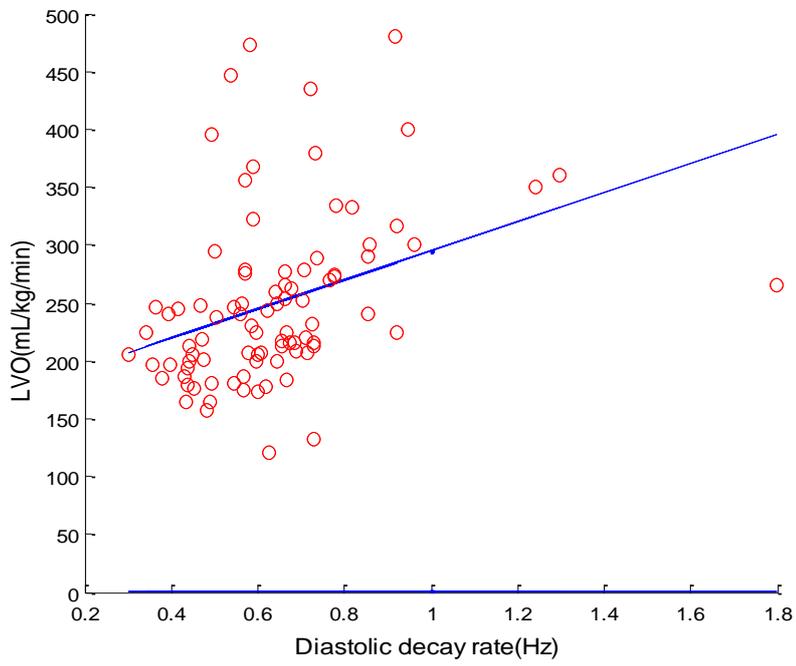

**Figure 5-4 :** Left ventricular outflow (LVO) versus diastolic decay rate (1/τ) in the cohort of preterm infants.

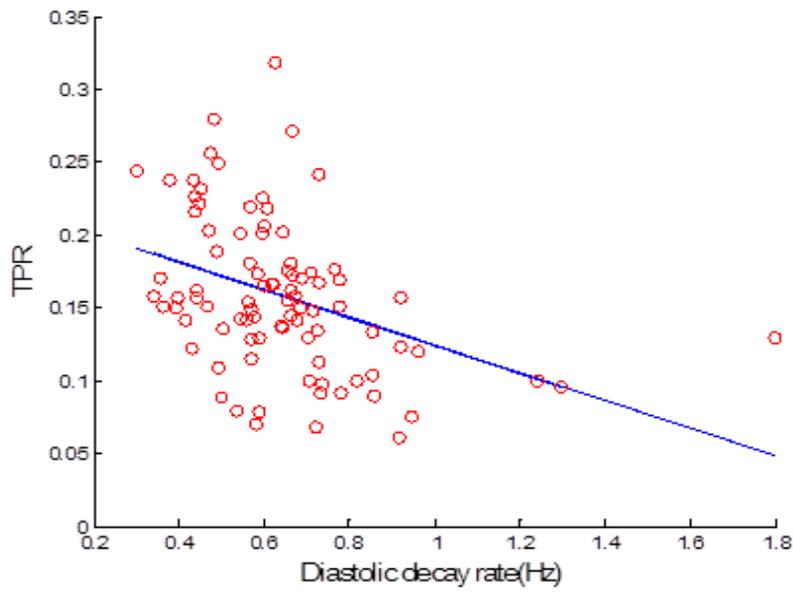

**Figure 5-5:** Total peripheral resistance (TPR) versus diastolic decay rate (1/τ) in the cohort of preterm infants.

The individual exponential fit to the ABP pulses displayed high $R^2$ (0.99 ± 0.01), which provided justification of the Windkessel model for the analysis of pressure waveforms in infants. While previous studies have demonstrated the utility of the Windkessel-model based approach for the hemodynamic monitoring of animals and adult patients [4, 5], the current study was the first to demonstrate its potential value for the assessment of preterm infants in the NICU setting. It has been suggested that in normal adults, the presence of arterial pulse wave reflection may affect the accuracy of the Windkessel model [5]. However, it appears that in the preterm infants, wave reflection is not prominent as the arterial pulses often do not display clear secondary peaks (Figure 5-2 and Figure 5-3). This may explain the good fit achieved by a simple Windkessel model.

The parameter MAP/τ (MAP × diastolic decay rate) had lower correlation with LVO (r = 0.266, P = 0.012) compared with 1/τ, as shown in Figure 5-6. If the outlier with a very high decay rate (1/τ = 1.8 Hz) was excluded, the correlation became higher (r = 0.299, P = 0.0046). According to the model, MAP/τ should have a stronger relationship with CO compared with 1/τ, but the data seemed to suggest otherwise which was unexpected. Another arterial waveform parameter PP × HR did not show any significant correlation with LVO (r = 0.066, P = 0.54). Thus this parameter is probably more suitable for monitoring intra-subject change [151] but not for estimating CO in a group of preterm infants.

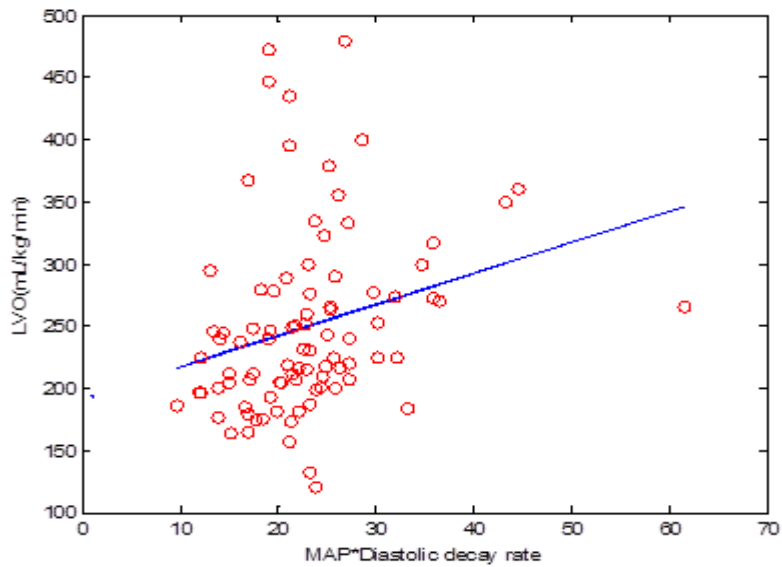

Figure 5-6: Left ventricular outflow (LVO) versus the product of mean arterial pressure and diastolic decay rate (MAP/τ) in the cohort of preterm infants.

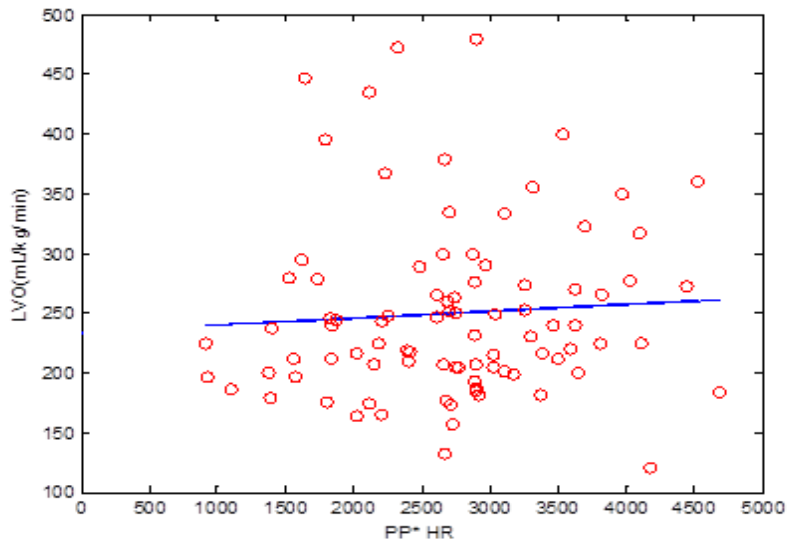

**Figure 5-7: Left ventricular outflow (LVO) versus the product of pulse pressure and heart rate (PP × HR) in the cohort of preterm infants.**

### 5.2.3 Discussion

Many extremely low birth weight infants have impaired cerebral autoregulation (some with a complete absence of autoregulation) and thus a pressure passive cerebral circulation. Intracranial haemorrhage is more likely in this state of pressure passive circulation. Routine methods of estimating cardiac function in these preterm infants are dependent on echocardiographic estimates of volumetric flow or M-Mode estimates of ejection fraction. There are obvious limitations with the need of appropriately trained clinicians to perform these examinations and those measurements are valid for the short time period of the examination. There is thus an urgent need to explore new methods of continuous estimates of cardiac function to support clinical decisions to treat hypotension and support vulnerable cerebral circulation.

## 5.3 Multivariate regression Model

### 5.3.1 Introduction

The preliminary results from this study demonstrated the potential utility of arterial pressure waveform analysis for estimating LVO and TPR in preterm infants. However, to further improve the accuracy and robustness of the estimation, more advanced multi-parameter models may be needed. Such models could include additional physiological parameters such as heart rate as well as incorporating possible nonlinear relationships between the parameters.

### 5.3.2 Multivariate regression

In this study, multivariate regression was used to model the relationship between the extracted features and the measurement of LVO and TPR, by fitting a linear least square equation. The LVO and TPR were estimated as a weighted sum of a chosen subset of k features, by means of linear least square modelling. In the following discussion, the unknown variable to be estimated is assumed to be LVO, but the methods are equally applicable to estimate TPR.

In this model, for each subject of LVO, $y_i$, i=1,2,…,N ( the number of subjects) is a linear function of all k input features.

$$y_i = \beta_1 x_{i1} + \cdots + \beta_k x_{ik} + \beta_{k+1} + \epsilon_i \tag{5.2}$$

The estimated output from the model $\hat{y}_i$ is assumed to be related to the extracted features $x_i = [x_{i1}, x_{i2}, \ldots, x_{ik}]^T$, where we write $x_{ij}$ for the $j^{th}$ feature variable measured for the $i^{th}$ subject, by a set of weight $\beta = [\beta_1, \beta_2, \ldots, \beta_k]^T$ and a constant $\beta_{k+1}$ is shown as follows:

$$\hat{y}_i = x_i^T \beta + \beta_{k+1} \tag{5.3}$$

In matrix form,

$$\hat{y} = X^T \beta + \beta_{k+1} \cdot 1$$

$$= [X^T | 1]\beta$$

$$= \tilde{X}\beta \tag{5.4}$$

where $\hat{y}$ is a column vector and the $j^{th}$ row of $\hat{y}$, $\hat{y}_i$, is the LVO of the $j^{th}$ subject; $\tilde{X}$ is a matrix with the $j^{th}$ row representing the k features of the j-th subject augmented with a column of ones, **1**; and $\beta = [\beta_1, \beta_2, \ldots, \beta_k, \beta_{k+1}]^T$.

The vector of weights, $\beta$, which defines the LVO least square model for the k features, can be calculated by the function as follows:

$$\beta = \tilde{X}^+ \hat{y} \tag{5.5}$$

where $\tilde{X}^+$ is the Moore-Penrose pseudoinverse of the matrix $\tilde{X}$.

The mean-squared error (MSE) between the measured and the estimated variables(LVO and TPR) of the model is calculated as follows:

$$MSE = \frac{1}{k}\sum_1^k (\hat{y}_i - y_i)^2 \tag{5.6}$$

where $y_i$ is the measurement of LVO and TPR for the $i^{th}$ subject. In the next section, the MSE will be used to estimate of how well the estimated LVO, $\hat{y}$, agrees with the LVO measurement.

### 5.3.3 Methods

#### *5.3.3.1 Database*

A group of 89 physiological data were collected from the infants at four different time periods were used in the modeling, as mentioned in section 5.2.1. Intra-arterial BP was measured via an umbilical or peripheral arterial catheter connected to a transducer calibrated to atmospheric barometric pressure and zeroed to the midaxillary point, as mentioned in section 5.2.1.2. The BP signal was continuously acquired via a bedside patient monitor (Philips Agilent Systems, Philip Healthcare, North Ryde, Australia), and recorded by a data acquisition system (ADInstruments, Sydney, Australia) at a sampling rate of 1 kHz. ABP was measured invasively using a radial artery catheter, and the LabChart software (ADInstruments, Sydney, Australia) provided automatic computation MAP and PP from the ABP. Spikes in the beat-to-beat time series due to occurrence of abnormal pulse intervals (e.g. prolonged heart period resulting from a missing beat), which constituted less than 3% of the total number of beats, were removed and replaced by linear interpolation of the beat values immediately preceding and following the replaced beats.

The derivation of various MAP features and their respective meanings have been explained in detail in the section 5.2.1.1.The diastolic decay rate of ABP (1/τ), MAP, PP, MAP × diastolic decay rate of ABP (MAP/τ), PP × HR，HR features are selected to build the model.

#### *5.3.3.2 Features selection*

The squared and cubed value, as well as the logarithms of each feature is added into the feature pool to extend the 6 existing features to a total of 24 features. These transformed features are included to ensure any non-linear relationship between the features and the LVO&TPR is captured by the multivariate model.

To study the usefulness of the diastolic decay rate to estimate LVO and TPR, features not directly derivable from the diastolic decay rate were removed from the feature pool when the feature selection algorithm was executed. The performances of the regression models selected in these two cases were compared to the model found from the complete feature pool.

A stepwise linear regression model was used to map an optimal feature subset to the LVO and TPR score, and the decision to add or remove features is based on their statistical significance in a regression model. The method starts with zero features selected. For each candidate feature that has not been selected, the p-value of a student t-test of its coefficient (null hypothesis is that the coefficient is equal to zero) is calculated. For those features whose p-value is less than the entrance tolerance (α= 0.05), the coefficient with the smallest p-value is selected. Of those features which have been selected, the feature that has the largest p-value and is also larger than the exit tolerance (α= 0.05) is removed. The process will be repeated until no improvement can be made, based on the significance test on the contribution of the candidate features.

### *5.3.3.3 K-fold cross-validation*

To select the best features subsets and validate the performance of the proposed regression model, the available data should be classified into three parts; one part for training, one part for feature selection and one part for testing the performance. However, because of the limited number of subjects available, it is not feasible to classify the data into three groups and expect a reliable measure of performance. If all of the subjects are used in the training phase to obtain the model (weight vector β) which is subsequently used to estimate $y_i$, the MSE in (5.7) will typically be underestimated. This is due to the reason that the model is optimized to reduce the MSE of the training points, but not on unseen data points.

K-fold cross validation is a widely used method for estimating generalised future performance of statistical classification or regression models. It was used to estimate the error rate of the regression model in this study. K-fold method divides all the samples in equal sizes (if possible). The prediction function is learned using k-1 folds, and the fold left out is used for test.

In this study, the data is divided into K=10 partitions containing an equal (or approximately equal) number of subjects. One of the partitions (the $k^{th}$ part, k = 1, 2, 3,…, K) is withheld as test subjects while the remaining K − 1 partitions are used to train the model. The MSE of the estimated $k^{th}$ partition, *MSE (k)*, is calculated. The procedure is repeated for K times and the cross validation error, CVerr, is calculated as the mean of the MSE for all the K parts,

$$CV_{err} = \frac{1}{K}\sum_{k=1}^{K} MSE(k) \tag{5.8}$$

There are N! ways to allocate the subjects before dividing them into K parts. If a proper order was selected, the error rate can be small and misleading. To obtain a fairer approximation of the true error rate, this K-fold cross validation was repeated many times using different randomly selected partitions and the error rate is averaged across all repeats[152]. In the analysis, the repartitioning and cross-validation was repeated 2000 times.

### *5.3.3.4 Model identification and verification*

The degree of agreement between the estimation method and the real measurements of CO and TPR was assessed using the Bland-Altman analysis method. The method has been proposed by Bland and Altman to compare the agreement of two clinical measurement methods in a series of publications [153-156] to determine whether these two methods can be used interchangeably or the new method can replace the established measurement. The basic concept of Bland-Altman's approach is the visualization of the difference of the measurements made by the two methods, then plotting the differences (Y-axis) versus the mean of the two readings (X-axis). The 95% limits of agreement are estimated as the mean difference (also known as bias) ± 1.96×standard deviation (s.d.) of the differences, where 95% of the differences between the measurements are expected to lie.

When the same subjects of two methods are repeatedly measured, the measurements by each method will be distributed about the expected measurement by that method for each subject. The difference between method means might change from subject to subject. In addition, the Bland-Altman analysis technique was enhanced to accommodate the analysis of repeated measurements of each clinical measurement method[156].When repeated measurements are taken using method X and Y, the variance of the difference (which directly affects the width of the limits of agreement) must be adjusted as follows:

$$\sigma_d^2 = s_{\bar{d}}^2 + \left(1 - \frac{1}{m_x}\right)s_{xw}^2 + \left(1 - \frac{1}{m_y}\right)s_{ym}^2 \qquad (5.9)$$

where $s_{\bar{d}}$ is the observed variance of the differences between the within-subject means, $m_x$ and $m_y$ is the number of measurements taken using methods X and Y, respectively, $s_{xw}^2$ and $s_{yw}^2$ are the within-subject variance (repeatability) of measurements taken using method X and Y, respectively. If the s.d. is estimated without adding the last two adjustment terms in equation, the s.d. will be underestimated.

In this study, the repeated runs (2000 times) of the 10-fold cross-validation are assumed to be repeatedly measured from the same subject. If method X and Y are the measurement of LVO&TPR and the multivariate model output, respectively, then $m_x$ =1 and $m_y$ = 2000. The within-subject variance of method Y, $s_{xw}^2$ is calculated by taking the average of the 89 subjects of the estimated LVO&TPR of each subject across 2000 runs. $s_{\bar{d}}$ is the variance of the difference between the average of the estimated LVO&TPR (across 2000runs) and the measurement of LVO&TPR. $\sigma_d^2$ is approximately the sum of $s_{\bar{d}}^2$ and $s_{ym}^2$ .Thus, $\sigma_d^2$ is expected to be large if the average of the 2000 estimations for each subject differs greatly from the real measurement of LVO and TPR or if the estimated output differs greatly across each of the 2000 runs for a single subject. The s.d. is estimated using the square root of the right hand side of equation 5.9, whereby the repeatability of the model is taken into account.

Furthermore, the performance of the model was compared using MSE and goodness of fit ($R^2$) criteria. The $R^2$ is given by:

$$R^2 = 1 - \frac{SSE_{reg}}{SSE_{total}} \qquad (5.10)$$

where SSE is the sum of square of error given by

$$SSE_{reg} = \sum_{i=1}^{N}(\hat{y}_i - \bar{y})^2 \qquad (5.11)$$

$$SSE_{total} = \sum_{i=1}^{N}(y_i - \bar{y})^2 \qquad (5.12)$$

where $\bar{y}$ is the mean value of the real measurement y, $\hat{y}_i$ is the estimated output from the model.

### 5.3.4 Results

Modelling was performed on the LVO and TPR estimations from the 89 measurements from preterm infants. Table 5-1 lists all the features in the feature pool and their associated number, which are used to reference the feature in the following discussions. Table 5-2 shows which features are selected by the feature search algorithm to estimate LVO and TPR, for each of the different starting feature subsets drawn from the entire feature pool. The bias (mean difference) and the precision (s.d. of the differences) of the estimation method when compared against the real measurements were calculated using the Bland-Altman analysis.

Figure 5-8 and Figure 5-9 have shown the plots of the estimated variable plotted against the measured variable, along with the line of equality, for the selected LVO and TPR models, respectively. The Bland-Altman plot of the best LVO and TPR estimations are depicted in Figure 5-10 and Figure 5-11.

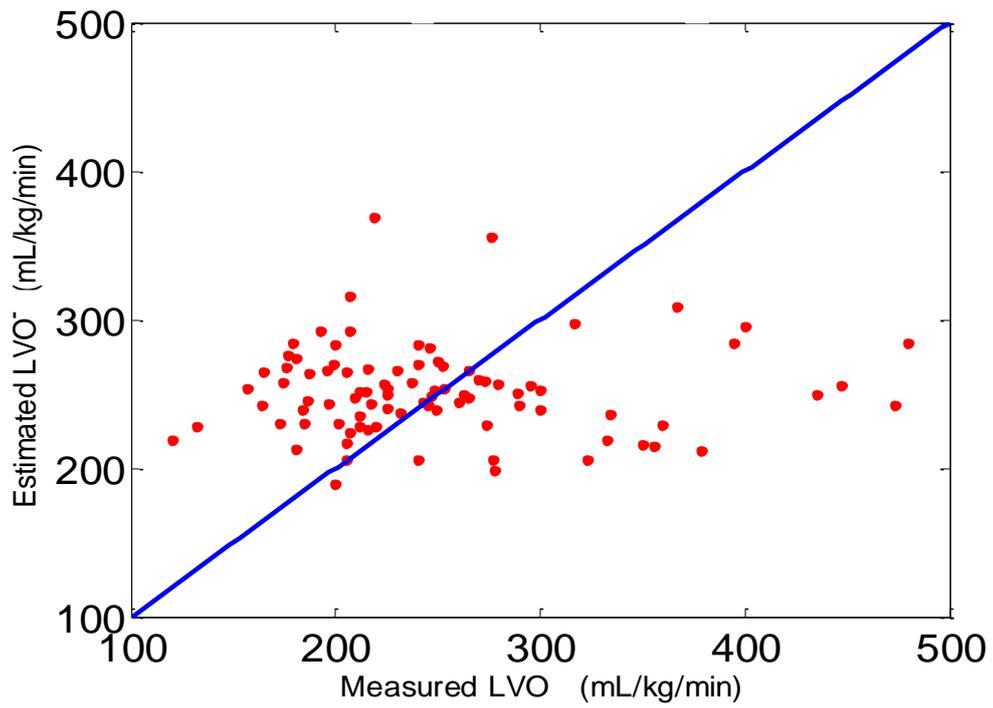

**Figure 5-8:** Plot of estimated LVO against measured LVO for all 89 subjects, with the line of equality. The value of the estimated LVO of each subject is represented by the average of the 2000 repeated estimations of the subject.

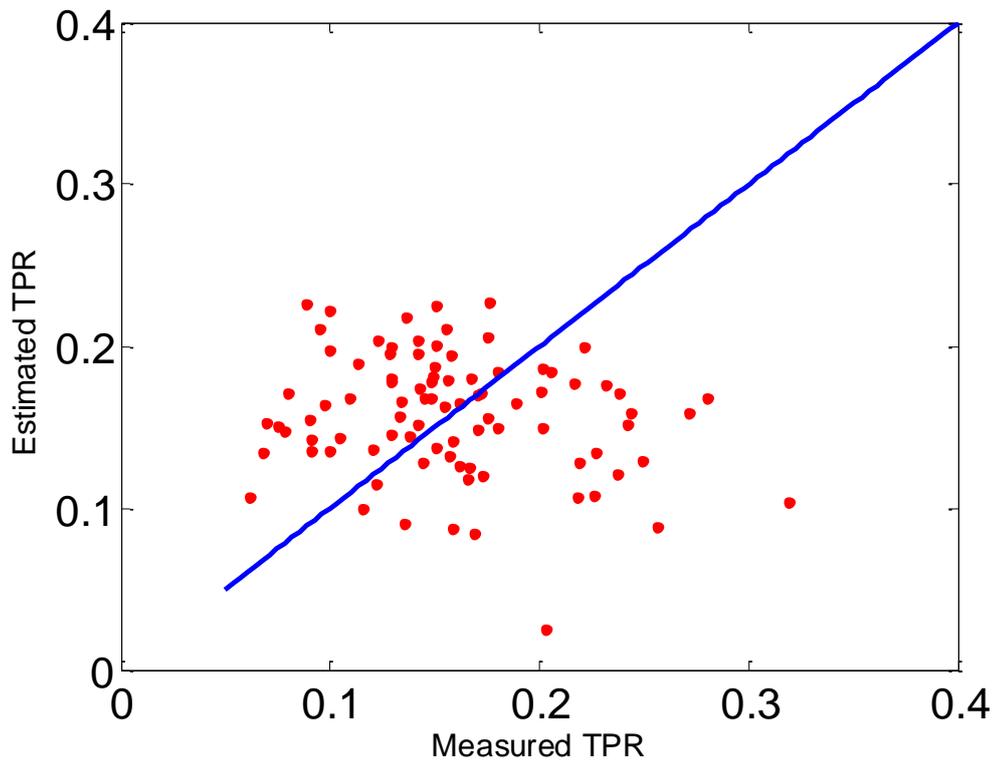

**Figure 5-9:** Plot of estimated TPR against measured TPR for all 89 subjects, with the line of equality. The value of the estimated TPR of each subject is represented by the average of the 2000 repeated estimations of the subject.

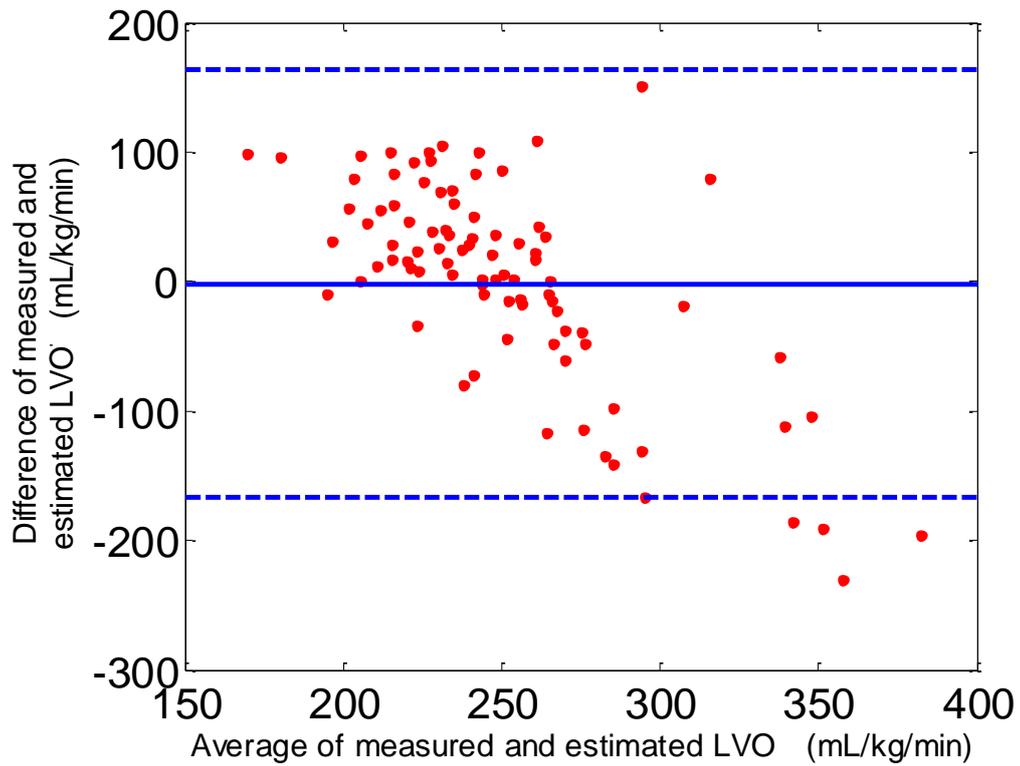

**Figure 5-10:** Bland-Altman plot of the LVO estimation.The solid line in the middle is the bias and the two lines above and below are the limits of agreement, calculated as bias±1.96×s.d.. Bias = -1.91m L/kg/ min, 1.96×s.d. = 161.89 mL/kg/ min.

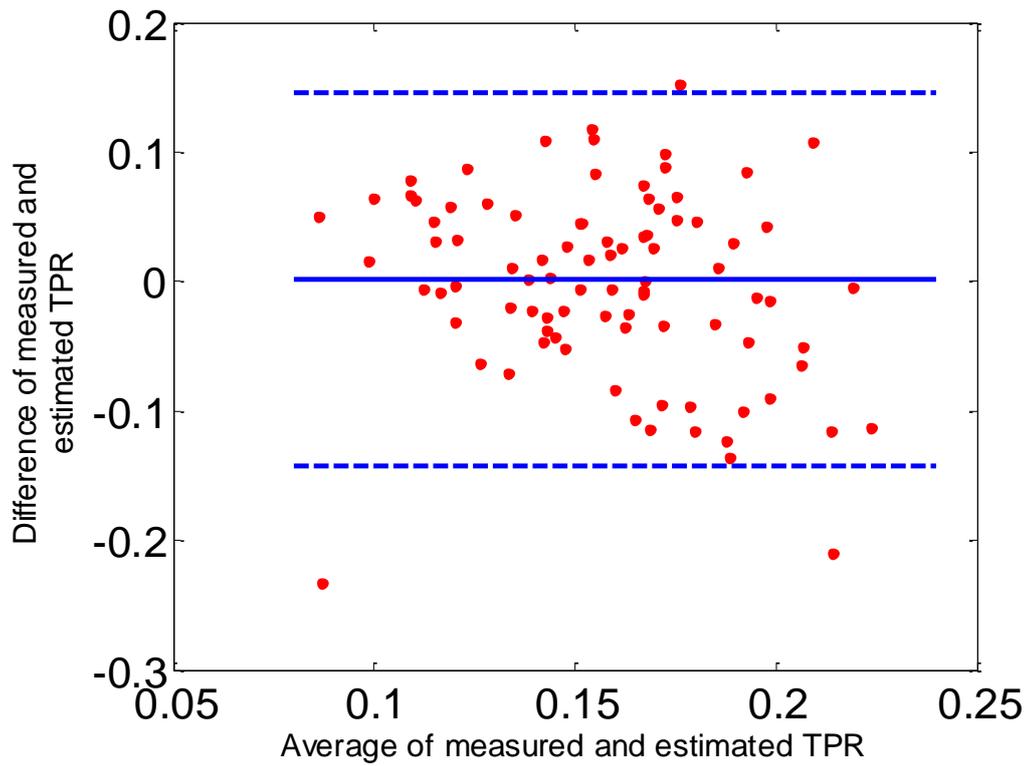

**Figure 5-11:** Bland-Altman plot of the TPR estimation.The solid line in the middle is the bias and the two lines above and below are the limits of agreement, calculated as bias±1.96×s.d.. Bias = 0.0002, 1.96×s.d. = 0.1411.

**Table 5-1:** Feature pool and their respective reference numbers

| Feature | x | $x^2$ | $x^3$ | Log(x) |
|---|---|---|---|---|
| Diastolic decay rate(1/τ) | 1 | 7 | 13 | 19 |
| MAP | 2 | 8 | 14 | 20 |
| HR | 3 | 9 | 15 | 21 |
| Pulse | 4 | 10 | 16 | 22 |
| MAP/τ | 5 | 11 | 17 | 23 |
| PP × HR | 6 | 12 | 18 | 24 |

$1/\tau$: the diastolic decay rate. Each number represents a feature and the transformation (top headings) that is used. For example, feature 2, 14 and 21 MAP, $MAP^2$ and log (HR) respectively.

**Table 5-2: LVO and TPR estimation performance and selected features**

| Estimated variable | $R^2$ | bias | s.d. | Feature pool |
|---|---|---|---|---|
| **LVO (mL/kg/min)** | 0.71 | -1.91 | 82.60 | All features |
|  | 0.72 | -1.77 | 82.80 | Diastolic decay rate features (1,7,13,19,5,11,17,23) |
|  | 0.71 | -1.13 | 83.01 | Only linear features (exclude 7-24) |
| **TPR** | 0.3680 | 0.0015 | 0.0720 | All features |
|  | 0.5013 | 0.0002 | 0.0700 | Diastolic decay rate features (1,7,13,19,5,11,17,23) |
|  | 0.3759 | 0.0007 | 0.0725 | Only linear features (exclude 7-24) |

The features are selected by the stepwise feature selection algorithm to estimate LVO and TPR. The correlation $R^2$, Bland-Altman bias and s.d. achieved is shown.

### 5.3.5 Discussion

In this study, multivariate regression models were developed to estimate LVO and TPR using a combination of features extracted from the PPG waveform, other routine cardiovascular measurements and the non-linear transformations of these features (quadratic and cubic powers and logarithm transform). The bias±precision (1.96×s.d.) for the best LVO model found -1.91±161.89 mL/kg/min, and for the TPR model, the best bias precision achieved was.

**LVO estimation**

For CO estimation, previous study by Wang et al. has found the significant relationship between CO and the features selected from the photoplethysmogram (PPG) and ECG [14]. It has been proposed by Awad et al. to use multi-linear regression analysis to estimate CO from PPG features [157], but the precision was not sufficiently high for the clinical use. More recently, the achievements in the study of CO estimation showed that a reasonable estimate of CO can be obtained using only the PPG waveform. From the results of section 5.2.2, diastolic decay rate of ABP was found to have mild but significant relationships with LVO (r = 0.383, P = 0.0002).It should be noted here this is a direct correlation coefficient with LVO not subjected to the linear regression model. It was also found from this study that the utility of non-linear transformation on the features slightly improved the LVO estimation performance from -1.91± 162.70 mL/kg/min to -1.13±161.90mL/kg/min.

Furthermore, this study revealed that excluding diastolic decay rate of APB related features, i.e. MAP/τ and (1/τ), did not severely reduce the LVO estimation bias and precision (-1.77±162.29). This means that a reasonable estimate of LVO can be obtained using diastolic decay rate related features.

**TPR estimation**

TPR is a useful diagnostic tool of critical illness, for example, an increase in TPR may be a symptom of hemorrhagic shock in ischemic and hemorrhagic stroke patients[131] while a depression in TPR may be observed in anaphylaxis patients. In the preterm infants, low TPR are symptoms of sepsis, which is a common cause of mortality and morbidity in preterm infants, particular those with very low birth weight [22].Therefore, continuous monitoring of TPR has been suggested as a diagnostic and research tool to assist the vasoactive drug administration[158].The previous study has demonstrated the feasibility of deploying a multivariate regression to estimate SVR in the clinical diagnosis, using a non-invasive PPG signal.

In this study, a quantitative estimation of TPR with bias±precision of 0.0007±0.0.1421 was achieved using the multivariate regression model. If only the related features of diastolic decay rate of ABP were used in the model, the bias±precision performance has been slightly improved to 0.0002±0.1372.

### 5.3.6 Limitations

One of the major limitations in this study is the ABP signal collection. The quality degradation caused by baseline drift, movement artefact and frequent ectopic beats accounting for the weak and unrecognizable BP pulses, which limited the length of data and the size of the data pool.

The small data size was challenged in developing the multivariate model and the validation for estimating generalised performance of the regression models. Ideally, the model should be trained using as many samples as possible to reflect the true population, and the model should be tested using samples not which cannot be used in the training phase. In this study, this problem was addressed by repeatedly running 10-fold cross-validation to obtain a fair estimate of the generalised model performance.

### 5.3.7 Summary

This study provides a preliminary indication of the potential usefulness of a method to non-invasively estimate LVO and TPR using ABP signal. The limited results obtained encourage further research to validate the method in a larger cohort, ultimately enabling a non-invasive, convenient and low-cost alternative for estimating or monitoring LVO and TPR.

# Chapter 6. Conclusions and Future Work

## 6.1 Summary

This chapter concludes all of the research work that has been accomplished in this PhD project, which includes the development of cardiac output monitoring tools and signal processing methods for the clinical investigation of intra-ventricular hemorrhage and other clinical risks of preterm infants. The performance of these new findings was analyzed based on data collected from actual patients, computer simulations and human experiments.

The overall structure of the paper consisted of two main parts: The first part was based on developing monitoring tools to detect the IVH and other clinical risks.The use of spectral indices and detrended fluctuation analysis results, which were derived from the heart rate variability signals, the blood pressure variability and the cerebral near-infrared spectroscopy measures, to monitor the pathophysiologic response in cerebral cardiovascular control mechanisms produced intriguing results, showing distinctively difference between preterm infants with and without IVH . In fact, clinical investigation on IVH infants provided further supports to the findings. The outcomes of this research project may represent an important step towards a more thorough understanding of IVH progression, and may have substantial clinical implications of preterm infants.

Based on the previous CO monitoring of the preterm infants, the second part of the paper developed an initial attempt to apply arterial blood pressure waveform analysis to continuously monitor clinical risks in the preterm infants by estimating LVO and TPR, which produced encouraging regression results.

The main findings of this PhD project are described in greater detail in the following section.

## 6.2 Major findings of the current work

- **Chapter 3**

    In this chapter, spectral and cross-spectral analysis was performed on HRV, BPV and NIRS measures such as HbO2, HHb, and tissue oxygenated index (TOI). In addition, indices related to cardiac baroreflex sensitivity and cerebral autoregulation were derived from the very low, low and mid frequency ranges (VLF, LF and MF). It was demonstrated that moderate correlations with CRIB II were identified from MAP normalised MF power, LF MAP-HHb coherence ,TOI VLF percentage power and LF baroreflex gain, with the latter two parameters also highly correlated with gestational age. The relationships between CRIB II and various spectral measures of arterial baroreflex and cerebral autoregulation functions have provided further justification for these measures as possible markers of clinical risks and predictors of adverse outcome in preterm infants.

    Also in the same demonstrated that the IVH infants were found to have significantly higher TOI, lower FTOE and lower coherence between arterial BP and HHb in the very low frequency range. These parameters can be utilised for non-invasive cerebral circulation monitoring of preterm infants and may help to completely understand the clinical utility of these NIRS measures for early identification of IVH.

- **Chapter 4**

To further investigate potential method for clinical risk assessment, this paper utilized the detrended fluctuation analysis to quantify the fractal correlation properties of physiological signals. It has been examined whether DFA of the ABP signal could derive markers for the identification of preterm infants who developed IVH. DFA method was performed on the beat-to-beat sequences of MAP, SBP and pulse interval, with short-term exponent ($α_1$, for time scale of 4-15 beats) and long-term exponent ($α_2$, for time scale of 15-50 beats) computed accordingly. The IVH infants were found to have higher short-term scaling exponents of both MAP and SBP compared to the non-IVH. These identified scaling exponents of MAP and SBP have important implications on that fractal dynamics embedded in the arterial pressure waveform could provide useful information that facilitates early identification of IVH in preterm infants.

- **Chapter 5**

Lastly, this study investigated whether ABP waveform analysis could be useful for estimating left ventricular outflow and total peripheral resistance in preterm infants. The diastolic decay rate ($1/τ$) was obtained via fitting an exponential function to the last one third of each arterial pulse, with the mean rate computed from 50 pulses selected from each infant. This decay rate was considered to be inversely related to TPR while positively related to LVO. The results of regression analysis have confirmed that the diastolic decay rate had significant positive and negative relationships with LVO and TPR respectively. These preliminary results demonstrated the feasibility and applicability of arterial pressure waveform analysis for continuously estimating LVO and TPR in preterm infants.

## 6.3 Future research directions

In this section, some directions for future work extending this paper are presented:

**Chapter 3**

- The cerebral NIRS measurement based monitoring tool proposed in the current work has shown a number of parameters that were potentially distinguishing between preterm infants with or without brain haemorrhage, further studies with larger sample size are warranted for a more complete understanding of the clinical utility of these NIRS measures for early identification of IVH infants.

- The relationships between CRIB II and various spectral measures of arterial baroreflex and cerebral autoregulation functions have provided further justification for these measures as possible markers of clinical risks and predictors of adverse outcome in preterm infants. Future studies can focus on relating more clinical measure to the CRIB II or other established clinical risk score to support the diagnosis of serious illness in preterm infants.

**Chapter 4**

- The results showed that the IVH infants had higher short-term scaling exponents of both MAP and SBP compared to the non-IVH infants. Thus, fractal dynamics in arterial BP could provide potentially useful information that facilitates early identification of IVH in preterm infants. Further studies should include more preterm infants with IVH to assess how the DFA exponents relate to the severity of haemorrhage.

- Based on the previous literature, it is assumed that other nonlinear analysis of NIRS measures, BP signals will provide potentially useful information to classify IVH in preterm infants.

- DFA methods can be applied in conjunction with the commonly used spectral analysis method to provide a more complete picture of the

underlying cardiovascular dynamics, as both methods can provide complementary information.

- In this study, with the limited number of infants without mechanical ventilation for statistical analysis, whether the ventilatory mode had any direct impact on the results would need to be addressed by further studies with larger sample size.

- The performance of the proposed cardiovascular indices from DFA method and multi-parameter classification model in the classification of IVH should be assessed in a prospective clinical trial, with the inclusion of larger sample size of preterm infants in our cohort.

- Advanced dynamical models can be used for estimation of transient behaviour of blood pressure and other physiological variables in real time [159-163], which might be potentially applied to prevent IVH in preterm infants.

**Chapter 5**

- Multivariate regression model in the estimation of LVO and TPR should be further assessed with the inclusion of larger sample size study, to validate the usefulness of monitoring in clinical decision-making.

- Improve the input features of the regression model, such as frequency analysis and DFA methods results to build the control system to manage the mechanical ventilation in the neonatal ICU.

- A promising new research direction is the estimation of LVO and TPR by applying a method of robust Kalman filtering [164-170].

The suggestions above are examples of possible future research based on the results presented in this paper.